\documentclass[useAMS, usenatbib]{mn2e}
\usepackage{color}
\usepackage{epsfig}
\usepackage{epstopdf}
\usepackage{amsmath, amssymb}
\usepackage{hyperref}
\usepackage{wasysym} 
\usepackage{subfigure}
\usepackage{multirow}
\usepackage{lipsum}
\usepackage{graphicx}
\usepackage{ulem}
\usepackage{mn2e-breakabs}
\graphicspath{{/Users/fbeutler/github/BOSS_DR12/DR12BAO/images/}}

\usepackage{setspace}  

\newlength{\plotwidth}
\newlength{\fullwidth}
\newcommand{\ihMpc}{h{\rm\;Mpc^{-1}}}
\newcommand{\hMpc}{h^{-1}{\rm\;Mpc}}

\newcommand{\FAP}{F_{\rm}}

\newcommand{\vd}{\mathbf{d}}
\newcommand{\vt}{\mathbf{t}}
\newcommand{\vC}{\mathbf{C}}
\newcommand{\vtheta}{\mathbf{\theta}}

\newcommand{\zeff}{z_{\rm eff}}
\def\vc#1{{\bf#1}}
\def\prd{Phys.\ Rev.\ D}
\def\apj{Astrophys.\ J.}
\def\mnras{Mon.\ Not.\ Roy.\ Astron.\ Soc.}
\renewcommand{\topmargin}{-1cm} 

\pagerange{\pageref{firstpage}--\pageref{lastpage}} \pubyear{2013}

\def\LaTeX{L\kern-.36em\raise.3ex\hbox{a}\kern-.15em
    T\kern-.1667em\lower.7ex\hbox{E}\kern-.125emX}

\title[BOSS: Fourier-space analysis of BAO]{The clustering of galaxies in the completed SDSS-III Baryon Oscillation Spectroscopic Survey: Baryon Acoustic Oscillations in Fourier-space}
\author[Florian Beutler et al.]
{\parbox{\textwidth}{Florian Beutler$^{1,2}$\thanks{E-mail: \texttt{florian.beutler@port.ac.uk}}, Hee-Jong Seo$^{3}$, Ashley J. Ross$^{4, 1}$, Patrick McDonald$^{2}$, Shun Saito$^{5,6}$, Adam S. Bolton$^{7, 8}$, Joel R. Brownstein$^{9}$, Chia-Hsun Chuang$^{10, 11}$, Antonio J. Cuesta$^{12}$, Daniel J. Eisenstein$^{13}$, Andreu Font-Ribera$^{2, 5}$, Jan Niklas Grieb$^{14, 15}$, Nick Hand$^{16}$, Francisco-Shu Kitaura$^{11}$, Chirag Modi$^{17}$, Robert C. Nichol$^{1}$, Will J. Percival$^{1}$, Francisco Prada$^{10,18,19}$, Sergio Rodriguez-Torres$^{10,18}$, Natalie A. Roe$^{2}$, Nicholas P. Ross$^{20}$, Salvador Salazar-Albornoz$^{14, 15}$, Ariel G. S\'anchez,$^{15}$, Donald P. Schneider$^{21, 22}$, An\v{z}e Slosar$^{23}$, Jeremy Tinker$^{24}$, Rita Tojeiro$^{25}$, Mariana Vargas-Maga\~na$^{26}$, Jose A. Vazquez$^{23}$}\vspace{0.4cm}\\
\parbox{\textwidth}{
$^{1}$Institute of Cosmology \& Gravitation, Dennis Sciama Building, University of Portsmouth, Portsmouth, PO1 3FX, UK\\
$^{2}$Lawrence Berkeley National Lab, 1 Cyclotron Rd, Berkeley CA 94720, USA\\
$^{3}$Department of Physics and Astronomy, Ohio University, 251B Clippinger Labs, Athens, OH 45701, USA\\
$^{4}$Department of Physics, Ohio State University, 140 West 18th Avenue, Columbus, OH 43210, USA\\
$^{5}$Kavli Institute for the Physics and Mathematics of the Universe (WPI),The University of Tokyo Institutes for Advanced Study, The University of Tokyo, Kashiwa, Chiba 277-8583, Japan\\
$^{6}$Max-Planck-Institut f\"{u}r Astrophysik, Karl-Schwarzschild-Strasse 1, D-85740 Garching bei M\"{u}nchen, Germany\\
$^{7}$Department of Physics and Astronomy, University of Utah, 115 South 1400 East, Salt Lake City, UT 84112 USA\\
$^{8}$National Optical Astronomy Observatory, 950 N Cherry Ave, Tucson, AZ 85719 USA\\
$^{9}$Department of Physics and Astronomy, University of Utah, 115 S 1400 E, Salt Lake City, UT 84112, USA\\
$^{10}$Instituto de F\'isica Te\'orica, (UAM/CSIC), Universidad Aut\'onoma de Madrid, Cantoblanco, E-28049 Madrid, Spain\\
$^{11}$Leibniz-Institut f\"ur Astrophysik Potsdam (AIP), An der Sternwarte 16, D-14482 Potsdam, Germany\\
$^{12}$Institut de Ci{\`e}ncies del Cosmos (ICCUB), Universitat de Barcelona (IEEC-UB), Mart{\'\i} i Franqu{\`e}s 1, E08028 Barcelona, Spain\\
$^{13}$ Harvard-Smithsonian Center for Astrophysics, 60 Garden St., Cambridge, MA 02138, USA\\
$^{14}$Universit\"ats-Sternwarte M\"unchen, Ludwig-Maximilians-Universit\"at M\"unchen, Scheinerstra\ss{}e 1, 81679 M\"unchen, Germany\\
$^{15}$Max-Planck-Institut f\"ur extraterrestrische Physik, Postfach 1312, Giessenbachstr., 85741 Garching, Germany\\
$^{16}$Department of Astronomy, University of California Berkeley, CA 94720, USA\\
$^{17}$Department of Physics, University of California Berkeley, CA 94720, USA\\
$^{18}$Campus of International Excellence UAM+CSIC, Cantoblanco, E-28049 Madrid, Spain\\
$^{19}$Instituto de Astrof\'{\i}sica de Andaluc\'{\i}a (CSIC), Glorieta de la Astronom\'{\i}a, E-18080 Granada, Spain \\
$^{20}$Institute for Astronomy, University of Edinburgh, Royal Observatory, Edinburgh EH9 3HJ, UK\\
$^{21}$Department of Astronomy and Astrophysics, The Pennsylvania State University, University Park, PA 16802\\
$^{22}$Institute for Gravitation and the Cosmos, The Pennsylvania State University, University Park, PA 16802\\
$^{23}$Brookhaven National Laboratory, Upton, NY 11973, USA\\
$^{24}$Center for Cosmology and Particle Physics, Department of Physics, New York University, 4 Washington Place, New York, NY 10003, USA\\
$^{25}$School of Physics and Astronomy, University of St. Andrews, St Andrews, Fife, KY16 9SS, UK\\
$^{26}$Instituto de Fisica, Universidad Nacional Autonoma de Mexico, Apdo. Postal 20-364, Mexico.\vspace{1cm}}}

\begin{document}

\label{firstpage}
\maketitle

\begin{abstract}
We analyse the Baryon Acoustic Oscillation (BAO) signal of the final Baryon Oscillation Spectroscopic Survey (BOSS) data release (DR12). Our analysis is performed in Fourier-space, using the power spectrum monopole and quadrupole. The  dataset includes $1\,198\,006$ galaxies over the redshift range $0.2 < z < 0.75$. We divide this dataset into three (overlapping) redshift bins with the effective redshifts $\zeff = 0.38$, $0.51$ and $0.61$. We demonstrate the reliability of our analysis pipeline using N-body simulations as well as $\sim 1000$ MultiDark-Patchy mock catalogues, which mimic the BOSS-DR12 target selection. We apply density field reconstruction to enhance the BAO signal-to-noise ratio. By including the power spectrum quadrupole we can separate the line-of-sight and angular modes, which allows us to constrain the angular diameter distance $D_A(z)$ and the Hubble parameter $H(z)$ separately. We obtain two independent $1.6\%$ and $1.5\%$ constraints on $D_A(z)$ and $2.9\%$ and $2.3\%$ constraints on $H(z)$ for the low ($\zeff=0.38$) and high ($\zeff=0.61$) redshift bin, respectively. We obtain two independent $1\%$ and $0.9\%$ constraints on the angular averaged distance $D_V(z)$, when ignoring the Alcock-Paczynski effect. The detection significance of the BAO signal is of the order of $8\sigma$ (post-reconstruction) for each of the three redshift bins. Our results are in good agreement with the Planck prediction within $\Lambda$CDM. This paper is part of a set that analyses the final galaxy clustering dataset from BOSS. The measurements and likelihoods presented here are combined with others in~\citet{Alam2016} to produce the final cosmological constraints from BOSS.
\end{abstract}


\section{introduction}
\label{sec:intro}

The baryon acoustic oscillation (BAO) signal in the distribution of galaxies is an imprint of primordial sound waves that have propagated in the very early Universe through the plasma of tightly coupled photons and baryons (e.g.~\citealt{Peebles:1970ag, Sunyaev:1970eu}). The corresponding BAO signal in photons has been observed in the Cosmic Microwave Background (CMB) and has revolutionised cosmology in the last two decades~\citep[e.g.,][]{Ade:2015xua}. 

The BAO signal has a characteristic physical scale that represents the distance that the sound waves have traveled before the epoch of decoupling.
In the distribution of galaxies, the BAO scale is measured in angular and redshift coordinates, and this observational metric is related to the physical coordinates through the angular diameter distances and Hubble parameters, which in turn depend on the expansion history of the Universe. Therefore, comparing the BAO scale measured in the distribution of galaxies with the true physical BAO scale, i.e., the sound horizon scale that is independently measured in the CMB, allows us to make cosmological distance measurements to the effective redshift of the distribution of galaxies. With this ``standard ruler'' technique one can map the expansion history of the Universe~\citep[e.g.,][]{Hu:1996,Eisen:2003,Blake:2003rh,Linder:2005in,Hu:2003,Seo:2003pu}.

While the BAO feature itself can be isolated from the broadband shape of any galaxy clustering statistic quite easily due to its distinct signature, it is still subject to several observational and evolutionary non-linear effects, which damp and shift the BAO feature, thereby biasing such a measurement if ignored~\citep[e.g.,][]{Meiksin:1999,Seo:2005,Crocce:2007dt,Seo:2009,Matsubara:2008wx,Mehta:2011xf,Taruya:2009ir}. In redshift space, the signal-to-noise ratio of the power spectrum is boosted along the line-of-sight due to the linear Kaiser factor $(1+\beta\mu^2)^2$~\citep{Kaiser:1987qv}, but also suffers the non-linear redshift-space distortion effects, which cause additional smearing of the BAO feature along the line of sight.

The BAO method has been significantly strengthened by~\citet{Eisenstein:2006nk}, who showed that non-linear degradation effects are reversible by undoing the displacements of galaxies due to bulk flow that are the very cause of the structure growth and  redshift-space distortions. This density field reconstruction technique has been tested against simulations and adopted in current galaxy survey data analyses (e.g.~\citealt{Padmanabhan:2012hf}). In this paper, we will apply this technique.

The galaxy BAO signal was first detected in the 
SDSS-LRG~\citep{Eisenstein:2005su} and 
2dFGRS~\citep{Percival:2001hw, Cole:2005sx} samples. The WiggleZ survey 
extended these early detections to higher 
redshifts~\citep{Blake:2011en,Kazin:2014qga}, while the 6dFGS survey measured 
the BAO signal at $z=0.1$~\citep{Beutler:2011hx}. Recently the BAO detection 
in the SDSS main sample at $z = 0.15$ was reported in~\cite{Ross:2014qpa}. The 
first analysis of the Baryon Oscillation Spectroscopic Survey (BOSS) dataset 
in DR9~\citep{Anderson:2012sa} presented a $1.7\%$ constraint on the angular 
averaged distance to $z=0.57$, which has been improved to $1\%$ with 
DR11~\citep{Anderson:2013zyy}. The LOWZ sample of BOSS has been used 
in~\citet{Tojeiro:2014eea} to obtain a $2\%$ distance constraint.

While the BAO technique has now been established as a standard tool for cosmology, the anisotropic Fourier-space analysis has been difficult to implement because of the treatment of the window function. To simplify the window function treatment it often has been assumed that the window function is isotropic, which simplifies its treatment considerably. However, the window functions of most galaxy surveys are anisotropic and this can introduce anisotropies by re-distributing power between the multipoles and potentially bias cosmological measurements. The first self-consistent Fourier-space analysis which does not put such assumptions on the window function was presented in~\citet{Beutler:2013yhm} using BOSS-DR11, which focused on constraining redshift-space distortions and the Alcock-Paczynski effect. Here we follow this earlier analysis with a few modifications and present a BAO-only analysis, marginalising over the broadband power spectrum shape. Our companion paper~\citep{Beutleretal2} (from now on B16) goes beyond BAO, studying the additional cosmological information of redshift-space distortions.

This paper uses the combined data from BOSS-LOWZ and BOSS-CMASS, covering the redshift range from $z=0.2$ to $z=0.75$. The dataset also includes additional data from the so called 'early regions', which have not been included before (see~\citealt{Alam2016} for details). BAO measurements obtained using the monopole and quadrupole correlation function are presented in~\citet{Ross2016}, while~\citet{Vargas-Magana2016} diagnoses the level of theoretical systematic uncertainty in the BOSS BAO measurements. Measurements of the rate of structure growth from the RSD
signal are presented in~\citet{Beutleretal2},~\citet{Grieb:2016},~\citet{Sanchez2016} and~\citet{Satpathy2016}. \citet{Alam2016} combines the results of these seven papers (including this work) into a single likelihood that can be used to test cosmological models.

The paper is organised as follows. In \S~\ref{sec:data}, we introduce the BOSS DR12 dataset. In \S~\ref{sec:estimator}, we present our anisotropic power spectrum estimator, followed by a description of our window function treatment in \S~\ref{sec:win}. In \S~\ref{sec:mocks} we present the MultiDark-Patchy mock catalogues, which are used to obtain a covariance matrix, and in \S~\ref{sec:model} we introduce our power spectrum model. In \S~\ref{sec:sys} we test our power spectrum model using N-body simulations and the MultiDark-Patchy mock catalogues. In \S~\ref{sec:analysis} we present the data analysis, followed by a discussion of the results in \S~\ref{sec:dis}. We conclude in \S~\ref{sec:conclusion}.

The fiducial cosmological parameters which are used to convert the observed angles and redshifts into co-moving coordinates and to generate linear power spectrum models as input for the power spectrum templates, follow a flat $\Lambda$CDM model with $\Omega_m=0.31$, $\Omega_bh^2=0.022$, $h=0.676$, $\sigma_8=0.824$, $n_s=0.96$, $\sum m_{\nu} = 0.06\,$eV and $r_s^{\rm fid} = 147.78\,$Mpc.

\section{The BOSS DR12 dataset}
\label{sec:data}

In this analysis we use the final data release (DR12) of the Baryon Oscillation Spectroscopic Survey (BOSS) dataset. The BOSS survey is part of SDSS-III~\citep{Eisenstein:2011sa,Dawson:2012va} and used the SDSS multi-fibre spectrographs~\citep{Bolton:2012hz,Smee:2012wd} to measure spectroscopic redshifts of $1\,198\,006$ million galaxies. The galaxies were selected from multicolour SDSS imaging~\citep{Fukugita:1996qt,Gunn:1998vh,Smith:2002pca,Gunn:2006tw,Doi:2010rf,Reid:2015gra} over $10\,252\deg^2$ divided in two patches on the sky and cover a redshift range of $0.2$ - $0.75$. In our analysis we split this redshift range into three overlapping redshift bins defined by $0.2 < z < 0.5$, $0.4 < z < 0.6$ and $0.5 < z < 0.75$ with the effective redshifts $\zeff = 0.38$, $0.51$ and $0.61$\footnote{The effective redshifts are calculated as the weighted average over all galaxies (see e.g. eq. 67 in~\citealt{Beutler:2013yhm})}, respectively. 

We include three different incompleteness weights to account for shortcomings of the BOSS dataset (see~\citealt{Ross:2012qm} and~\citealt{Anderson:2013zyy} for details): a redshift failure weight, $w_{\rm rf}$, a fibre collision weight, $w_{\rm fc}$ and a systematics weight, $w_{\rm sys}$, which is a combination of a stellar density weight and a seeing condition weight. Each galaxy is thus counted as 
\begin{equation}
w_c = (w_{\rm rf} + w_{\rm fc} - 1)w_{\rm sys}.
\end{equation} 
More details about these weights and their effect on the DR12 sample can be found in~\citet{Ross2016}.

\section{The power spectrum estimator}
\label{sec:estimator}

We employ the Fast Fourier Transform (FFT)-based anisotropic power spectrum estimator suggested by~\citet{Bianchi:2015oia} and~\citet{Scoccimarro:2015bla}. This method divides the power spectrum estimate in its spatial components, which can then be calculated using 3D Fourier transforms. While this technique requires multiple FFTs, it still provides the computational complexity of $\mathcal{O}(N\log(N))$, significantly faster than a straight forward pair counting analysis, which would result in $\mathcal{O}(N^2)$, where $N$ is the number of cells in the 3D Cartesian grid in which the data and random galaxies are binned.

\begin{figure*}
\begin{center}
\epsfig{file=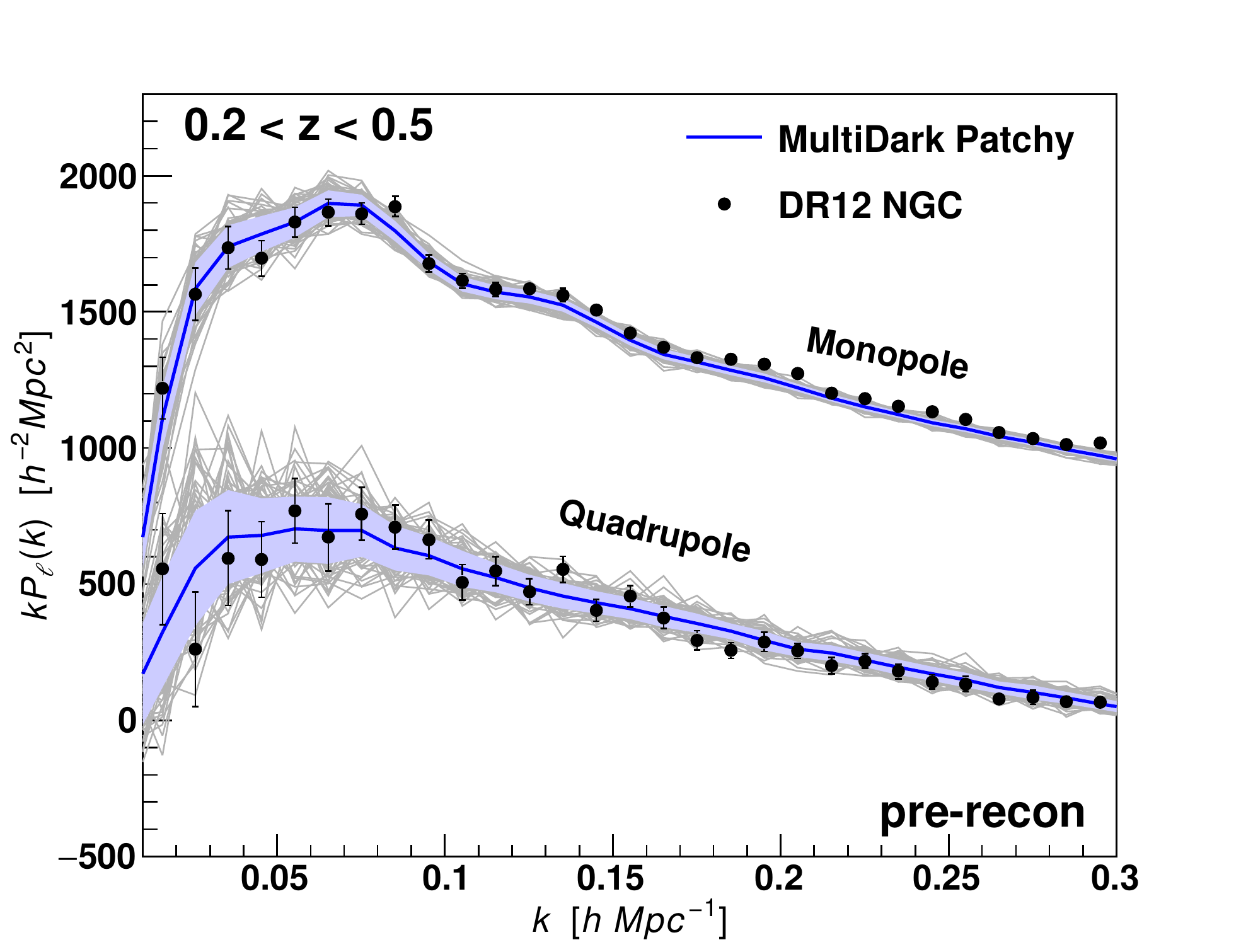,width=5.8cm}
\epsfig{file=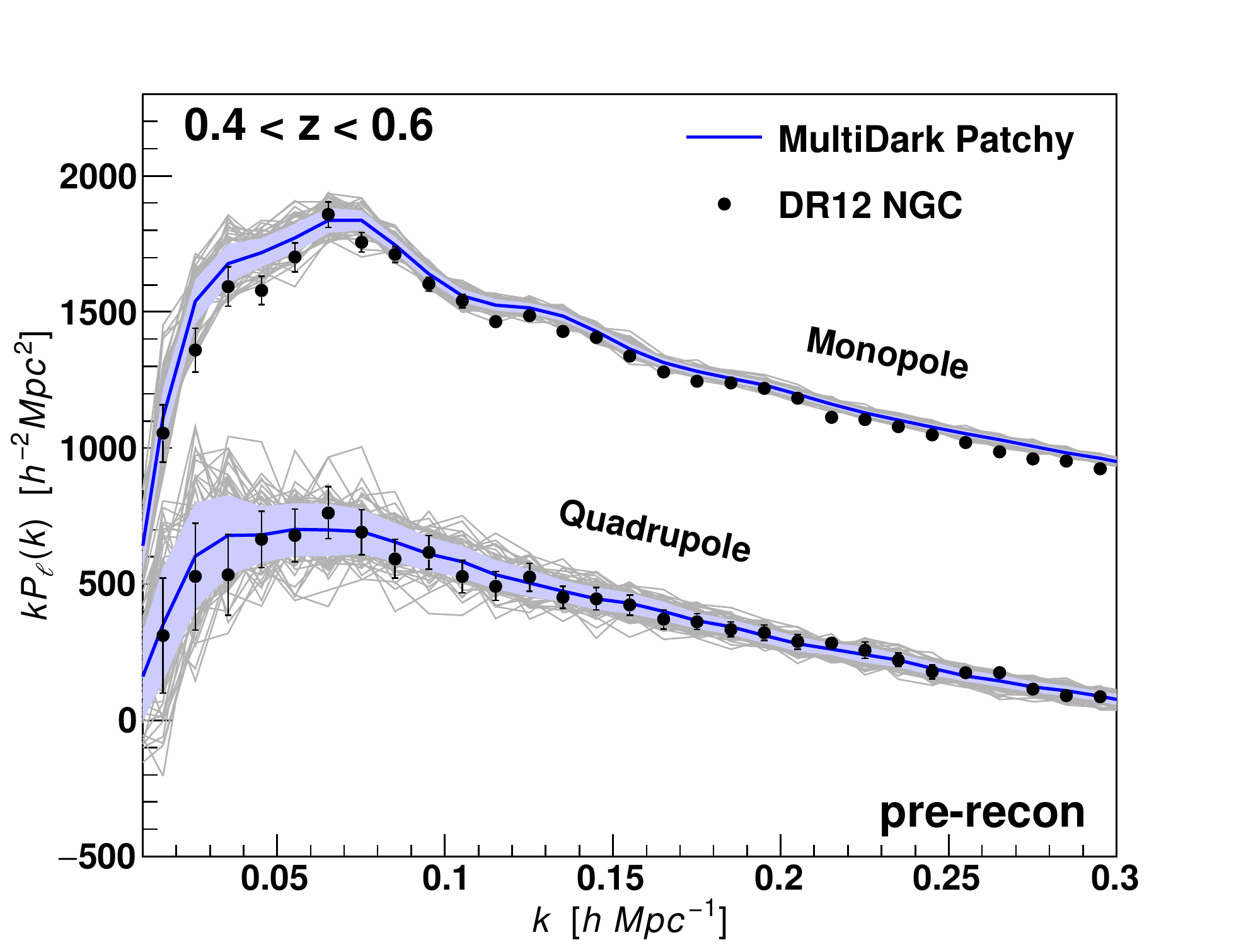,width=5.8cm}
\epsfig{file=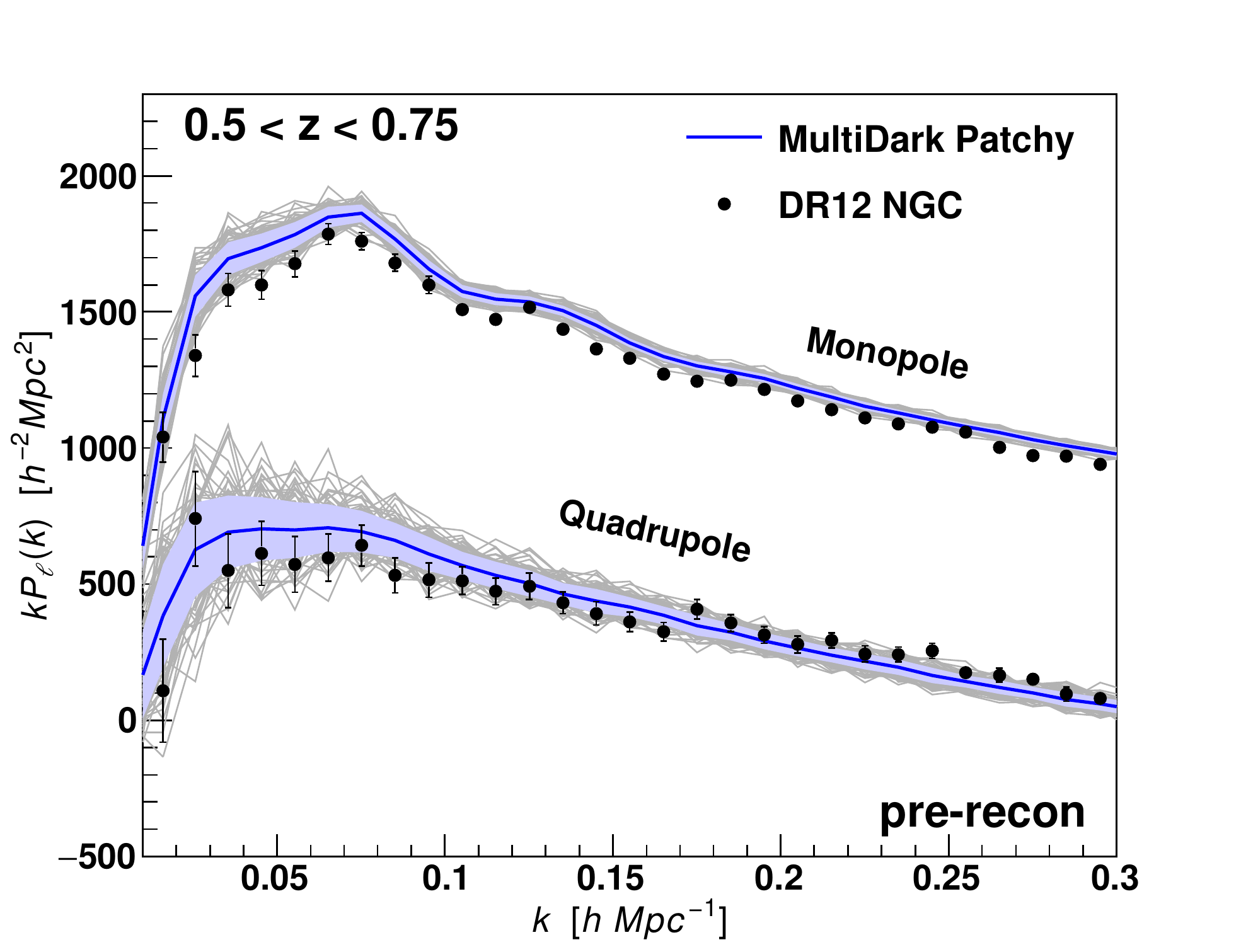,width=5.8cm}\\
\epsfig{file=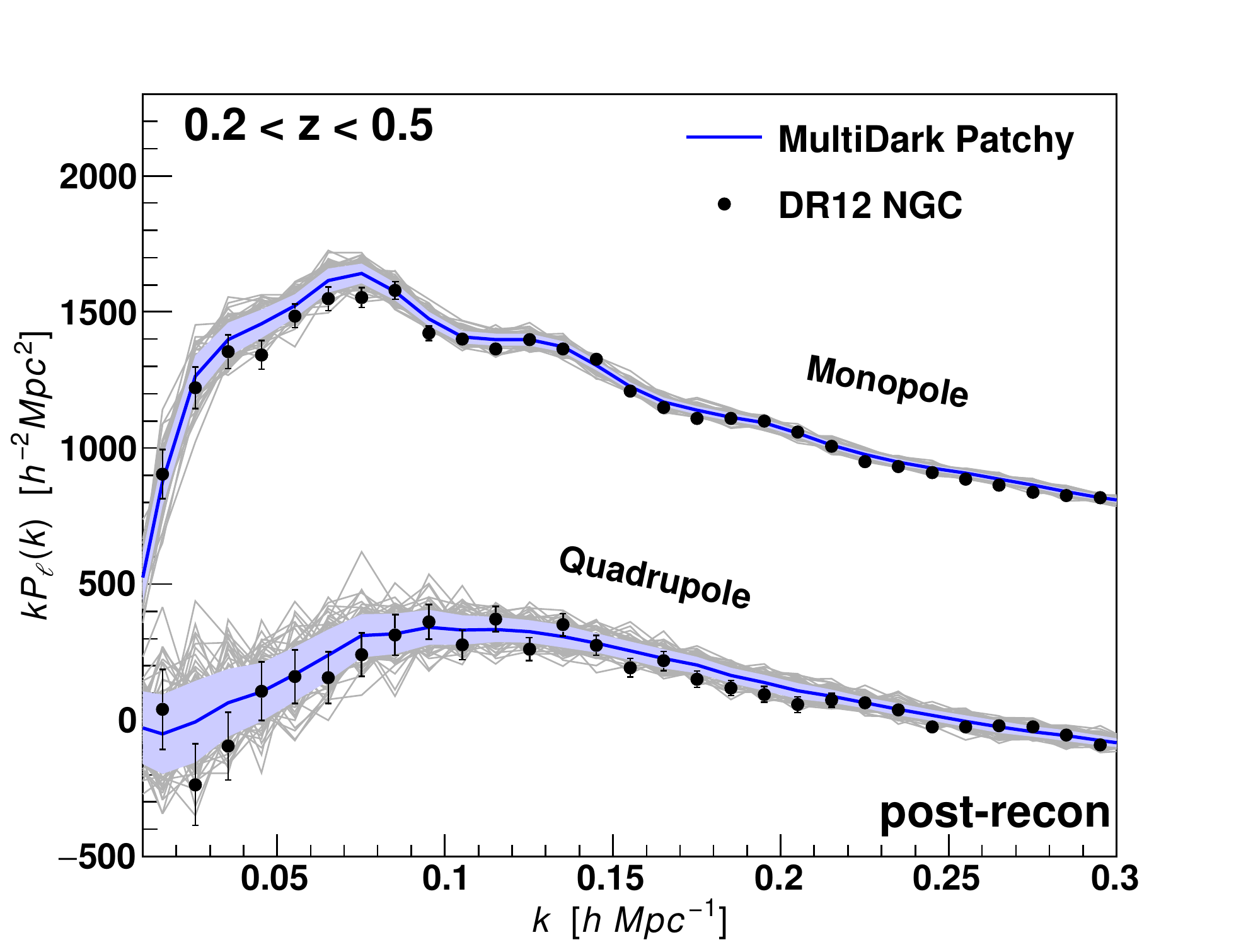,width=5.8cm}
\epsfig{file=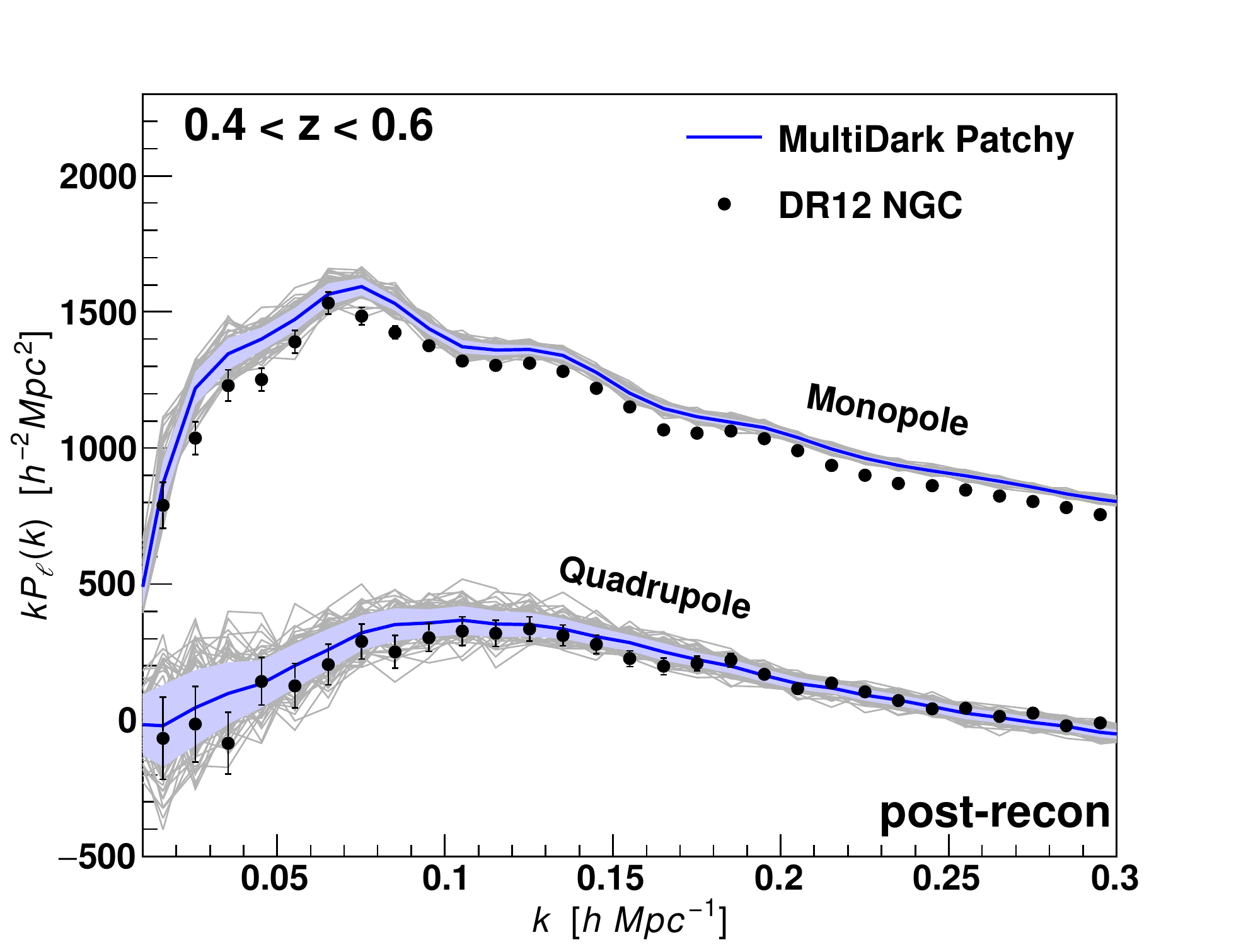,width=5.8cm}
\epsfig{file=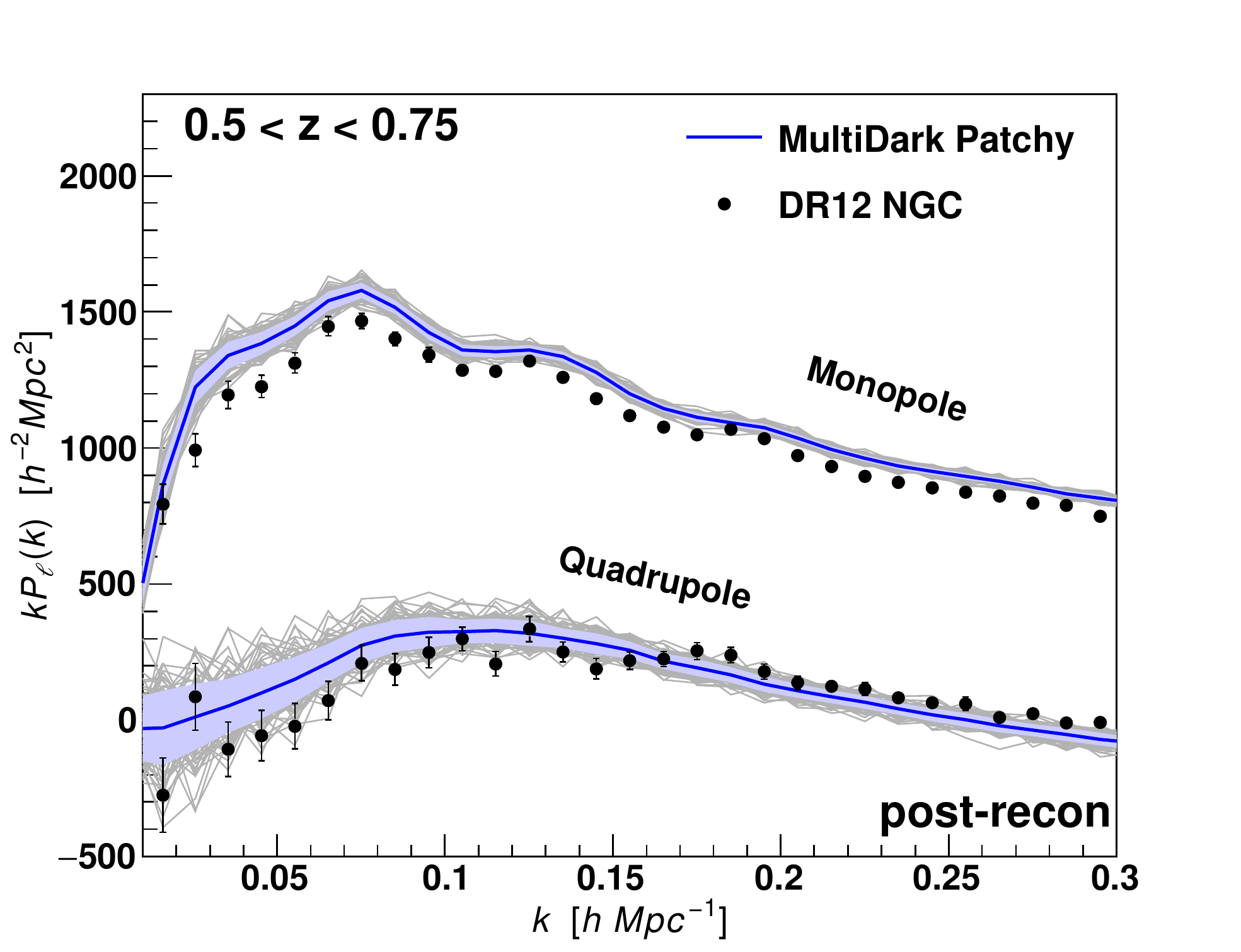,width=5.8cm}
\caption{BOSS DR12 power spectra in the North Galactic Cap (NGC) for the three redshift bins used in this analysis. The panels in the top row show the power spectra before density field reconstruction, while the bottom row displays the power spectra after density field reconstruction. The blue line indicates the mean of the $2045$ (pre-recon) and $996$ (post-recon) MultiDark-Patchy mock catalogues, while the blue shaded area shows the r.m.s. between them. The errors on the data points are the diagonal of the covariance matrix.}
\label{fig:psNGC}
\end{center}
\end{figure*}

\begin{figure*}
\begin{center}
\epsfig{file=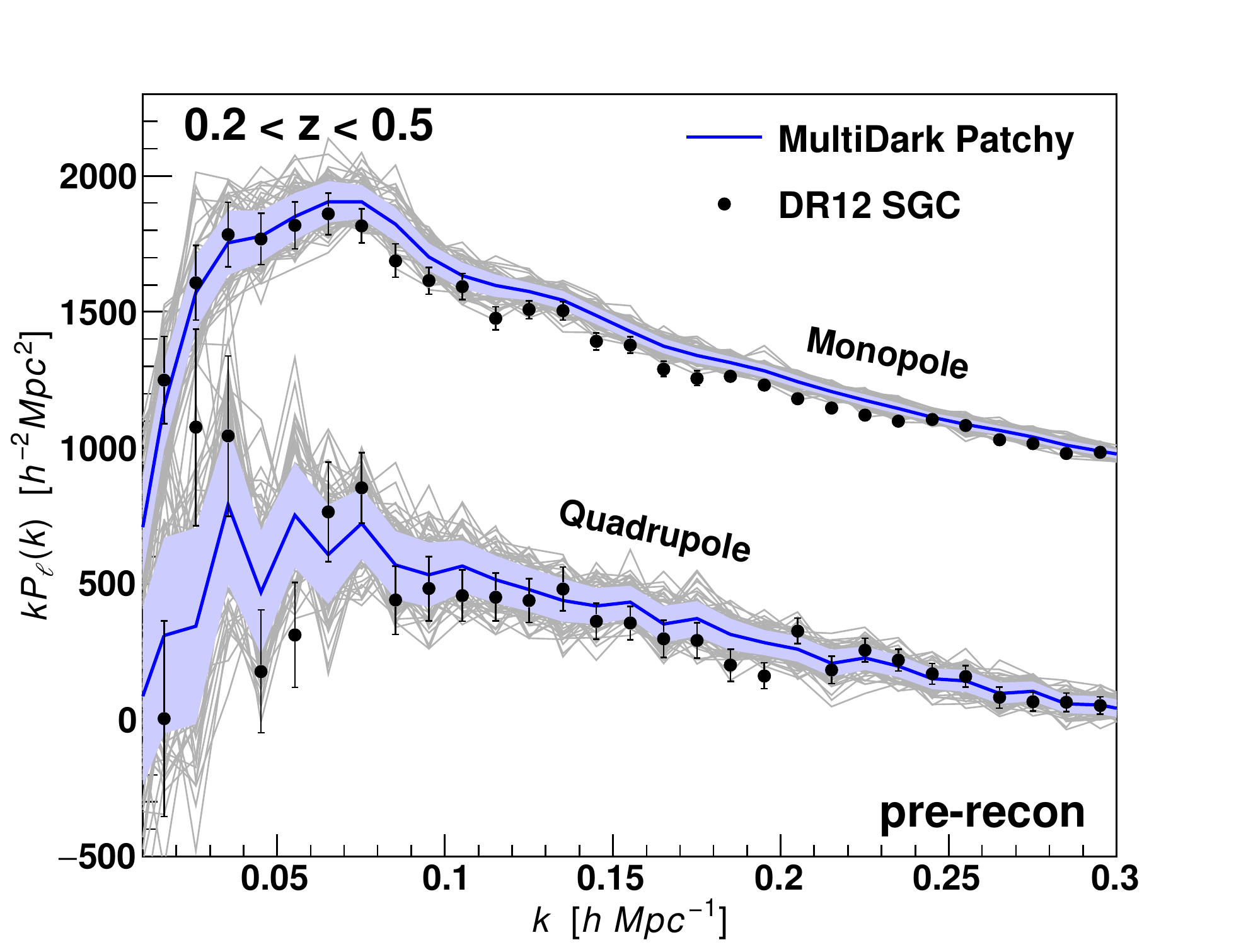,width=5.8cm}
\epsfig{file=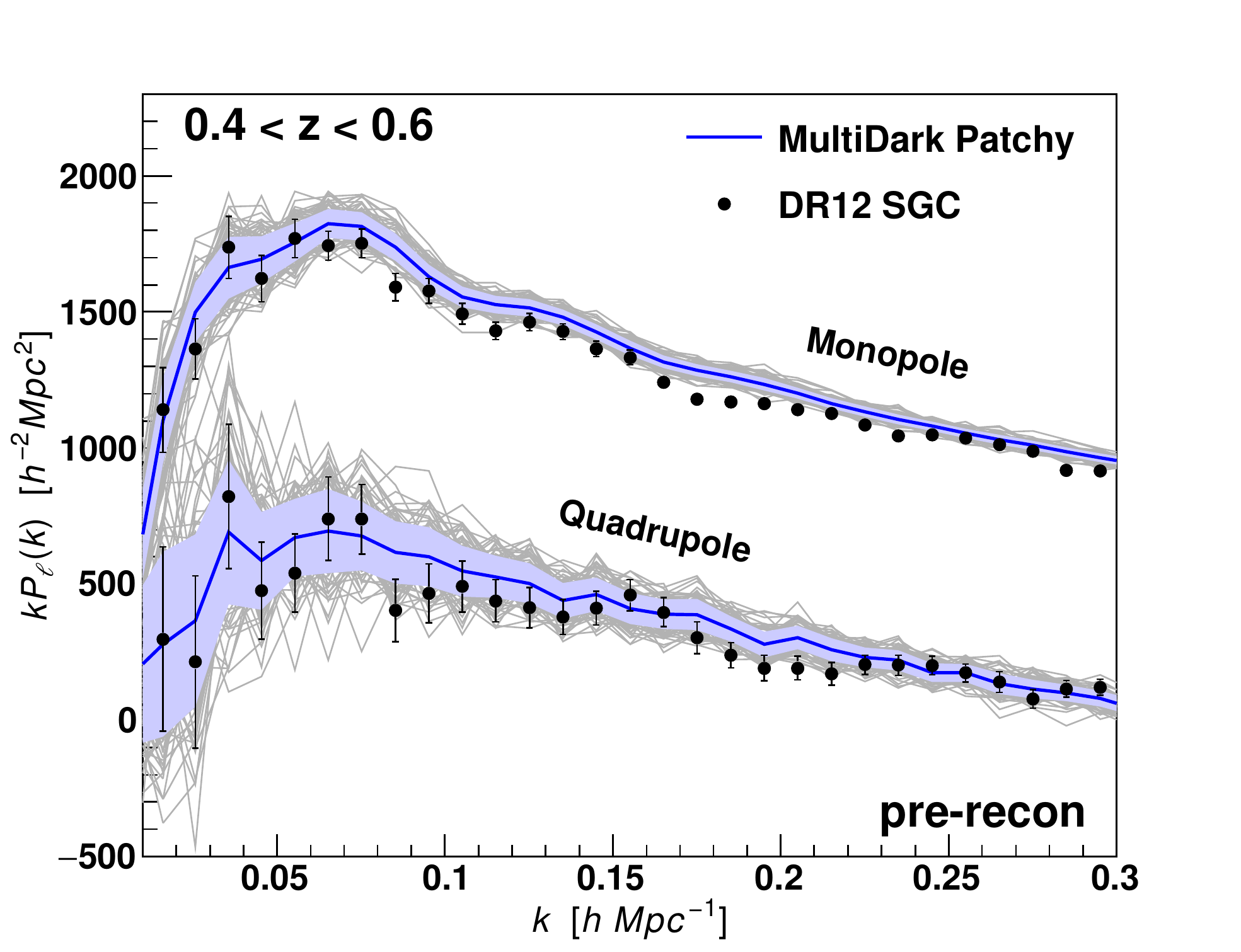,width=5.8cm}
\epsfig{file=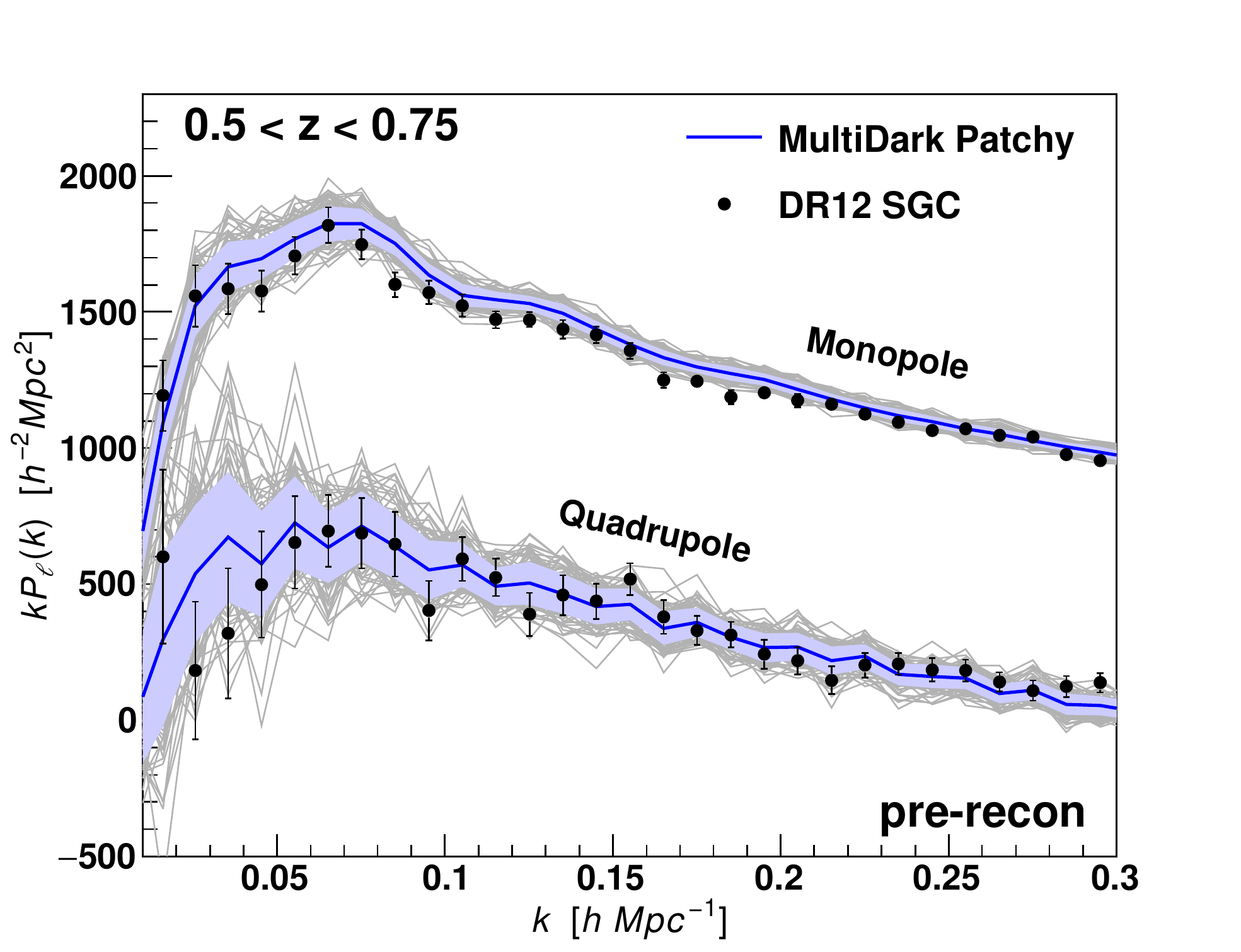,width=5.8cm}\\
\epsfig{file=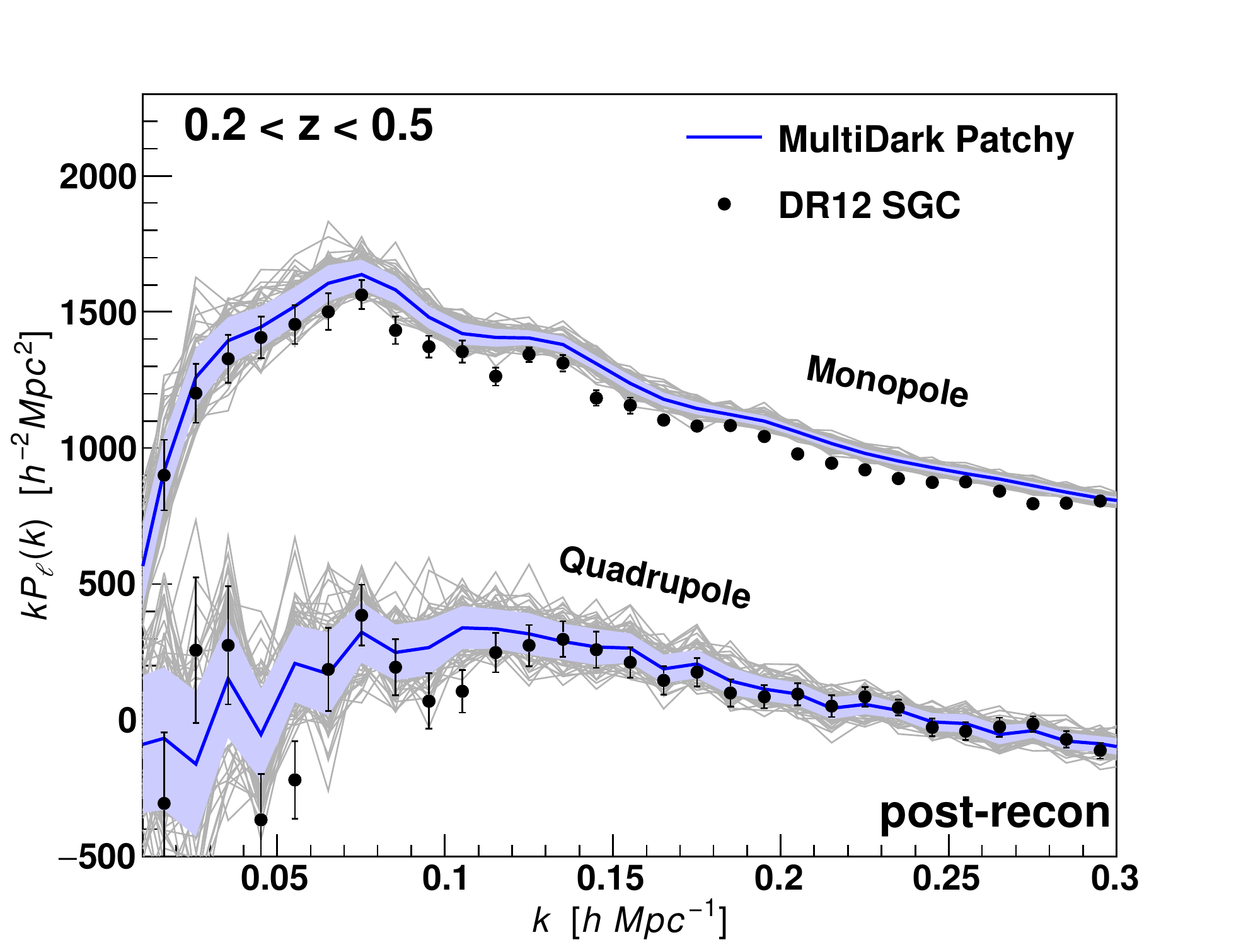,width=5.8cm}
\epsfig{file=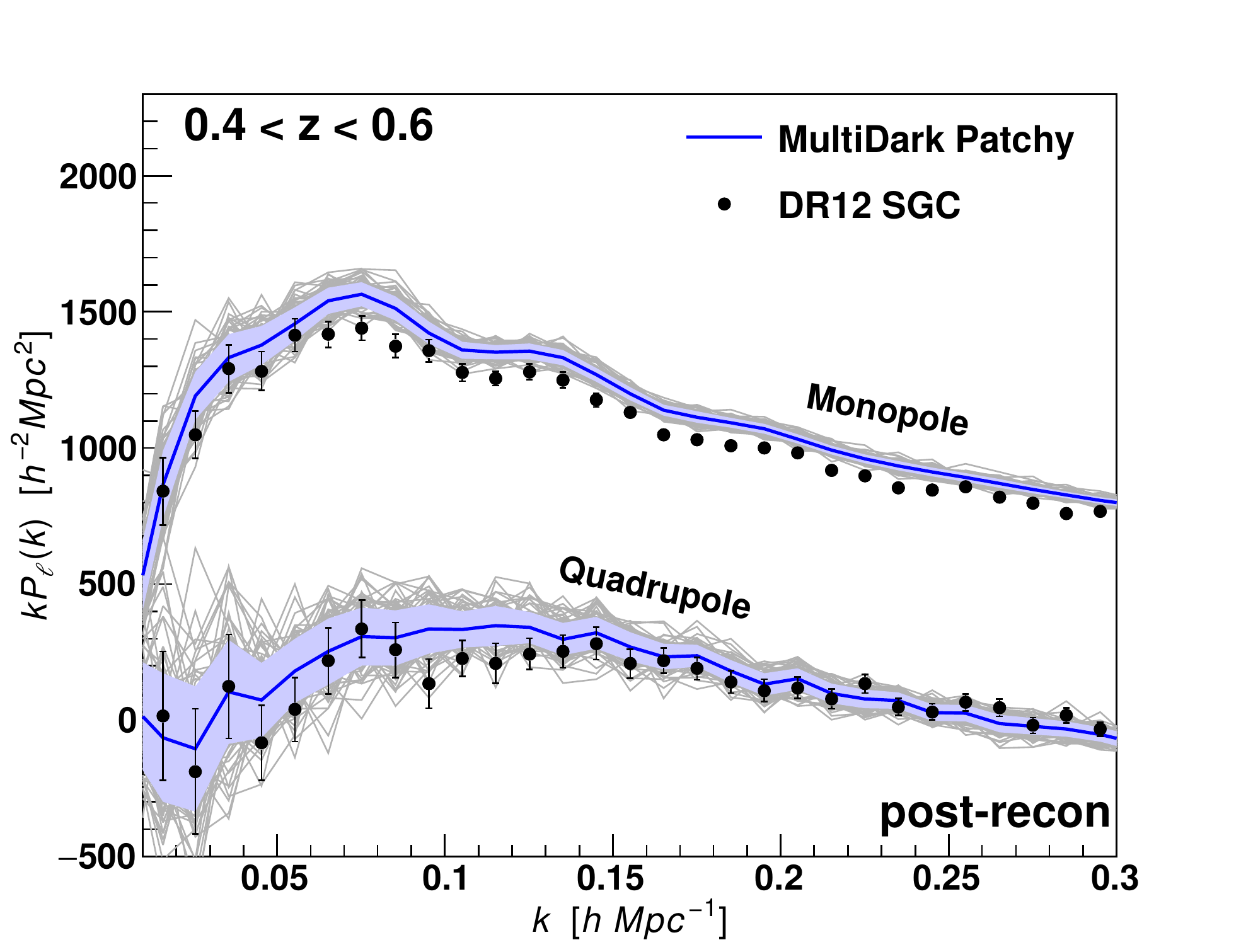,width=5.8cm}
\epsfig{file=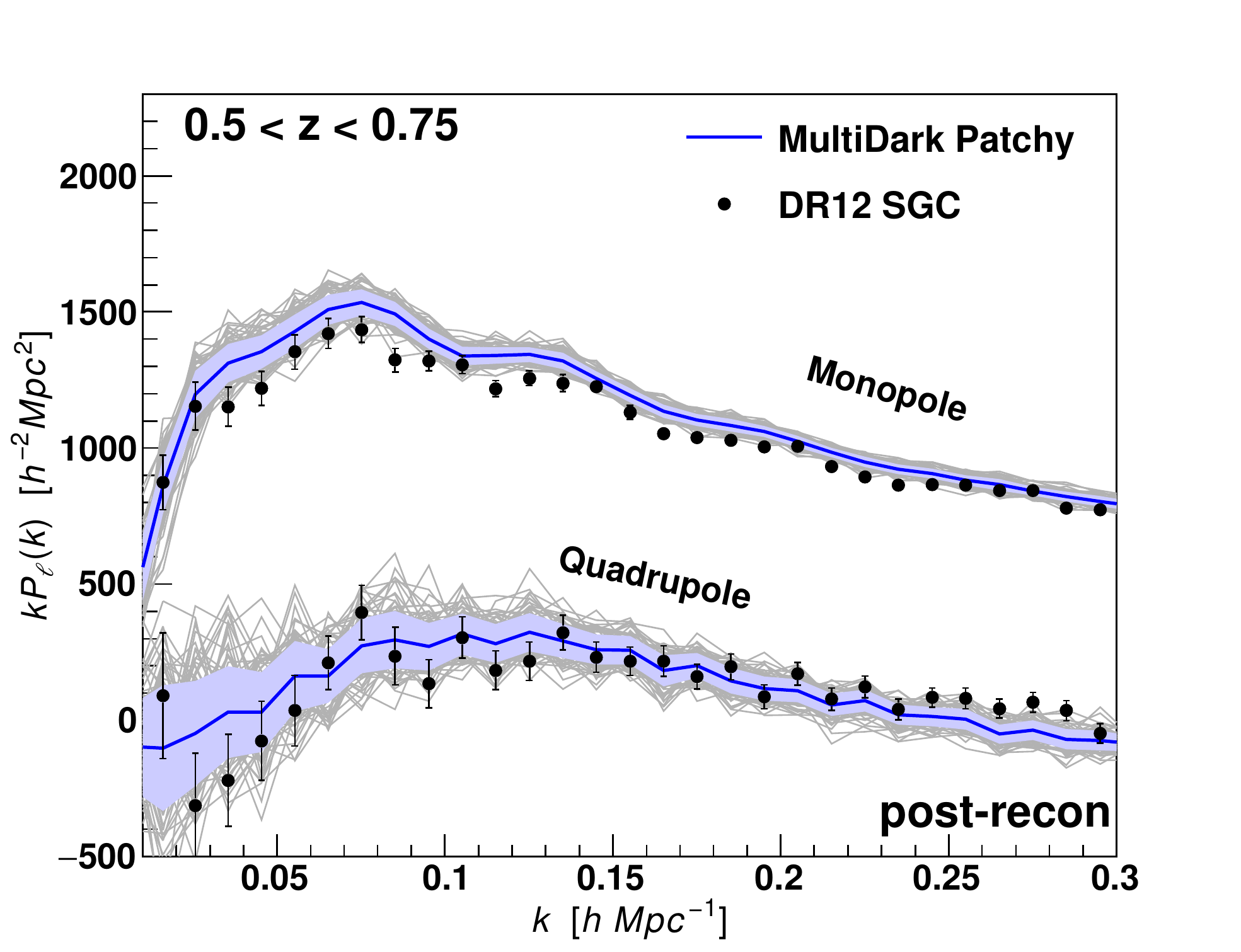,width=5.8cm}
\caption{BOSS DR12 power spectra in the South Galactic Cap (SGC) for the three redshift bins used in this analysis. The panels in the top row show the power spectra before density field reconstruction, while the bottom row displays the power spectra after density field reconstruction. The blue line indicates the mean of the $2048$ (pre-recon) and $999$ (post-recon) MultiDark-Patchy mock catalogues, while the blue shaded area shows the r.m.s. between them. The errors on the data points are the diagonal of the covariance matrix.}
\label{fig:psSGC}
\end{center}
\end{figure*}

The first two non-zero power spectrum multipoles can be calculated as~\citep{Feldman:1993ky,Yamamoto:2005dz,Bianchi:2015oia,Scoccimarro:2015bla}
\begin{align}
P_{0}(\vc{k}) &= \frac{1}{2A}\bigg[F_{0}(\vc{k})F_{0}^*(\vc{k})  - S\bigg]\label{eq:ps0_eq}\\
P_{2}(\vc{k}) &= \frac{5}{4A}F_{0}(\vc{k})\bigg[3F_2^*(\vc{k}) - F_0^*(\vc{k})\bigg]\label{eq:ps2_eq}
\end{align}
with 
\begin{align}
F_{0}(\vc{k}) &= A_0(\vc{k})\\
\begin{split}
F_{2}(\vc{k}) &= \frac{1}{k^2}\bigg[k_x^2 B_{xx} + k_y^2 B_{yy} + k_z^2B_{zz}\\
&+2\bigg(k_zk_y B_{xy} + k_xk_zB_{xz} + k_yk_zB_{yz}\bigg)\bigg]
\end{split}
\end{align}
and
\begin{align}
A_0(\vc{k}) &= \int d\vc{r} D(\vc{r})e^{i\vc{k}\cdot \vc{r}},
\label{eq:ps_eq1}\\
B_{xy}(\vc{k}) &= \int d\vc{r} \frac{r_xr_y}{|\vc{r}|^2}D(\vc{r})e^{i\vc{k}\cdot \vc{r}}.
\label{eq:ps_eq2}
\end{align}
The over-density field $D$ is defined on a 3-dimensional Cartesian grid $\vc{r}$:
\begin{equation}
D(\vc{r}) = G(\vc{r}) - \alpha'R(\vc{r}),
\label{eq:overdensity}
\end{equation}
where $G(\vc{r})$ represents the number of data galaxies at $\vc{r}$ and $R(\vc{r})$ is the number of random galaxies at $\vc{r}$. The normalisation of the random fields is given by $\alpha' = N'_{\rm gal}/N'_{\rm ran}$, where $N'_{\rm gal}$ and $N'_{\rm ran}$ are the total number of weighted data and random galaxies, respectively. All the integrals in eq.~\ref{eq:ps_eq1} and~\ref{eq:ps_eq2} can be solved with Fast Fourier Transforms~\citep{Bianchi:2015oia, Scoccimarro:2015bla}. The normalisation in eq.~\ref{eq:ps0_eq} and \ref{eq:ps2_eq} is
\begin{equation}
A = \alpha'\sum_i^{N_{\rm ran}}n'_g(\vc{r}_i)w^2_{\rm FKP}(\vc{r}_i),
\label{eq:norm}
\end{equation}
where $n'_g$ is the weighted galaxy density. The shot noise term is only relevant for the monopole and is given by
\begin{equation}
\begin{split}
S &= \sum^{N_{\rm gal}}_i \bigg[f_cw_{c}(\vec{x}_i)w_{\rm sys}(\vc{x}_i)w_{\text{\tiny{FKP}}}^2(\vc{x}_i)\\
&+ (1-f_c)w_c^2(\vc{x}_i)w^2_{\rm FKP}(\vc{x}_i)\bigg]\\
&+ \alpha'^2\sum^{N_{\rm ran}}_i w_{\text{\tiny{FKP}}}^2(\vc{x}_i),
\end{split}
\end{equation}
where $f_c$ is the probability of the fibre collision correction being successful, which we set to $0.5$ based on the study by~\citet{Guo:2011ai}. Even though this definition of the shot noise deviates from the one used in~\citet{Beutler:2013yhm}, the difference does not actually impact our analysis since we marginalise over any residual shot noise (see section~\ref{sec:model}).

The final power spectrum is then calculated as the average over spherical k-space shells
\begin{equation}
P_{\ell}(k) = \langle P_{\ell}(\vc{k})\rangle = \frac{1}{N_{\rm modes}}\sum_{k-\frac{\Delta k}{2} < |\vc{k}| < k+\frac{\Delta k}{2}}P_{\ell}(\vc{k}),
\label{eq:averaging}
\end{equation}
where $N_{\rm modes}$ is the number of $\vc{k}$ modes in that shell. In our analysis we use $\Delta k = 0.01\ihMpc$. 

We employ a Triangular Shaped Cloud method to assign galaxies to the 3D grid and correct for the aliasing effect following~\citet{Jing:2004fq}. The setup of our grid implies a Nyqvist frequency of $k_{\rm Ny}=0.6\ihMpc$, twice as large as the largest scale used in our analysis ($k_{\rm max} = 0.3\ihMpc$) and the expected error on the power spectrum monopole at $k = 0.3\ihMpc$ due to aliasing is $< 0.1\%$~\citep{Sefusatti:2015aex}.

The measured power spectrum multipoles for the three redshift bins are presented in Figure~\ref{fig:psNGC} for NGC and Figure~\ref{fig:psSGC} for the SGC. 

\section{The survey window function}
\label{sec:win}

\begin{figure*}
\begin{center}
\epsfig{file=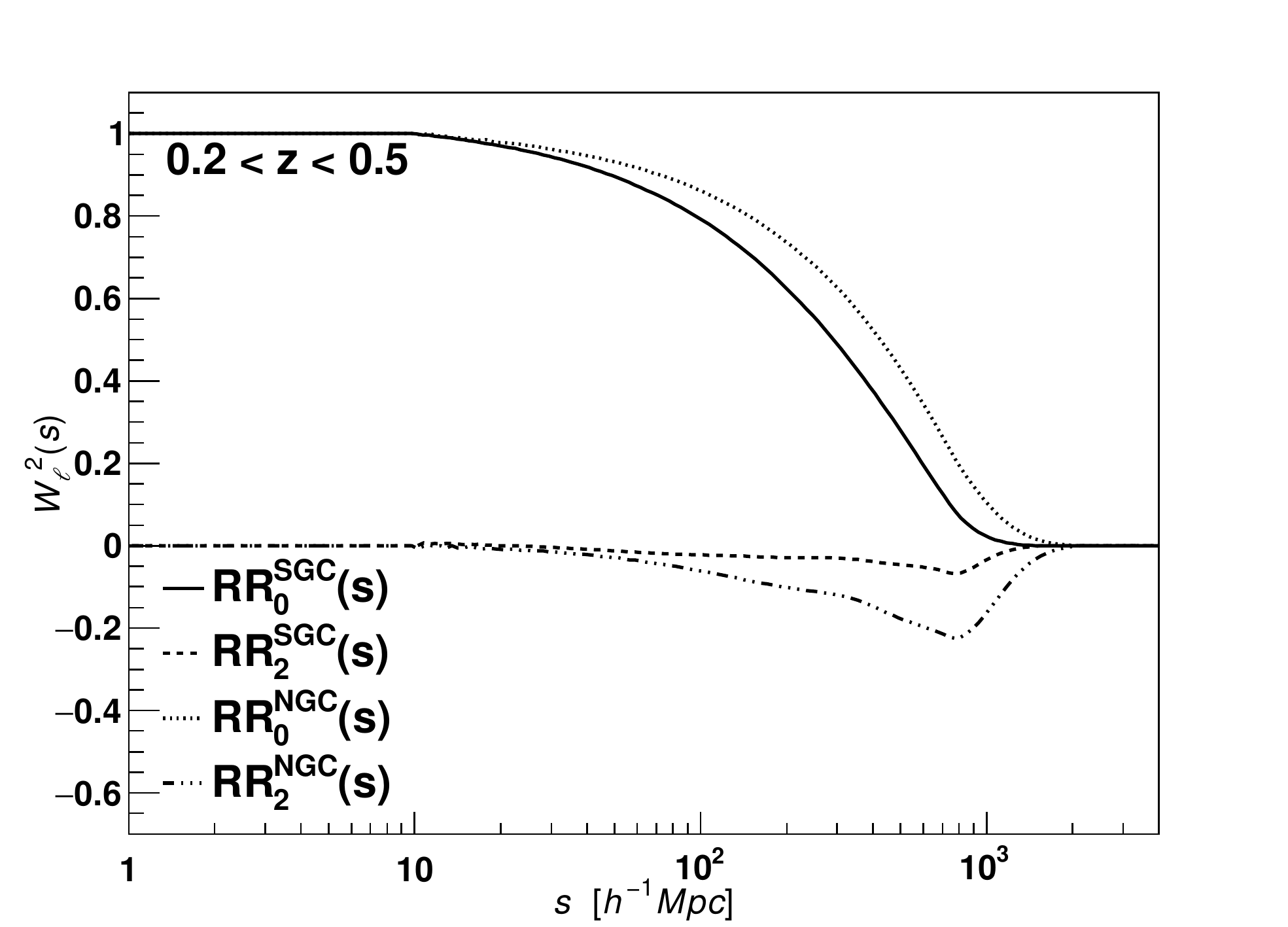,width=5.8cm}
\epsfig{file=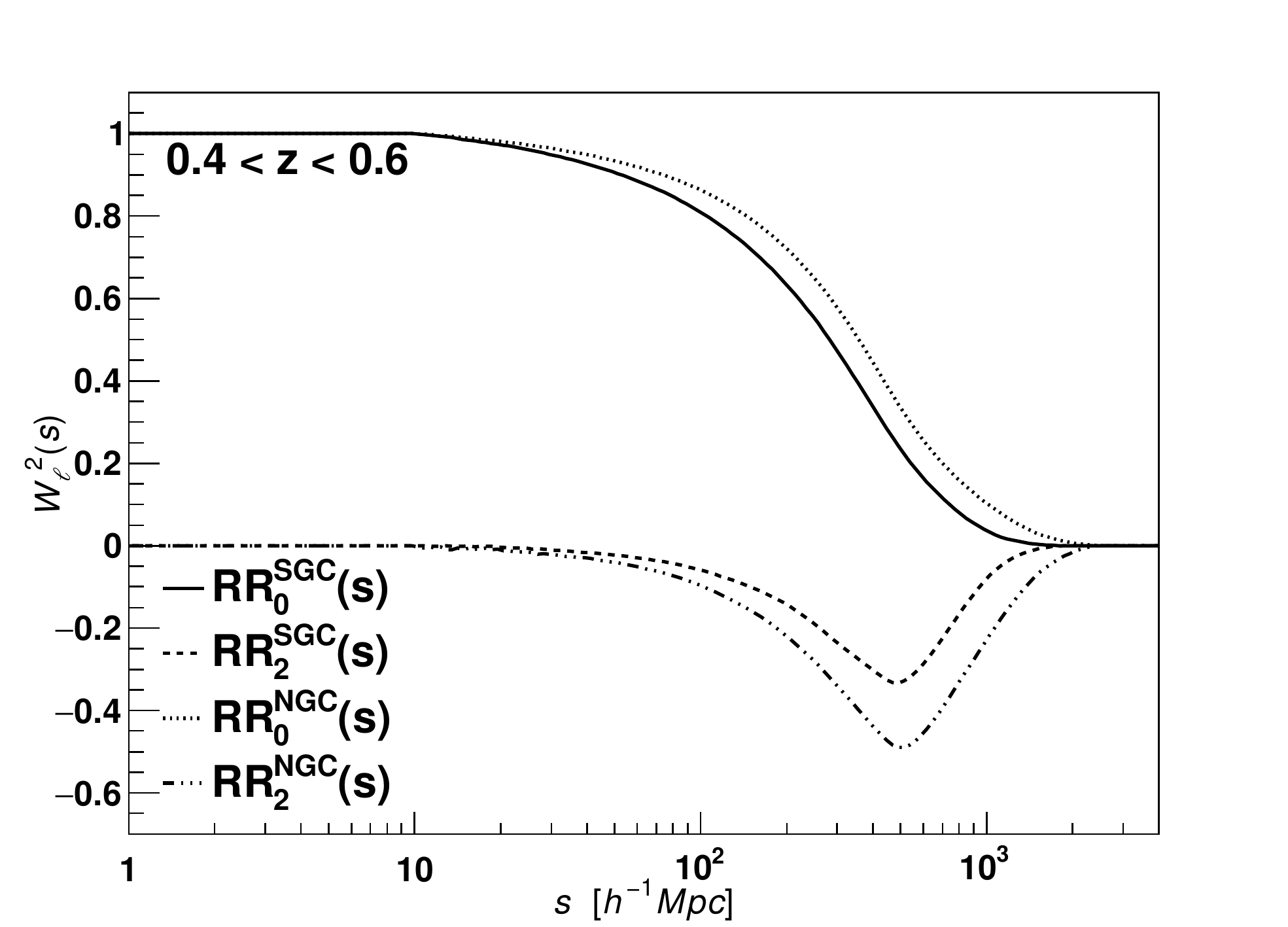,width=5.8cm}
\epsfig{file=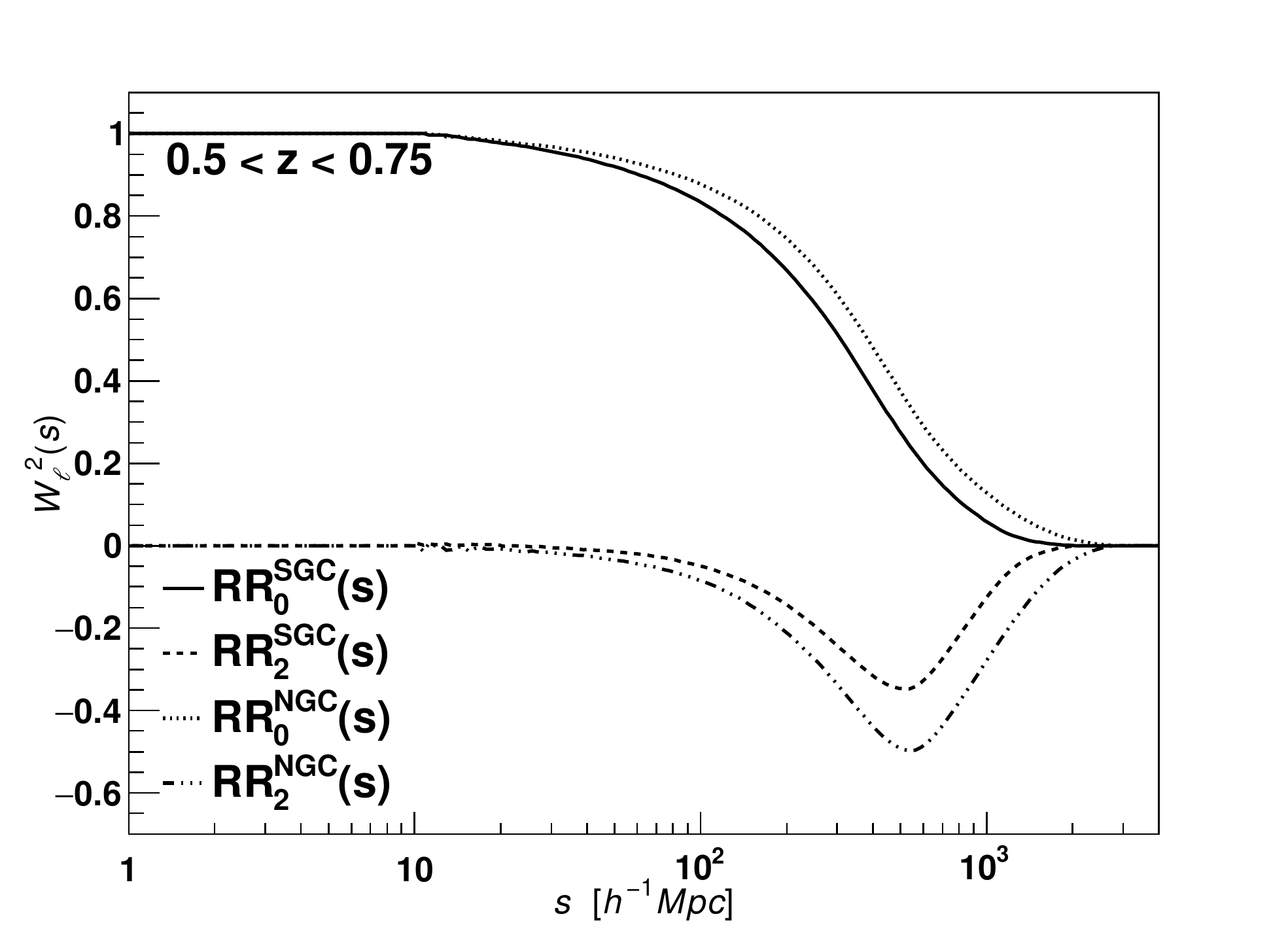,width=5.8cm}
\caption{Window function monopole and quadrupole for the three redshift bins of BOSS DR12 as given in eq.~\ref{eq:Well} and used for the convolved correlation functions in eq.~\ref{eq:conv1} and~\ref{eq:conv2}. As expected, the NGC window function extends to larger scales, because of the larger volume of the NGC compared to the SGC.}
\label{fig:win}
\end{center}
\end{figure*}

The survey mask is defined as the multiplicative term which turns the Poisson sampled galaxy density field in the observed galaxy density field. In Fourier-space this multiplicative term becomes a convolution. The broad extent of the window function in Fourier space makes the convolution process computationally expensive. Conversely, applying the window function in configuration space is easy and straightforward. Here we follow a method suggested by~\citet{Wilson:2015} which can be summarised in three steps:
\begin{enumerate}
\item Calculate the model power spectrum multipoles and Fourier-transform them to obtain the correlation function multipoles $\xi_{L}^{\rm model}(s)$.
\item Calculate the ``convolved'' correlation function multipoles $\hat{\xi}^{\rm model}_{\ell}(s)$ by multiplying the correlation function with the window function multipoles. 
\item Conduct 1D FFTs~\citep[][FFTlog]{Hamilton:2000} to transform the convolved correlation function multipoles back into Fourier space to obtain the convolved power spectrum multipoles, $\hat{P}^{\rm model}_{\ell}(k)$. This result becomes our model to be compared with the observed power spectrum multipoles.
\end{enumerate}
In~\citet{Wilson:2015} the formalism following these three steps is derived within the global plane parallel approximations, meaning that a global line-of-sight, $\hat{\eta}$, is defined for all galaxies in the sample. B16 demonstrates that this method can be derived within the local plane parallel approximation, which means that it is applicable to wide angle surveys like BOSS. Here we will summarise the formalism and refer to B16 for more details.

The convolved correlation function multipoles can be expressed as
\begin{equation}
\hat{\xi}_{\ell}(s) = (2\ell + 1)\sum_{L}\xi_{L}(s) \sum_p\frac{1}{2p + 1}W_{p}^2(s) a^{\ell}_{Lp}.
\label{eq:multixi}
\end{equation}
with the window function multipoles $W_{p}^2(s)$
\begin{equation}
W_{p}^2(s)  =\frac{2p+1}{2} \int d\mu_s  \int d\vc{x}_1 W(\vc{x}_1)W(\vc{x}_1+\vc{s})\mathcal{L}_{p}(\mu_s),
\label{eq:Wp}
\end{equation}
where $\mathcal{L}_{\ell}$ is the Legendre polynomial of order $\ell$, $\vc{s} = \vc{x}_2-\vc{x}_1$ is the pair separation vector, and $\mu_s$ is the cosine angle of the separation vector relative to the line of sight, i.e., $\mu_s = \hat{\vc{s}} \cdot \hat{\vc{x}}_h$. To calculate the coefficients $a^{\ell}_{Lp}$ we use
\begin{equation}
\mathcal{L}_{\ell}\mathcal{L}_{p} = \sum_t a^{\ell}_{p t}\mathcal{L}_{t},
\label{eq:multicol}
\end{equation}
to multiply the polynomial expressions for the Legendre polynomials on the left and apply 
\begin{equation}
\mu^n = \sum_{\ell = n,(n-1),...}\frac{(2\ell + 1)n!\mathcal{L}_{\ell}(\mu_s)}{2^{(n-\ell)/2}(\frac{1}{2}(n-\ell))!(\ell + n + 1)!!}.
\label{eq:mupower}
\end{equation}
The convolved power spectrum multipoles are given by 
\begin{equation}
\hat{P}_{\ell}(k) = 4\pi i^\ell \int ds\,s^2\hat{\xi}_{\ell}(s) j_{\ell}(sk).
\label{eq:PWell}
\end{equation}
For any real survey dataset the window function is calculated from the random pair counts $RR(s, \mu_s)$ as 
\begin{equation}
W_{\ell}^2(s) \propto \sum_{\vc{x}_1} \sum_{\vc{x}_2}   RR(s,\mu_s)\mathcal{L}_{\ell}(\mu_s),
\label{eq:Well}
\end{equation}
with the normalisation $W_{0}^2(s\rightarrow 0) = 1$.

We are interested in the monopole and quadrupole power spectra and therefore, in eq.~\ref{eq:PWell},  the convolved correlation function multipoles relevant for our analysis are given by  
\begin{align}
\hat{\xi}_{0}(s) &= \xi_{0}W_{0}^2  + \frac{1}{5}\xi_2W^2_2 + ...
\label{eq:conv1}\\
\begin{split}
\hat{\xi}_{2}(s) &= \xi_{0} W_{2}^2  + \xi_2\left[W^2_0 + \frac{2}{7}W^2_2\right]\\
&\;\;\;\;\;\;\;\;\;\;\;\;\;\,+...
\label{eq:conv2}
\end{split}
\end{align}
where we ignored all terms beyond the quadrupole $\xi_{\ell}$. In B16 we find that the hexadecapole contribution to the monopole and quadrupole due to the window function effect can be neglected. The two different window function multipoles included in the equations above are shown in Figure~\ref{fig:win}. We assume that the window function is the same for pre- and post reconstruction.

We account for the integral constraint bias by correcting the model power spectrum as 
\begin{equation}
P^{\rm ic-corrected}_{\ell}(k) = \hat{P}_{\ell}(k) - P_{0}W_{\ell}^2(k),
\end{equation}
where the window functions $W(k)$ can be obtained from $W_{\ell}(s)$ defined in eq.~\ref{eq:Well} as
\begin{equation}
W^2_{\ell}(k) = 4\pi\int ds\; s^2W_{\ell}^2(s)j_{\ell}(sk).
\end{equation}
The integral constraint correction in BOSS only affects modes $k\lesssim 0.005\ihMpc$ and does not affect any of the results in this analysis.

\section{Mock catalogues}
\label{sec:mocks}

\begin{figure*}
\begin{center}
\epsfig{file=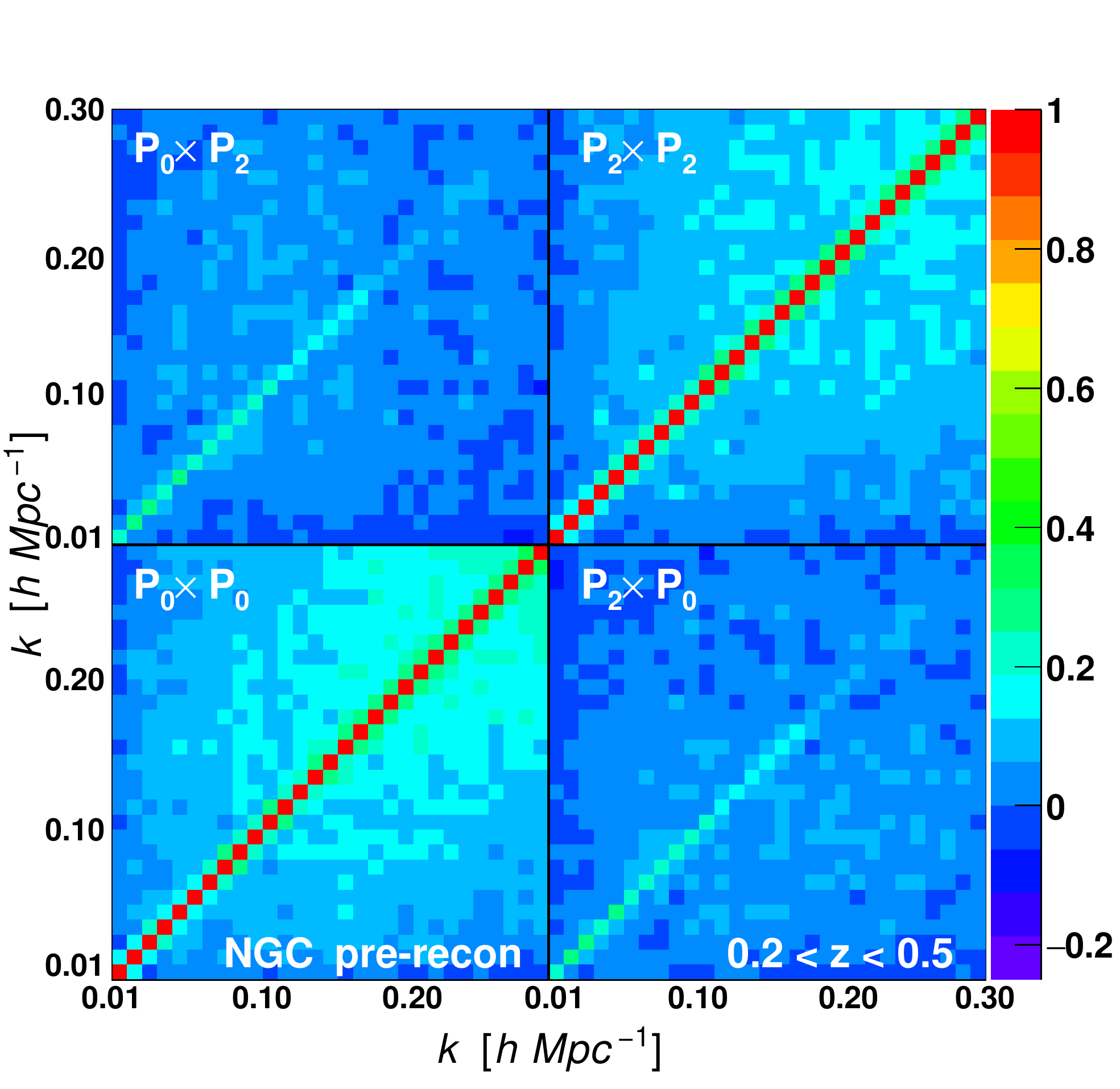,width=5.8cm}
\epsfig{file=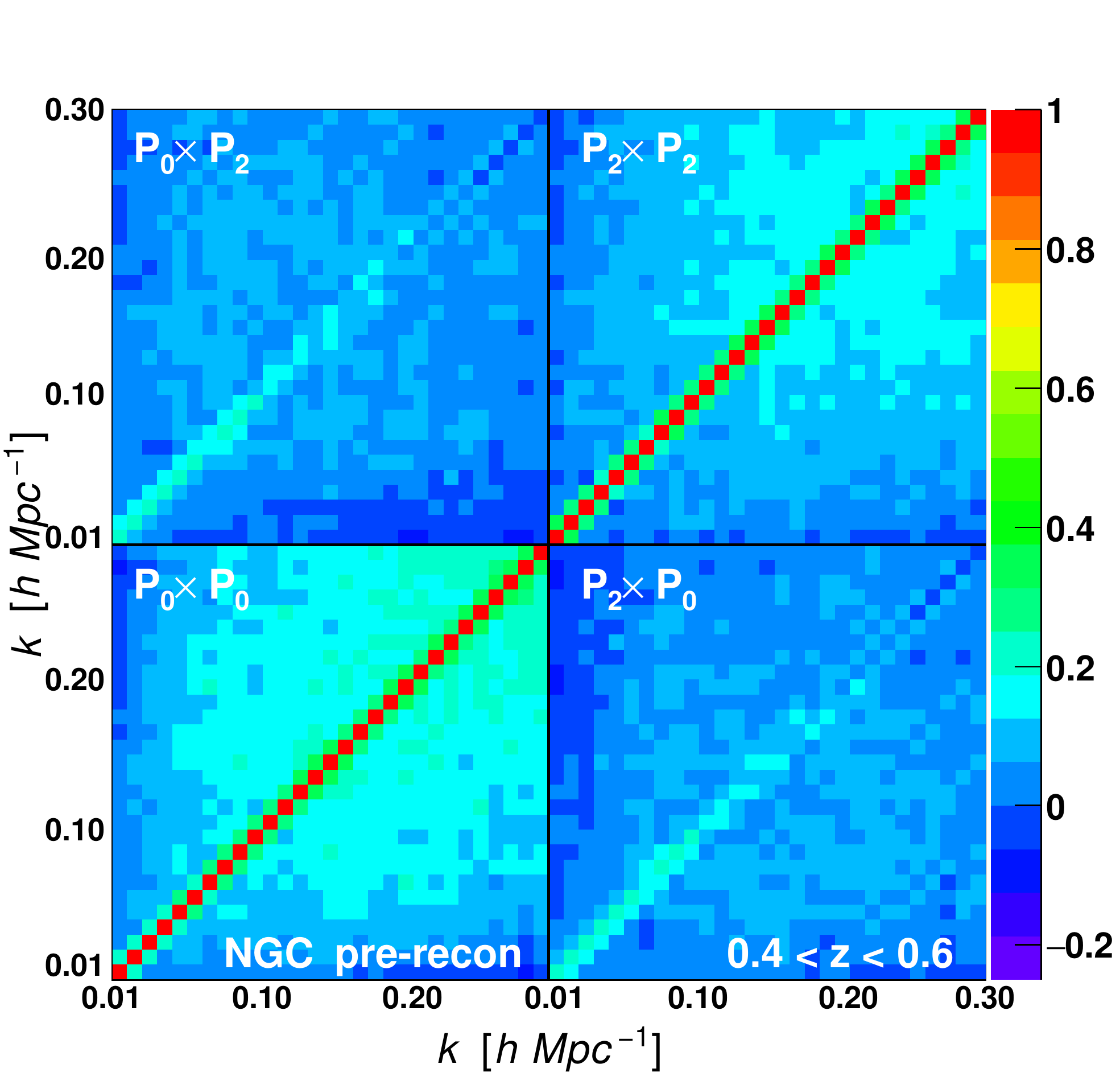,width=5.8cm}
\epsfig{file=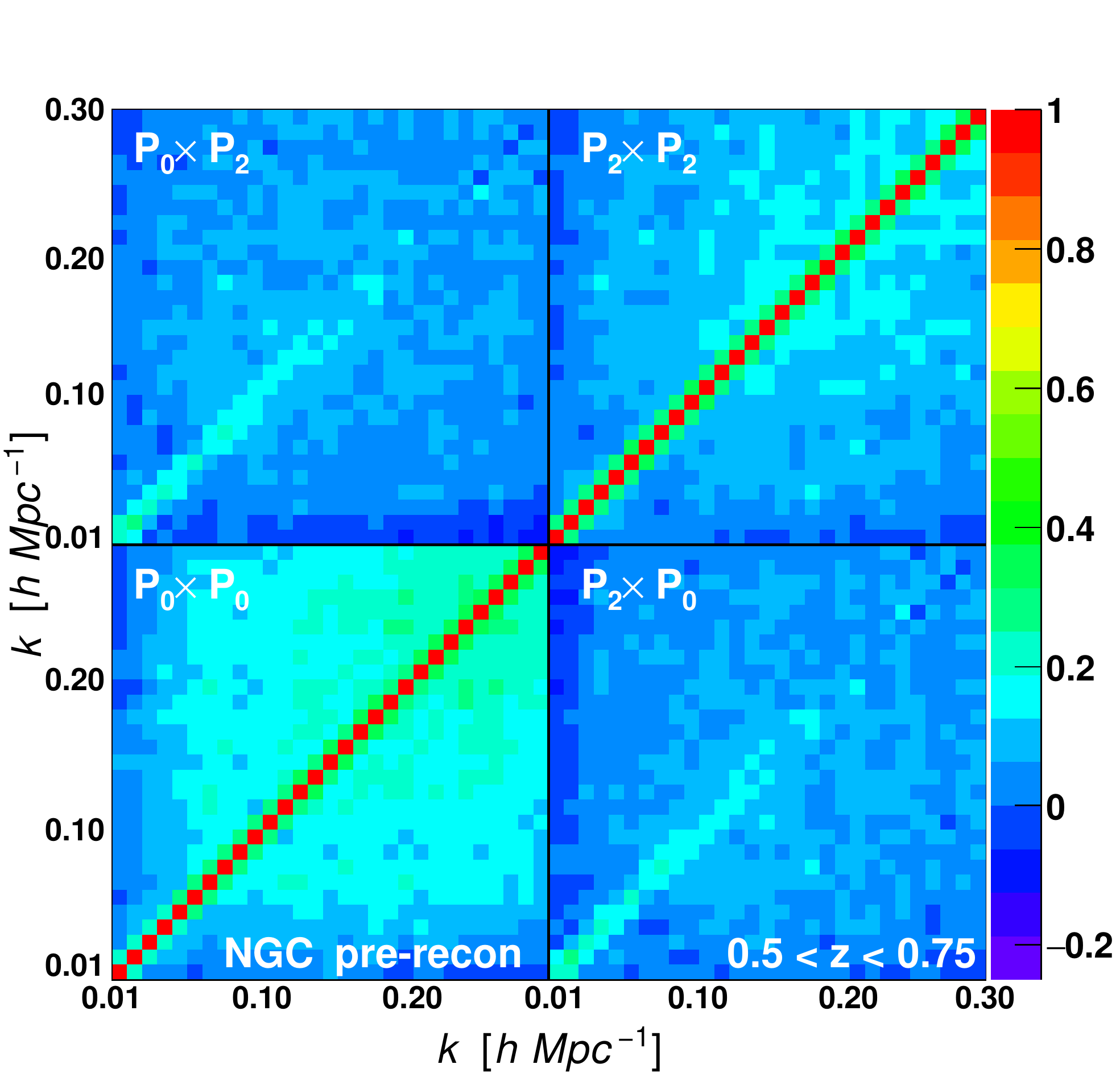,width=5.8cm}\\
\epsfig{file=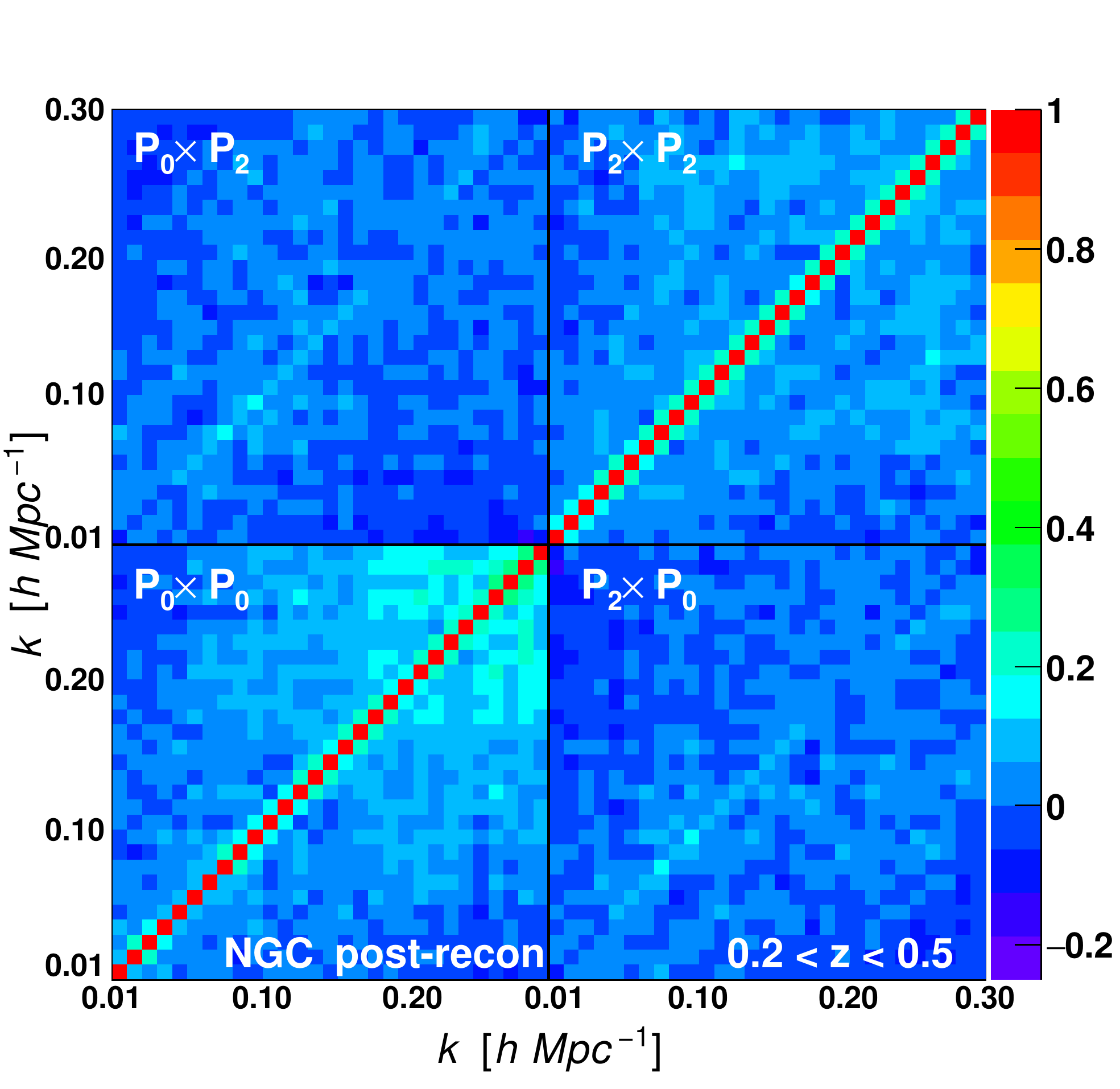,width=5.8cm}
\epsfig{file=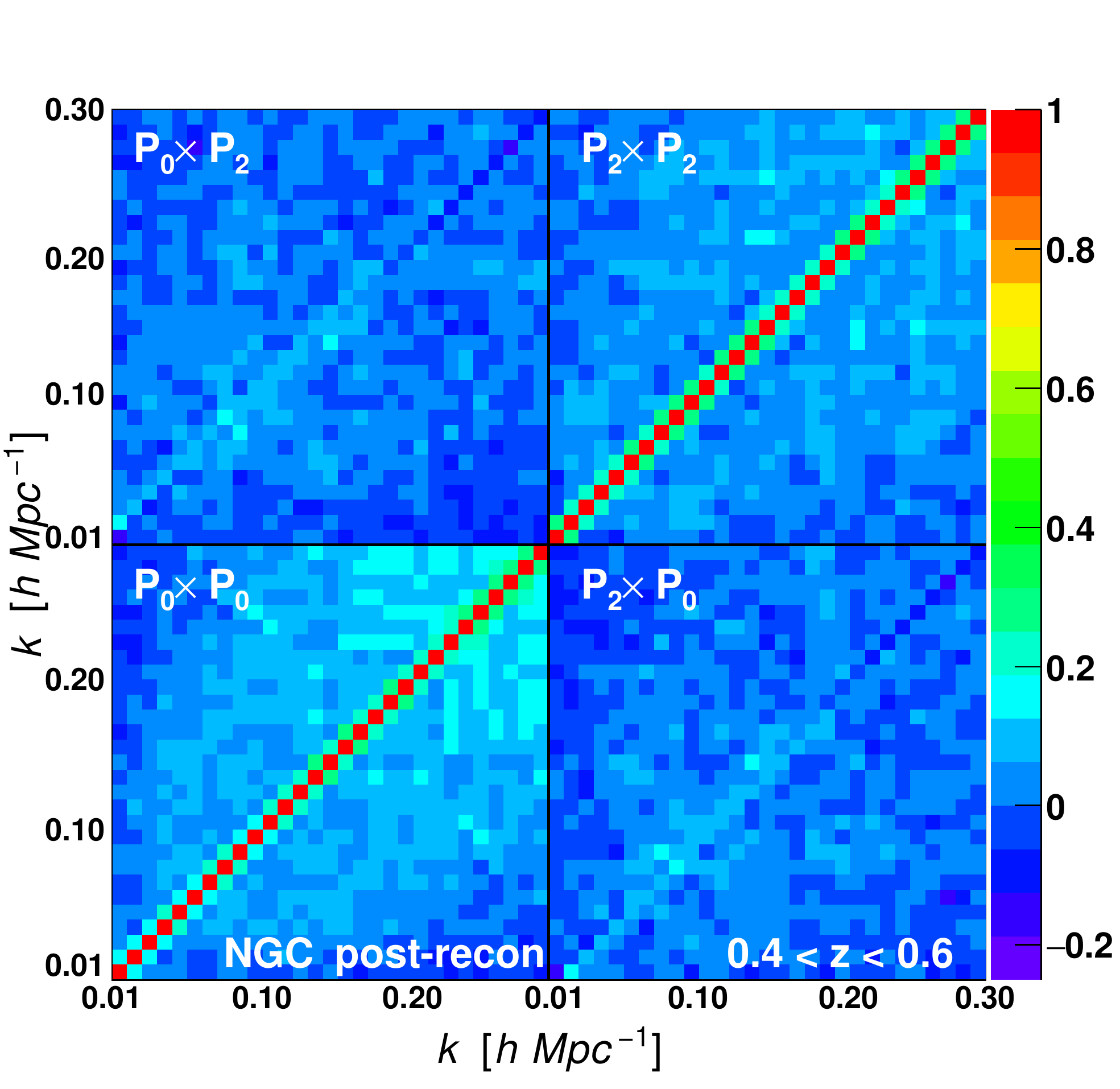,width=5.8cm}
\epsfig{file=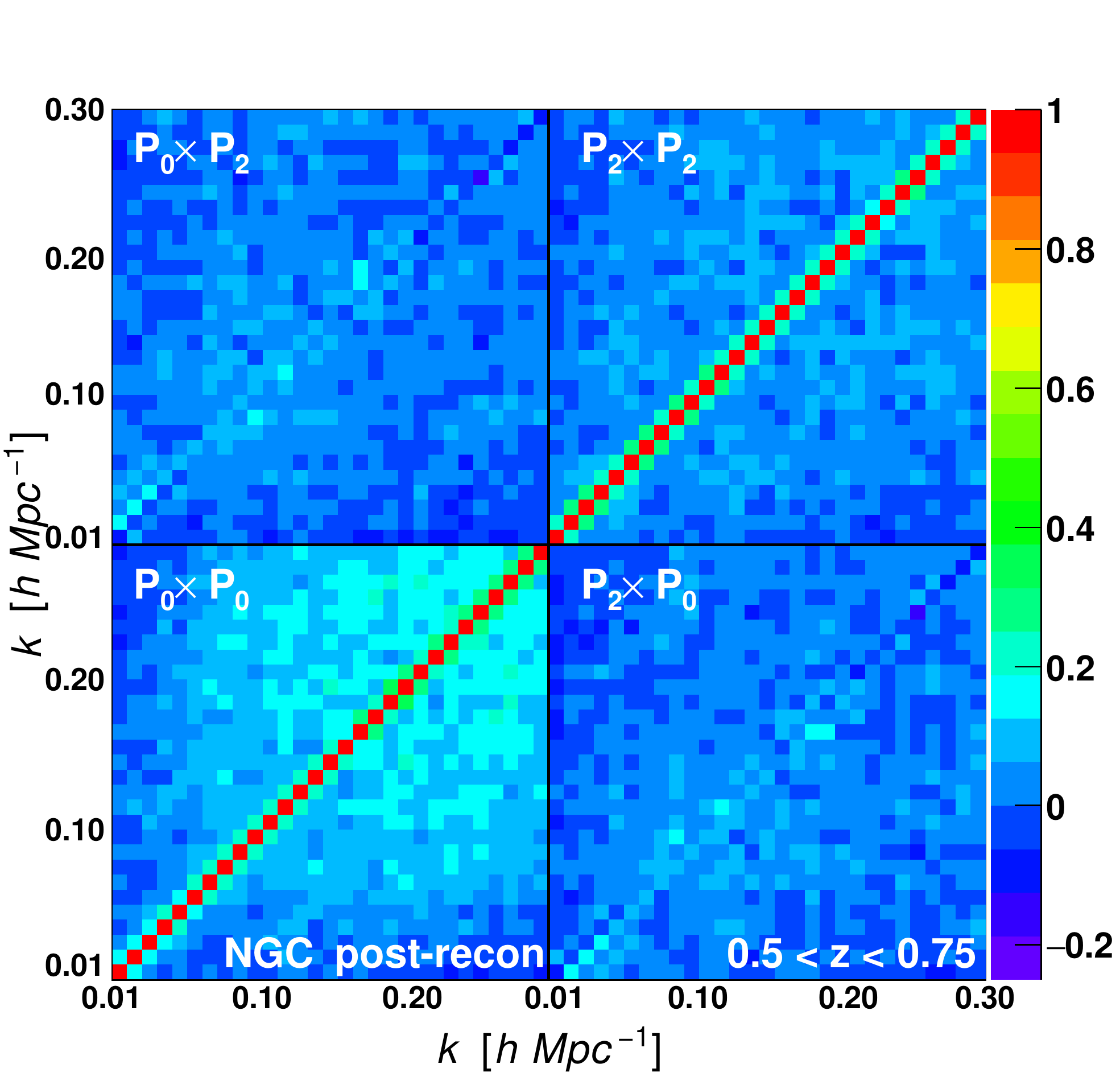,width=5.8cm}
\caption{Correlation matrix before (top) and after (bottom) density field reconstruction for the North Galactic Cap (NGC) in the three redshift bins used in this analysis. The matrices include the monopole (bottom left corner) and quadrupole (top right corner) as well as their correlation (top left and bottom right). The pre-reconstruction matrices contain $2045$ mock catalogues, while the post-reconstruction results contain $996$ mock catalogues. The colour indicates the level of correlation, with red corresponding to $100\%$ correlation and magenta corresponding to $-25\%$ anti-correlation (there are not many fields lower than $-25\%$). After reconstruction there is less correlation between different $k$ modes and between the multipoles.}
\label{fig:covNGC}
\end{center}
\end{figure*}

\begin{figure*}
\begin{center}
\epsfig{file=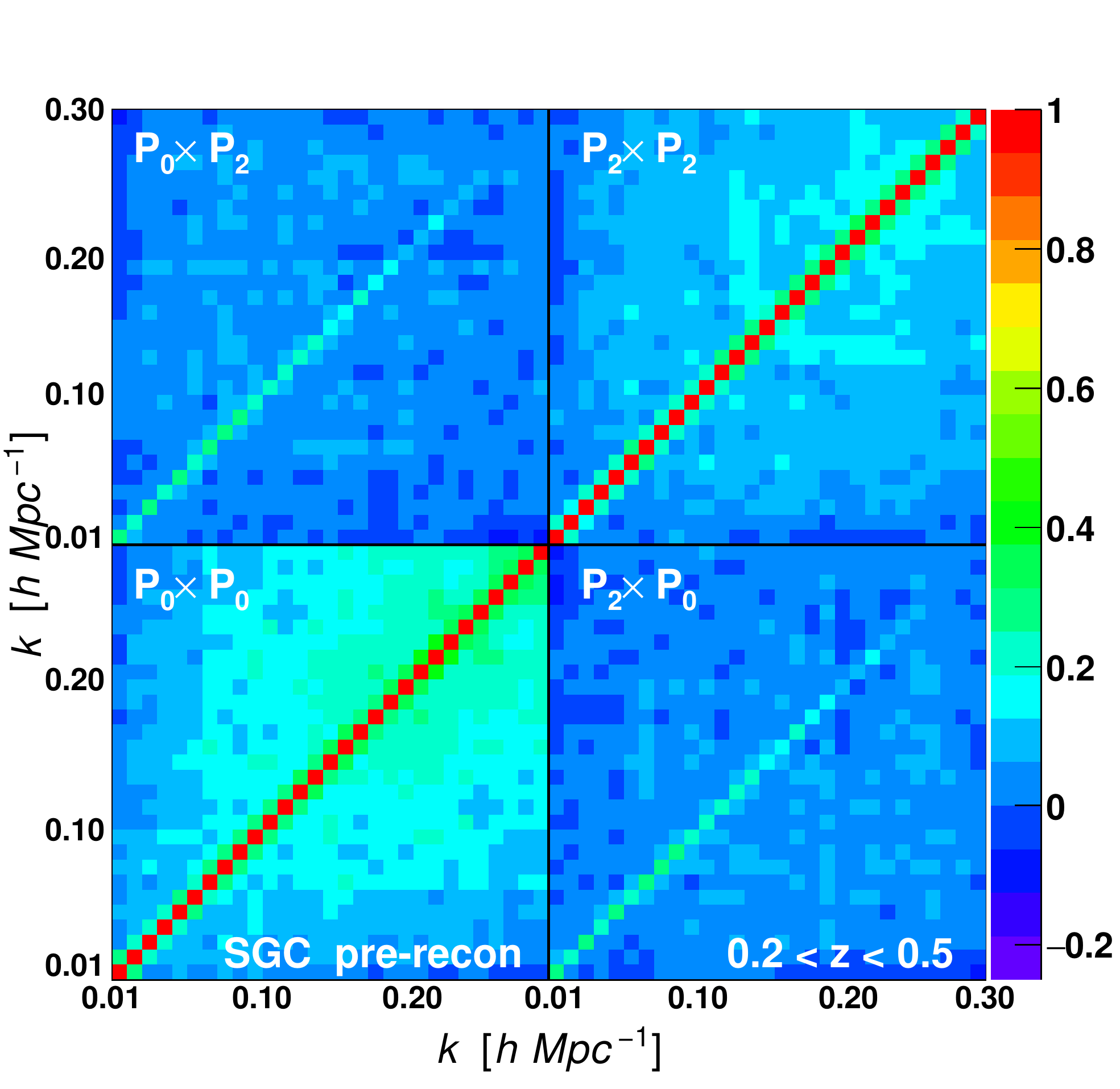,width=5.8cm}
\epsfig{file=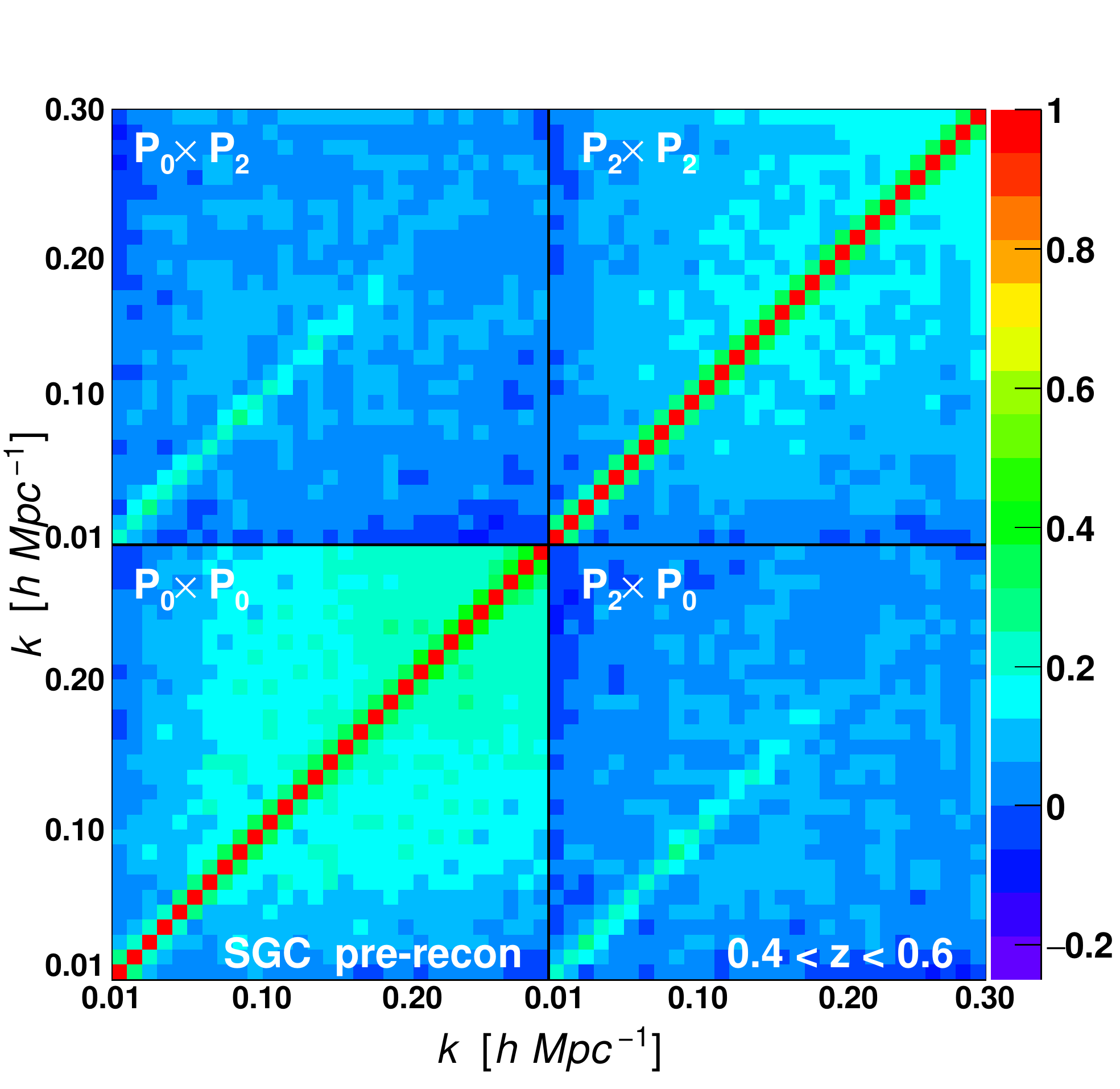,width=5.8cm}
\epsfig{file=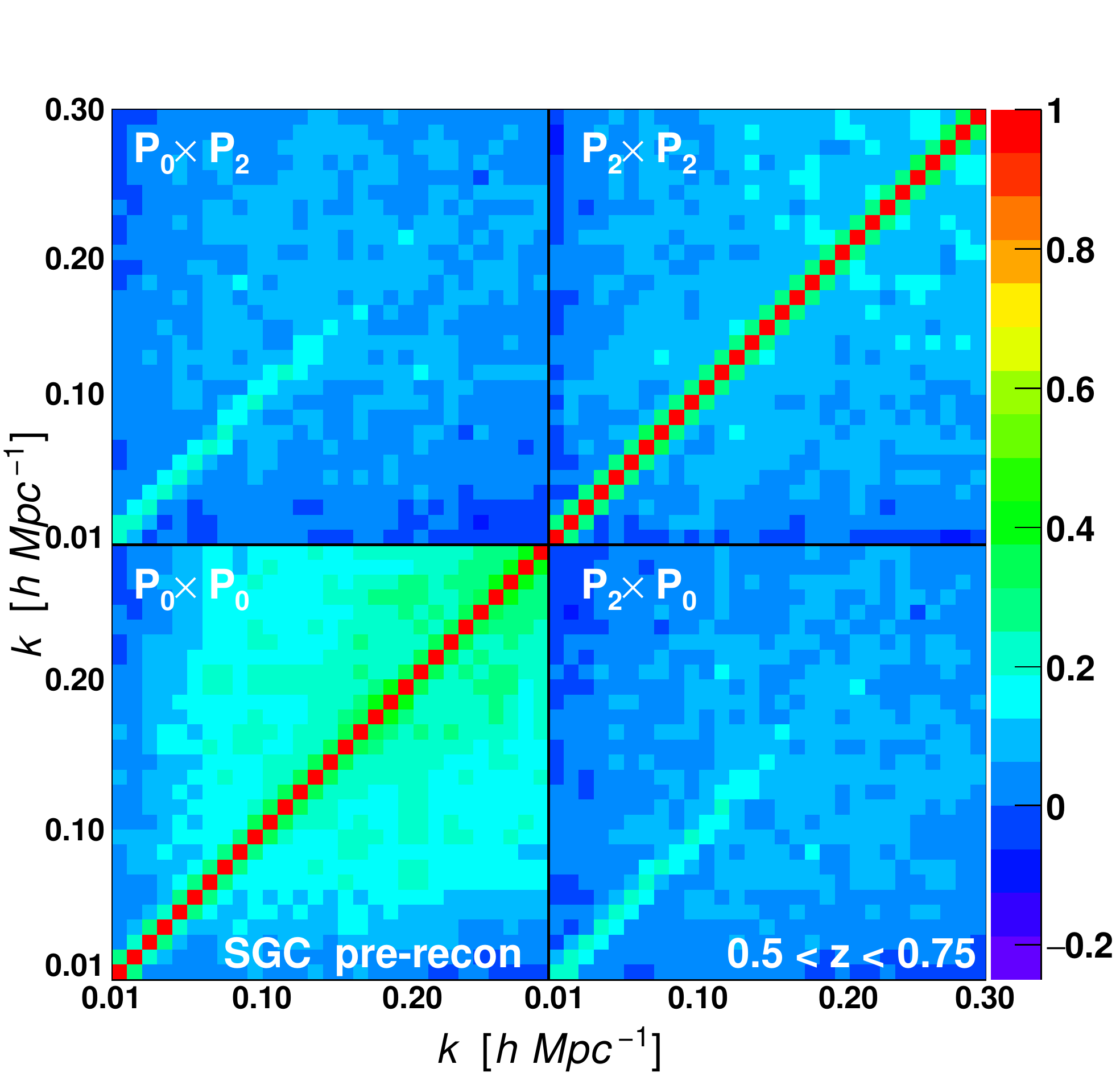,width=5.8cm}\\
\epsfig{file=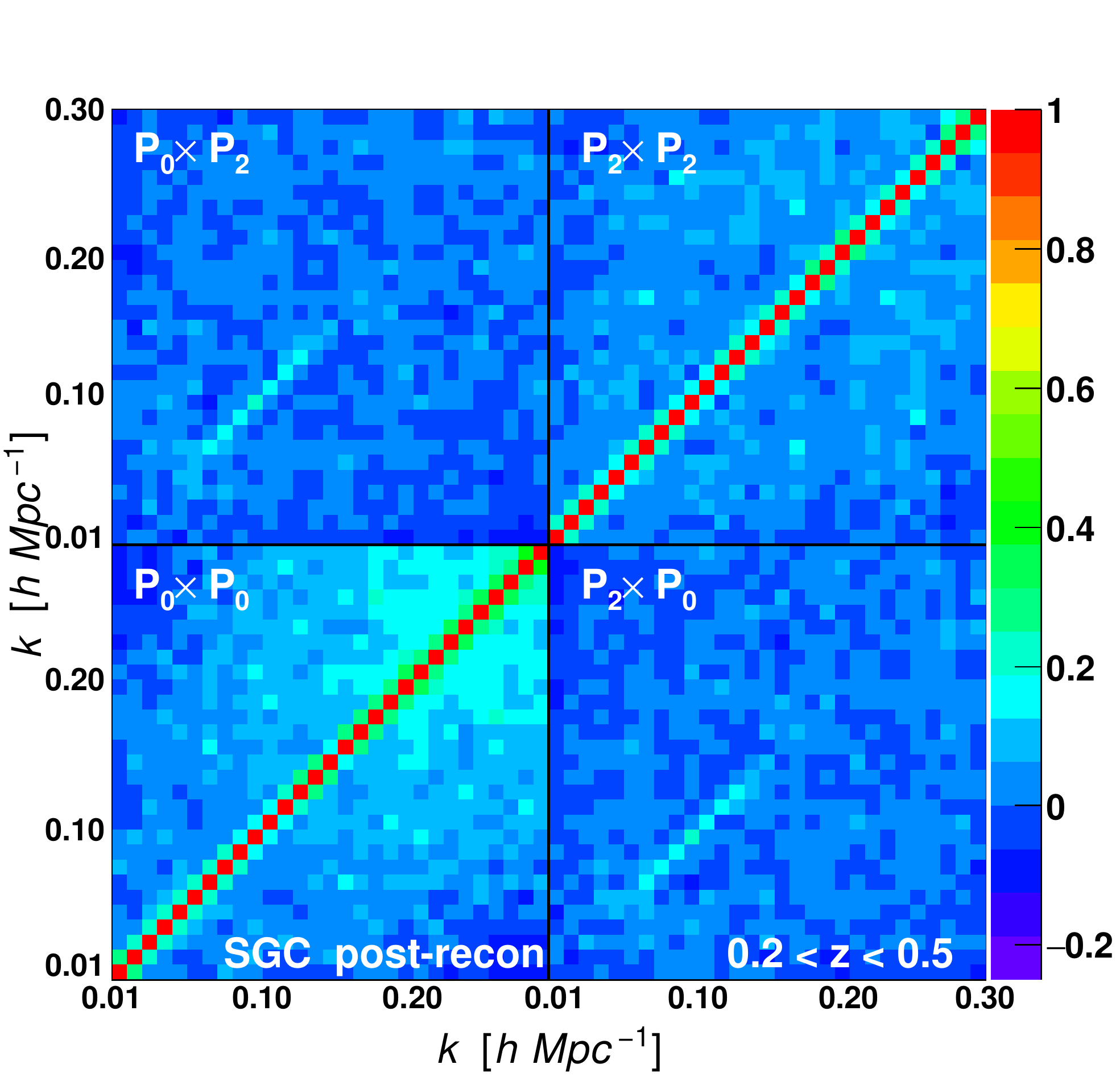,width=5.8cm}
\epsfig{file=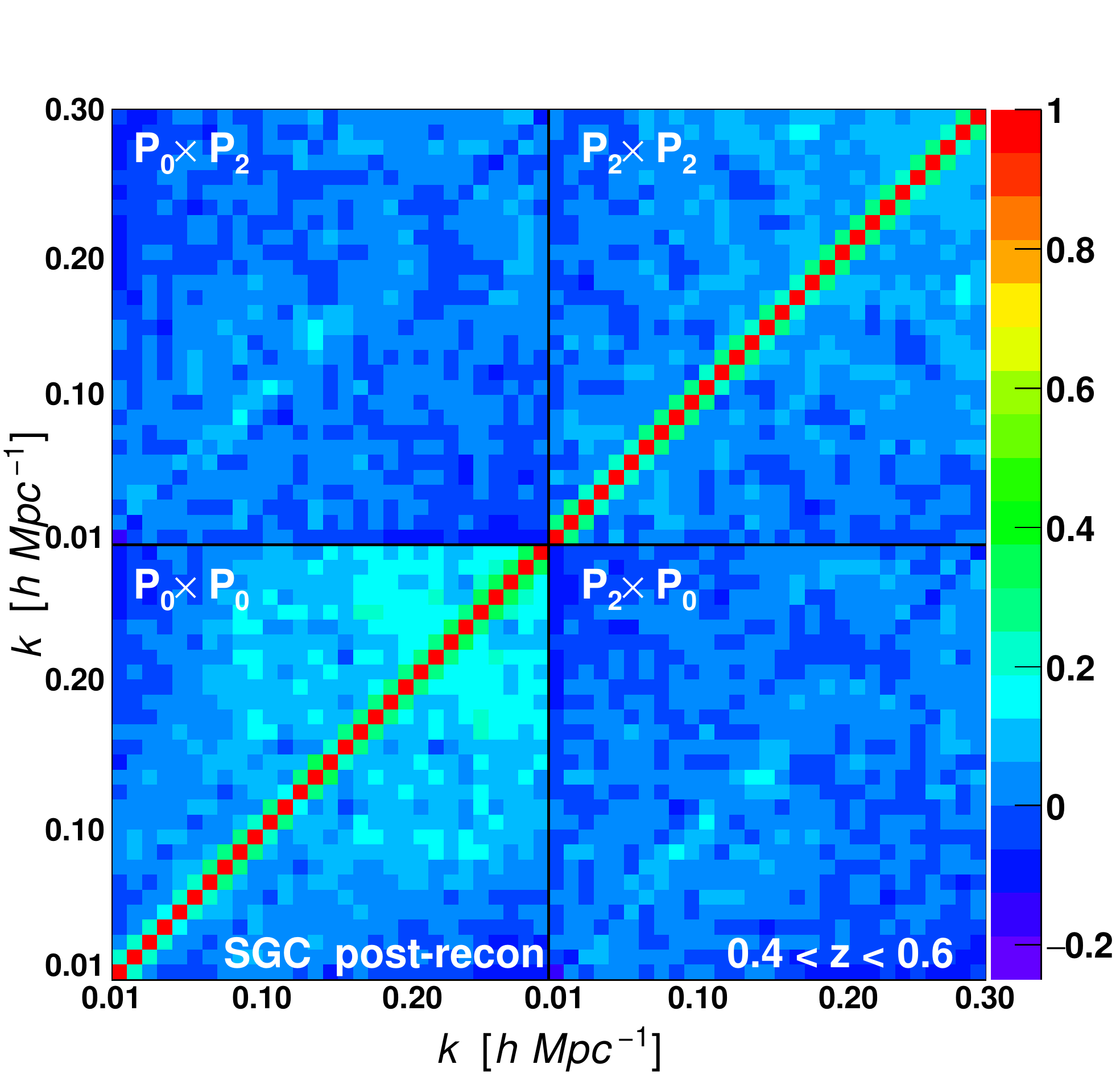,width=5.8cm}
\epsfig{file=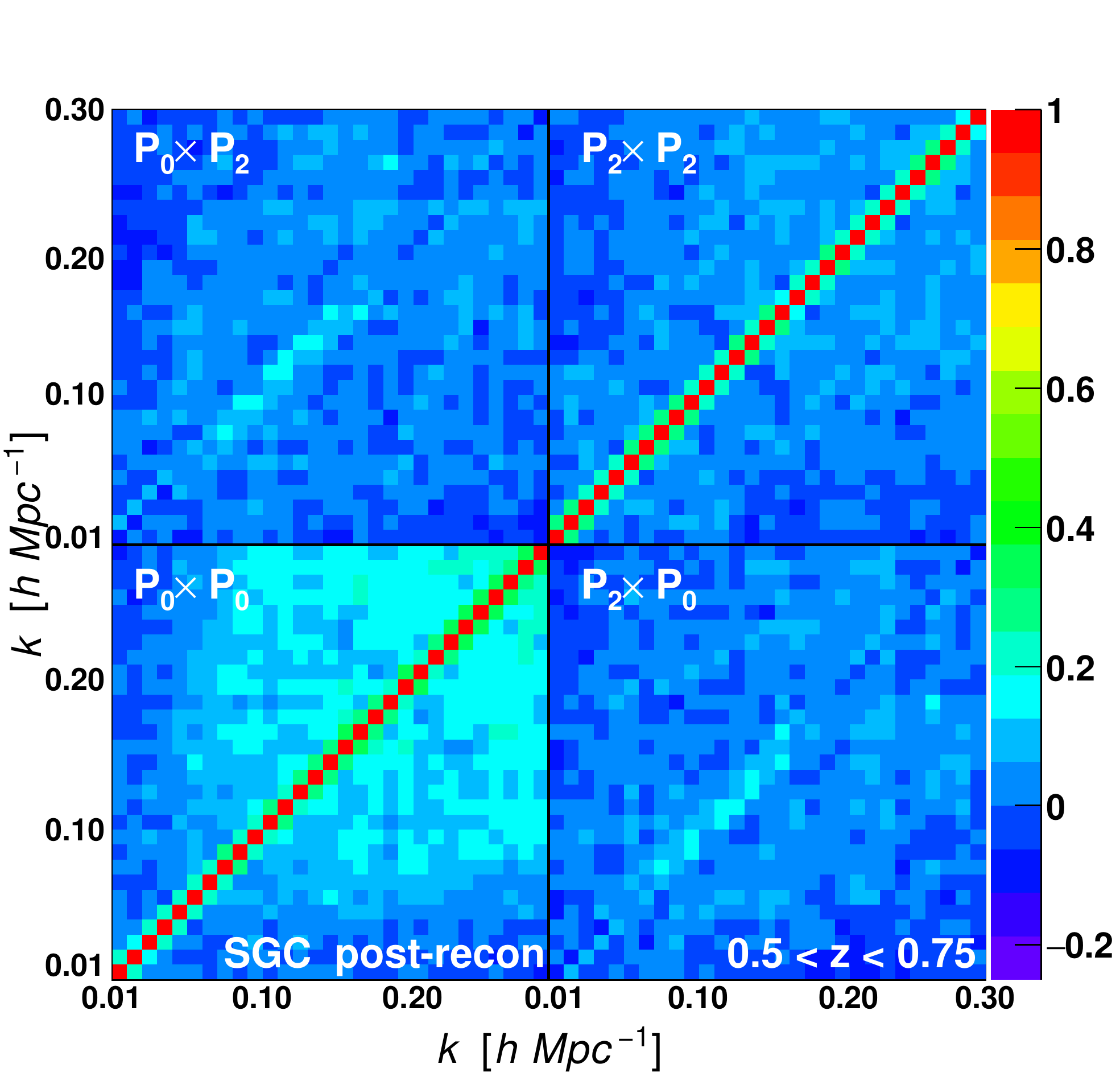,width=5.8cm}
\caption{Correlation matrix before (top) and after (bottom) density field reconstruction for the South Galactic Cap (SGC) in the three redshift bins used in this analysis. The matrices include the monopole (bottom left corner) and quadrupole (top right corner) as well as their correlation (top left and bottom right). The pre-reconstruction matrices contain $2048$ mock catalogues, while the post-reconstruction results contain $999$ mock catalogues. The colour indicates the level of correlation, with red corresponding to $100\%$ correlation and magenta corresponding to $-25\%$ anti-correlation (there are not many fields lower than $-25\%$). After reconstruction there is less correlation between different $k$ modes and between the multipoles.}
\label{fig:covSGC}
\end{center}
\end{figure*}

To derive a covariance matrix for the power spectrum monopole and quadrupole we use the MultiDark-Patchy mock catalogues~\citep{Kitaura:2013cwa, Kitaura:2015}. These mock catalogues  have been produced using approximate gravity solvers and analytical-statistical biasing models. The catalogues have been calibrated to a $N$-body based reference sample with higher resolution. The reference catalogue is extracted from one of the BigMultiDark simulations~\citep{Klypin:2014kpa}, which was performed using gadget-2~\citep{Springel:2005mi} with $3\,840^3$ particles on a volume of ($2.5h^{-1}$Mpc)$^3$ assuming a $\Lambda$CDM cosmology with $\Omega_M = 0.307115$, $\Omega_b = 0.048206$, $\sigma_8 = 0.8288$, $n_s = 0.9611$, and a Hubble constant of $H_0 = 67.77\,$km\,s$^{-1}$Mpc$^{-1}$. 

Halo abundance matching is used to reproduce the observed BOSS two and three-point clustering measurements~\citep{Rodriguez-Torres:2015vqa}. This technique is applied at different redshift bins to reproduce the BOSS DR12 redshift evolution. These mock catalogues are combined into light cones, also accounting for the selection effects and survey mask of the BOSS survey. In total we have $2045$ mock catalogues available for the NGC and $2048$ mock catalogues for the SGC. The BAO reconstruction procedure~\citep{Eisenstein:2006nk} has only been applied to $996$ NGC catalogues and $999$ SGC catalogues.

The mean power spectrum multipoles for the MultiDark-Patchy mock catalogues are shown in Figure~\ref{fig:psNGC} for the NGC and Figure~\ref{fig:psSGC} for the SGC together with the BOSS measurements (black data points). The mock catalogues closely reproduce the data power spectrum multipoles for the entire range of wave-numbers relevant for this analysis. 

\subsection{The covariance matrix}
\label{sec:covs}

We can derive a covariance matrix from the set of mock catalogues described in the last section as
\begin{equation}
\begin{split}
C_{xy} = \frac{1}{N_s - 1} \sum^{N_s}_{n=1}&\left[P_{\ell,n}(k_i) - \overline{P}_{\ell}(k_i)\right]\times\\
&\left[P_{\ell',n}(k_j) - \overline{P}_{\ell'}(k_j)\right]
\end{split}
\label{eq:cov}
\end{equation}
with $N_s$ being the number of mock catalogues. Our covariance matrix at each redshift bin contains the monopole as well as the quadrupole, and the elements of the matrices are given by $(x, y) = (\frac{n_b\ell}{2} + i, \frac{n_b\ell'}{2} + j)$, where $n_b$ is the number of $k$ bins in each multipole power spectrum. Our k-binning yields $n_b = 29$ for the fitting range $k = 0.01$ - $0.30\hMpc$, and hence the dimensions of the covariance matrices become $58\times 58$. The mean of the power spectra is defined as
\begin{equation}
\overline{P}_{\ell}(k_i) = \frac{1}{N_s}\sum^{N_s}_{n=1}P_{\ell, n}(k_i).
\end{equation}
Given that the mock catalogues follow the same selection as the data, they incorporate the same window function as we separately match the randoms to each hemisphere.

Figure~\ref{fig:covNGC} and~\ref{fig:covSGC} present the correlation matrices for BOSS NGC and SGC for the three redshift bins, where the correlation coefficient is defined as
\begin{equation}
r_{xy} = \frac{C_{xy}}{\sqrt{C_{xx}C_{yy}}}.
\end{equation}
For each panel in Figure~\ref{fig:covNGC} and~\ref{fig:covSGC}, the lower left hand corner shows the correlation between bins in the monopole, the upper right hand corner displays the correlations between the bins in the quadrupole and the upper left hand corner and lower right hand corner show the correlation between the monopole and quadrupole. After reconstruction there is less correlation between different $k$ modes. Reconstruction not only sharpens the BAO feature, but also removes some of the correlation between different $k$-modes and between the multipoles, making the covariance matrix more diagonal.

\begin{figure*}
\begin{center}
\epsfig{file=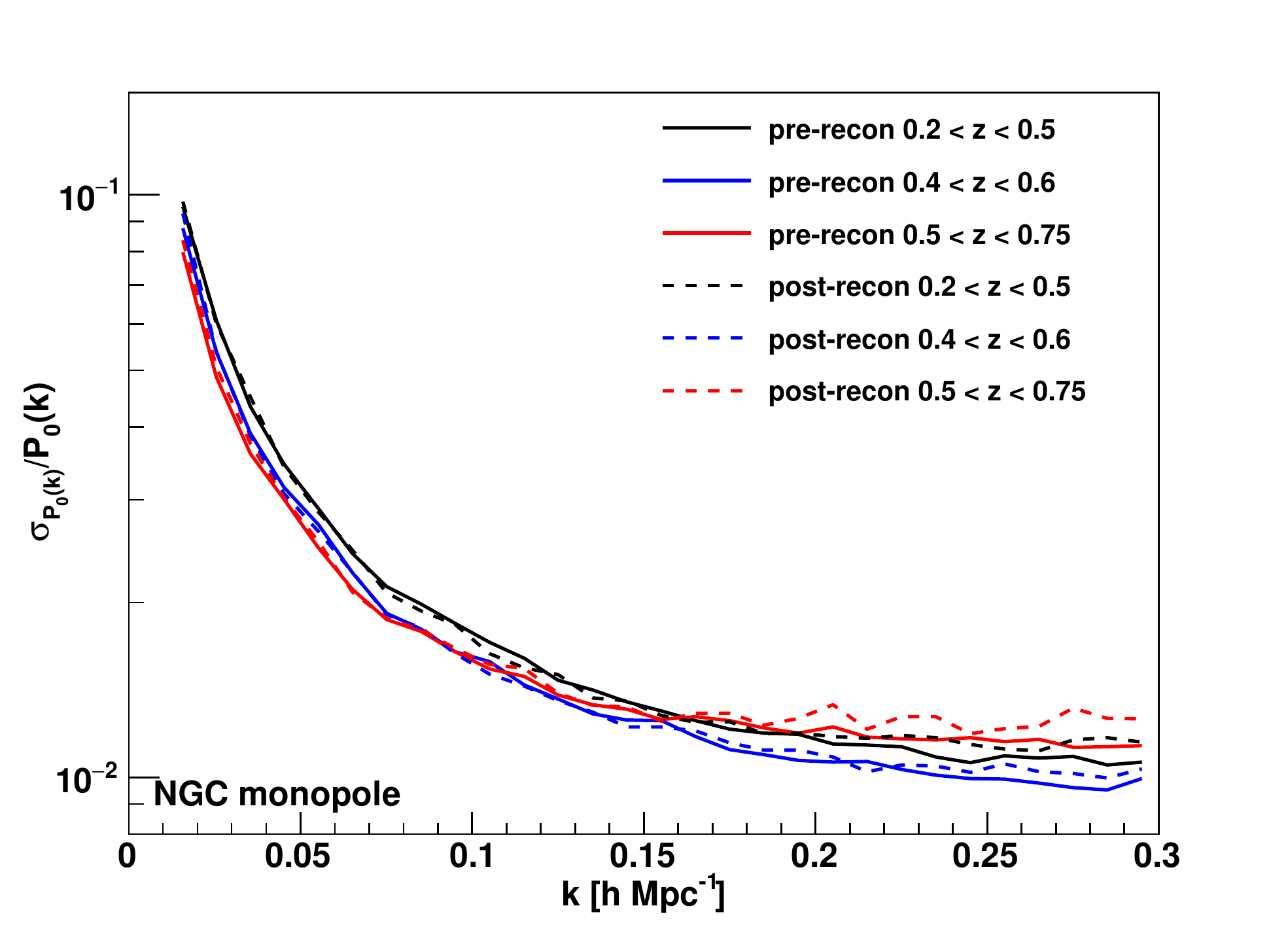,width=8.8cm}
\epsfig{file=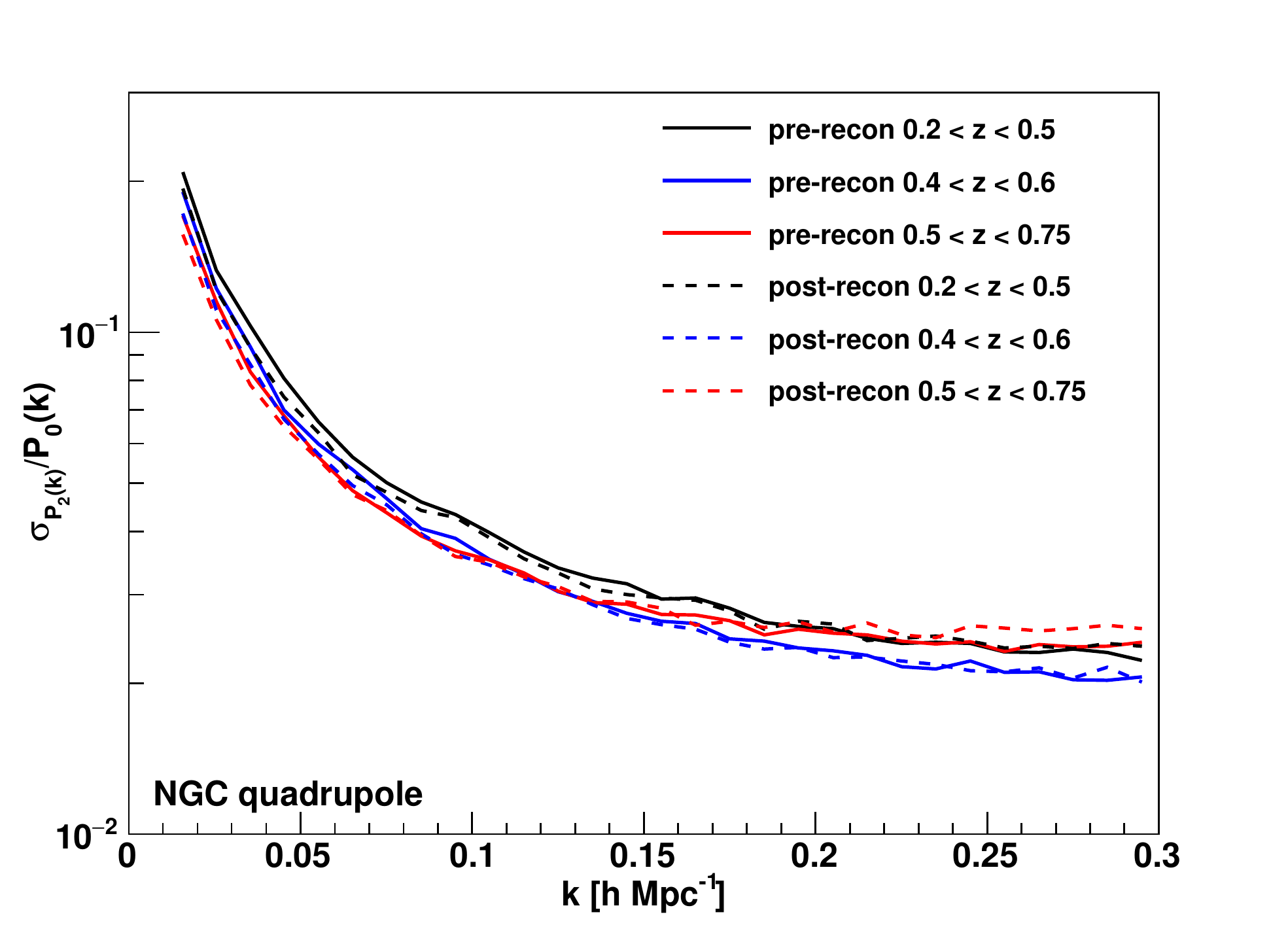,width=8.8cm}\\
\epsfig{file=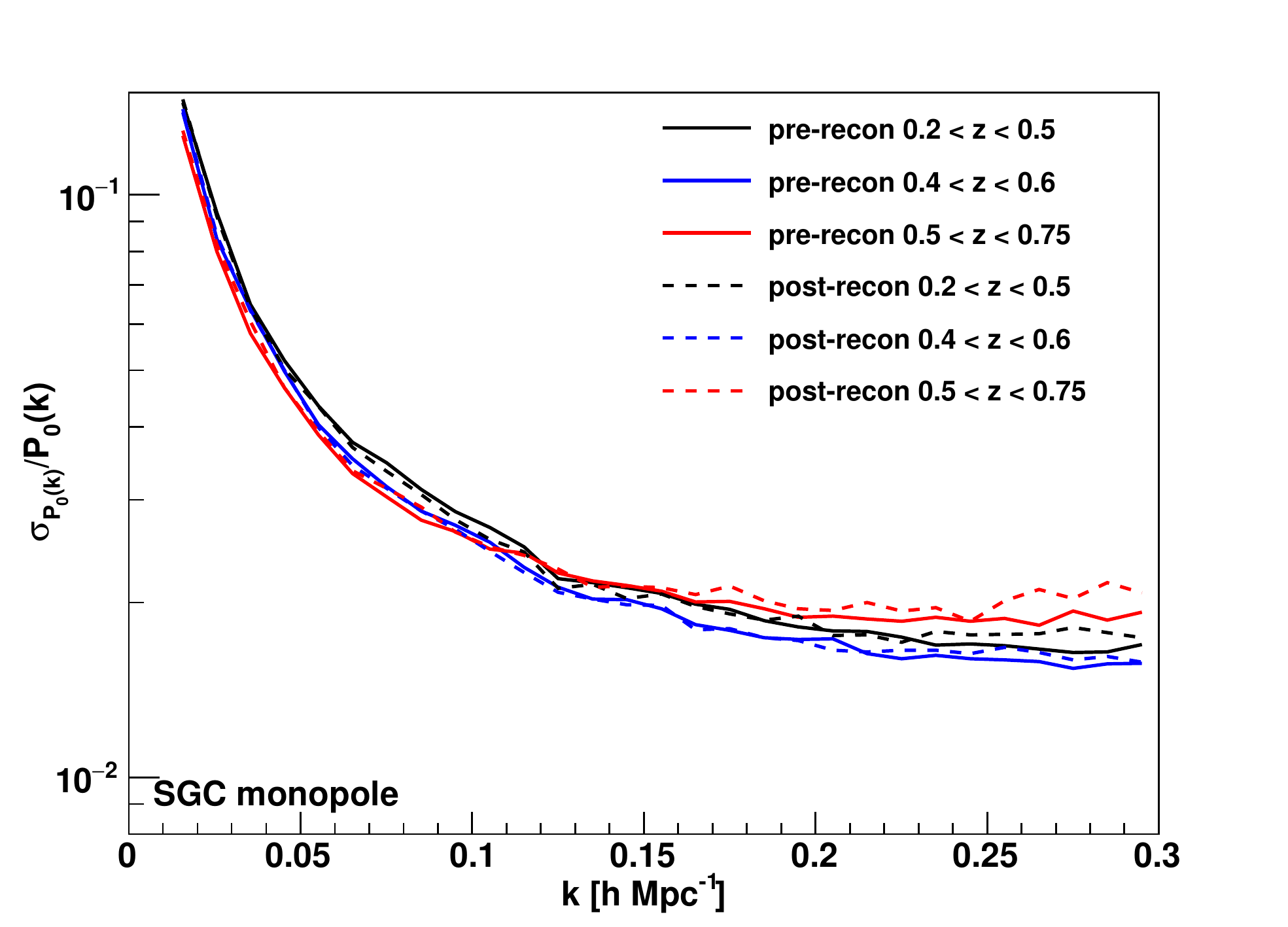,width=8.8cm}
\epsfig{file=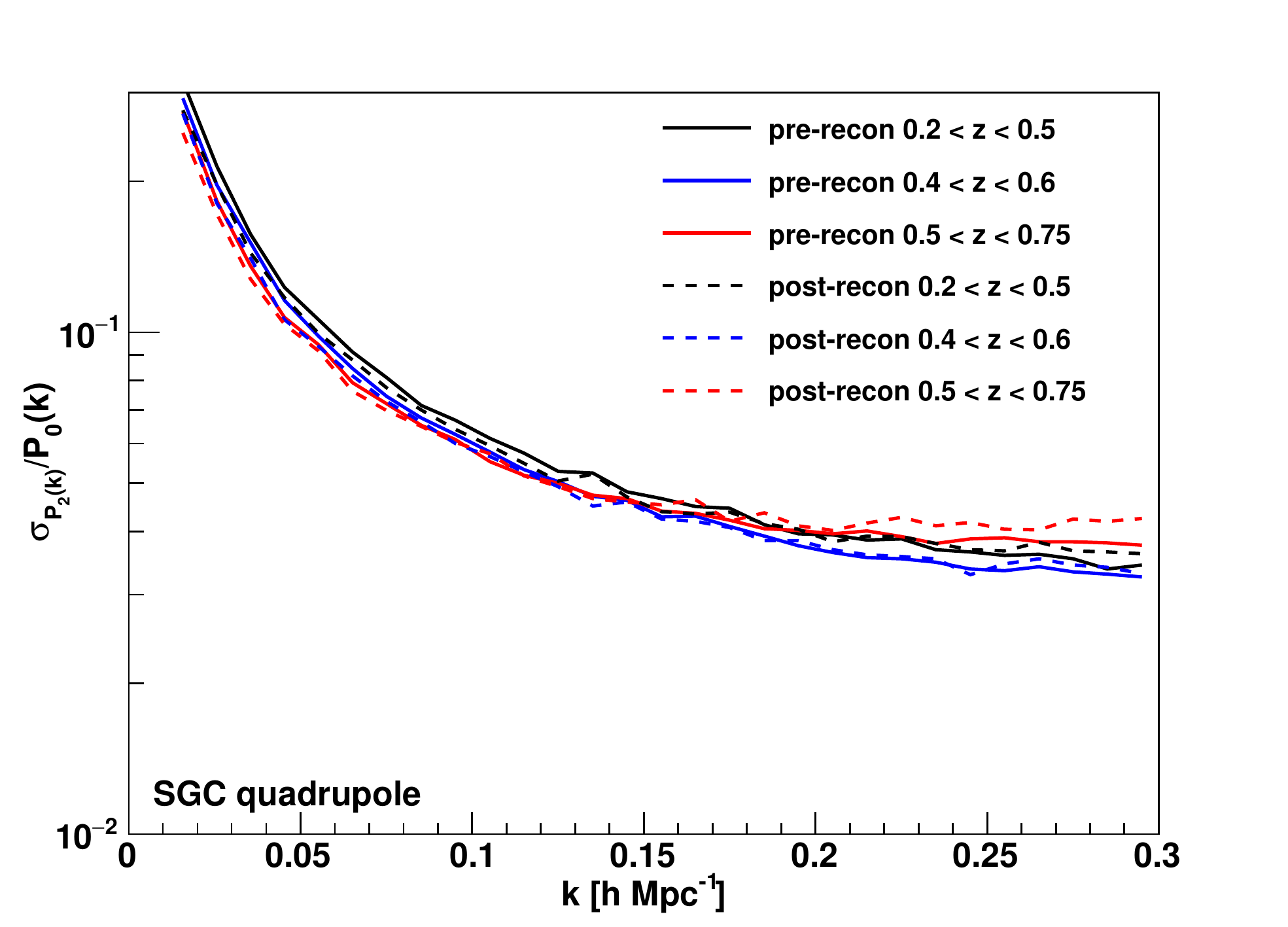,width=8.8cm}
\caption{The relative uncertainty of the NGC power spectrum monopole (left) and
quadrupole (right) before (solid lines) and after (dashed lines) density field 
reconstruction. The power spectrum monopole in the denominator does have the 
shot noise subtracted (we use the monopole in the denominator of the 
quadrupole plot because the quadrupole is often nearly zero).}
\label{fig:diag}
\end{center}
\end{figure*}

Figure~\ref{fig:diag} shows the diagonal elements of the covariance matrix for the monopole and quadrupole power spectrum. We find an error of $\sim1.5\%$ in the monopole and $\sim10\%$ in the quadrupole at $k = 0.15\ihMpc$. This result represents the most precise measurements of the galaxy power spectrum to date. 

Figure~\ref{fig:diag} shows the fractional errors for the monopole and quadrupole pre- and post-reconstruction. In the Gaussian limit the fractional errors depend on the number of independent $k$-modes and the shot noise contribution (on small scales). Because the error decreases by the same factor as the power spectrum decreases, the fractional errors of the monopole are almost identical before and after reconstruction on large scales. Since reconstruction cannot remove the shot noise contribution, the error on small scales cannot decrease significantly, while non-linearity on the shot-noise subtracted monopole is reduced by a small amount. This situation makes the fractional errors of the monopole after reconstruction (dashed lines) slightly larger than the monopole before reconstruction (solid lines) on small scales. The leading contribution to the quadrupole error is produced by the monopole power spectrum, not the quadrupole power spectrum~\citep[see appendix of][]{Taruya:2010mx,Yoo:2013zga}. On small scales, the non-linear effects on the quadrupole decrease after reconstruction, while again the shot noise contribution to the quadrupole error remains the same. As a result, the fractional error on the quadruple on small scales also becomes larger after reconstruction. This result does not contradict the observed improved information content~\citep[e.g.,][]{Ngan2012} after reconstruction since the information content accounts for the entire covariance between different modes and different multipoles, which is reduced after reconstruction. Also, the main signal-to-noise ratio improvement from BAO reconstruction is produced by the sharpening of the BAO.

\subsection{Inverting the mock covariance matrix}

As the estimated covariance matrix $C$ is inferred from mock catalogues, its inverse, $C^{-1}$, provides a biased estimate of the true inverse covariance matrix,~\citep{Hartlap:2006kj}. To correct for this bias we rescale the inverse covariance matrix as
\begin{equation}
C^{-1}_{ij,\rm Hartlap} = \frac{N_s - n_b - 2}{N_s - 1}C^{-1}_{ij},
\label{eq:hartlap}
\end{equation}
where $n_b$ is the number of power spectrum bins. This scaling assumes a Gaussian error distribution and a uncorrelated data vector, which is not strictly true for our dataset (see Figure~\ref{fig:covNGC} and~\ref{fig:covSGC}). We therefore produce many random simulations to keep this scaling factor small. For our post-reconstruction case with $n_b = 58$ and $N_s = 999$ for the SGC ($N_s = 996$ for the NGC) the correction of eq.~\ref{eq:hartlap} increases the parameter variance by about $6\%$. With these covariance matrices we can then perform a standard $\chi^2$ minimisation to find the best fitting parameters.

\section{Density field reconstruction}
\label{sec:recon}

The main complication of studying BAO in the distribution of galaxies compared to similar studies in the CMB arises due to non-linear structure evolution; the damping of the BAO feature lessens the precision of the BAO measurements, and potential shifts in the BAO scale can introduce a systematic bias on the resulting cosmology. Redshift-space distortions enhance such complications along the line of sight. 

Density field reconstruction~\citep{Eisenstein:2006nk} is a technique to enhance the signal-to-noise ratio of the BAO signature by partly undoing non-linear effects of structure formation and redshift-space distortions, i.e., by bringing the information that leaked to the higher order statistics of the galaxy distribution back to the two-point statistic~\citep{Schmittfull:2015mja}. The main steps of density field reconstruction are:
\begin{enumerate}
\item Estimate the displacement field due to structure growth and redshift-space distortions based on the observed galaxy density field.
\item Displace the observed galaxies and a sample of randomly distributed particles with this estimated displacement field.
\item Subtract the data and random displaced density fields.
\end{enumerate}
In this analysis we follow the method of~\citet{Padmanabhan:2012hf}. The observed redshift-space galaxy density field is calculated as 
\begin{equation}
\delta(\vc{s}) = \frac{G(\vc{s})}{\alpha'R(\vc{s})} - 1,
\end{equation}
where $G$ and $R$ are defined in eq~\ref{eq:overdensity}. We  smooth this field with a Gaussian filter of the form
\begin{equation}
S(k) = \exp\left[-(k\Sigma_{\rm smooth})^2/2\right],
\end{equation}
where we chose $\Sigma_{\rm smooth} = 15\hMpc$, which is close to the optimal smoothing scale given the signal-to-noise ratio of the BOSS data~\citep{Xu:2012hg,Burden:2014cwa,Vargas-Magana:2015rqa,Seo:2015eyw}. In linear perturbation theory, the real-space displacement field $\vc{\Psi}(x)$ is related to the redshift-space density field by
\begin{equation}
	\nabla \cdot \vc{\Psi}(\vc{s}) + \beta\nabla\cdot(\vc{\Psi}\cdot\hat{\vc{s}}_{\parallel})\hat{\vc{s}}_{\parallel} = -\frac{\delta(\vc{s})}{b},
\label{eq:bias}
\end{equation} 
where $\hat{\vc{s}}_{\rm los}$ is the unit vector along the line of sight~\citep{Nusser:1993sx}. Assuming the $\vc{\Psi}$ is irrotational, we write $\vc{\Psi} = \nabla \phi$ and solve for the scalar potential $\phi$. To do this we convert all the derivatives to their finite difference counterparts and solve the resulting linear equation~\citep{Padmanabhan:2012hf}. Once $\phi$ is derived, $\vc{\Psi}$ can be calculated using finite differences.

We then apply the displacement to our galaxies by shifting their line-of sight and angular position following 
\begin{align}
s^{\rm new}_{\parallel} &= s_{\parallel}^{\rm old} - (1+f)\Psi_{\parallel}(\vc{s}^{\rm old})
\label{eq:los}\\
s^{\rm new}_{\perp} &= s_{\perp}^{\rm old} - \Psi_{\perp}(\vc{s}^{\rm old}),
\end{align}
where we multiply the derived displacement with $(1+f)$ when displacing the galaxies along the line of sight in order to remove linear redshift-space distortions. Our reconstruction convention therefore substantially removes redshift-space distortions on large scales. The remaining redshift-space distortions are well modelled by a damping term which will be discussed in the next section (see eq~\ref{eq:smani}).

The procedure of reconstruction outlined above does rely on a fiducial cosmological model providing the growth rate $f(z)$, needed in eq.~\ref{eq:los} as well as the bias parameter in eq.~\ref{eq:bias}. We refer to~\citet{Mehta:2011xf} and~\citet{Vargas-Magana:2015rqa} for a detailed study of how these initial assumptions influence the reconstructed BAO results.

This procedure leads to a shifted galaxy, $G^s(\vc{r})$, and shifted random catalogue, $R^{s}(\vc{r})$, where the positions of all galaxies are modified based on the estimated displacement field. The over-density field $D(\vc{r})$, required for the power spectrum estimate can be obtained in an analogous way to eq.~\ref{eq:overdensity} and is given by
\begin{equation}
D^{s}(\vc{r}) = G^s(\vc{r}) - \alpha'R^s(\vc{r}).
\end{equation}

\section{The power spectrum model}
\label{sec:model}

Here we introduce the anisotropic and isotropic power spectrum model used to extract the BAO information by fitting to the measurements. The method used in this paper follows~\citealt{Anderson:2012sa,Anderson:2013zyy} with small modifications as discussed in~\citet{Seo:2015eyw}.

\subsection{The anisotropic case}
\label{sec:ani}

Our anisotropic power spectrum model is given by 
\begin{equation}
\begin{split}
P(k,\mu) &= P_{\rm sm}(k,\mu)\times\\
&\left[ 1 + \left(O_{\rm lin}(k) - 1\right)e^{-\left[k^2\mu^2\Sigma_{\parallel}^2 + k^2(1-\mu^2)\Sigma_{\perp}^2\right]/2}\right],
\end{split}
\label{eq:BAOmodelfirst}
\end{equation}
where $\mu$ is the cosine angle to the line of sight, $O_{\rm lin}(k)$ represents the oscillatory part of the fiducial linear power spectrum, and $P_{\rm sm}(k,\mu)$ is the smooth anisotropic power spectrum. We use two damping scales to model the anisotropic non-linear damping on the BAO feature, one for modes along the line-of-sight, $\Sigma_{\parallel}$ and one for modes perpendicular to the line-of-sight, $\Sigma_{\perp}$. To obtain $O_{\rm lin}(k)$ we fit the fiducial linear power spectrum, $P_{\rm lin}(k)$, with an~\citet{Eisenstein:1997ik} no-Wiggle power spectrum, $P_{\rm nw}(k)$, together with five polynomial terms and derive a smooth fit, $P_{\rm sm,lin}(k)$.
The oscillatory part is then given by 
\begin{equation}
O_{\rm lin}(k) = \frac{P_{\rm lin}(k)}{P_{\rm sm,lin}(k)}.
\end{equation}
The smooth anisotropic power spectrum, $P_{\rm sm}(k,\mu)$, is given by 
\begin{equation}
P_{\rm sm}(k,\mu) = B^2(1 + \beta\mu^2R)^2P_{\rm sm,lin}(k)F_{\rm fog}(k,\mu,\Sigma_s),
\label{eq:smani}
 \end{equation}
where $R = 1$ before density field reconstruction and $R = 1-\exp\left[-(k\Sigma_{\rm smooth})^2/2\right]$ after reconstruction. The parameter $B$ is used to marginalise over the power spectrum amplitude. \citet{Seo:2015} demonstrates that this $R$-term after reconstruction depends on the conventions used in the reconstruction process. The $R$ term we use accounts for the removal of redshift-space distortions on large scales during reconstruction (`Rec-Iso' convention in \citealt{Seo:2015}), and the smoothing scale $\Sigma_{\rm smooth} = 15\hMpc$ used when deriving the displacement field (see section~\ref{sec:recon} for details).
The damping term $F_{\rm fog}(k,\mu, \Sigma_s)$ due to the non-linear velocity field (`Finger-of-God') is given by 
\begin{equation}
F_{\rm fog}(k,\mu,\Sigma_s) = \frac{1}{(1 + k^2\mu^2\Sigma_s^2/2)^2}.
\label{eq:lorents}
\end{equation}
We add extra polynomial terms to marginalise over the angle-dependent overall shape of the power spectrum. The power spectrum monopole and quadrupole are
\begin{align}
P_0(k) &= \frac{1}{2}\int^1_{-1}P(k,\mu)d\mu + A_0(k),\\
P_2(k) &= \frac{5}{2}\int^1_{-1}P(k,\mu)\mathcal{L}_2(\mu)d\mu + A_2(k),
\end{align}
where
\begin{align}
A^{\rm pre-recon}_{\ell}(k) &= \frac{a_{\ell,1}}{k^3} + \frac{a_{\ell,2}}{k^2} + \frac{a_{\ell,3}}{k} + a_{\ell,4} + a_{\ell,5}k.\\
A^{\rm post-recon}_{\ell}(k) &= \frac{a_{\ell,1}}{k^3} + \frac{a_{\ell,2}}{k^2} + \frac{a_{\ell,3}}{k} + a_{\ell,4} + a_{\ell,5}k^2.
\label{eq:BAOmodellast}
\end{align}
The decision which polynomial to use is based on the $\Delta\chi^2$ achieved by each term. The data prefers a linear polynomial in the pre-recon case and a $k^2$ polynomial post-recon leading to a different set of polynomials in the two cases.

Given that the galaxies in the NGC and SGC follow slightly different selections~\citep{Alam2016}, we use two separate parameters to describe the clustering amplitude in the two samples: $B_{\rm SGC}$ and $B_{\rm NGC}$. In our analysis we fix $\Sigma_{\parallel} = 4h^{-1}$Mpc and $\Sigma_{\perp} = 2h^{-1}$Mpc for the post-reconstruction case and $\Sigma_{\parallel} = 8h^{-1}$Mpc and $\Sigma_{\perp} = 4h^{-1}$Mpc for the pre-reconstruction case~\citep{Eisenstein:2006nj}. The exact choice for $\Sigma_{\perp}$ and $\Sigma_{\parallel}$ does not affect our analysis. This approach leads to $14$ free nuisance parameters ($B_{\rm SGC}, B_{\rm NGC}, \beta, a_{\ell,1-5}, \Sigma_s$). In \S~\ref{sec:alcock} we will introduce the two BAO scale parameters $\alpha_{\perp}$ and $\alpha_{\parallel}$, which will complete our set of $16$ free fitting parameters (we fit the NGC and SGC power spectra simultaneously).

\subsection{The anisotropic standard ruler test} 
\label{sec:alcock}

If the fiducial cosmological parameters used to convert galaxy redshifts into physical distances and angular separations into the physical separations deviate from the true cosmology, the observed BAO scale will deviate from the true BAO scale, i.e., the sound horizon scale $r_s(z_d)$. In the full (i.e., anisotropic) standard ruler test, we can measure the deviations along the line-of-sight and perpendicular to the line of sight separately, thereby deriving constraints on the true Hubble parameter and angular diameter distance relative to the fiducial relations.
We parameterise the observed BAO scales along and perpendicular to the line-of-sight relative to the BAO scale in the power spectrum template using two scaling parameters 
\begin{align}
\alpha_{\parallel} &= \frac{H^{\rm fid}(z)r^{\rm fid}_s(z_d)}{H(z)r_s(z_d)},\\
\alpha_{\perp} &= \frac{D_A(z)r^{\rm fid}_s(z_d)}{D^{\rm fid}_A(z)r_s(z_d)},
\end{align}
where $H^{\rm fid}(z)$ and $D^{\rm fid}_A(z)$ are the fiducial values for the Hubble parameter and angular diameter distance at the effective redshift of the sample, and $r^{\rm fid}_s(z_d)$ is the fiducial sound horizon assumed in the template power spectrum. The sound horizon scale $r^{\rm fid}_s(z_d)$ is considered here to correct for the fiducial location of the BAO feature assumed in the template.
Alternatively we can use the values 
\begin{align}
\alpha &= \alpha_{\parallel}^{1/3}\alpha_{\perp}^{2/3}
\label{eq:ep_alpha}\\
\epsilon &= \left( \frac{\alpha_{\parallel}}{\alpha_{\perp}}\right)^{1/3} - 1,
\label{eq:ep_epsilon}
\end{align}
where $\alpha$ describes an isotropic shift (radial dilation) in the BAO scale and $\epsilon$ captures any anisotropic warping. We will employ both expressions for the rest of this paper.

The true wave-numbers ($k_{\parallel}'$ and $k_{\perp}'$) are related to the observed wave-numbers by $k_{\parallel}' = k_{\parallel}/\alpha_{\parallel}$ and $k_{\perp}' = k_{\perp}/\alpha_{\perp}$. Transferring this information into scalings for the absolute wavenumber $k = \sqrt{k^2_{\parallel} + k^2_{\perp}}$ and the cosine of the angle to the line-of-sight $\mu$, we can relate the true and observed values by
\begin{align}
k' &= \frac{k}{\alpha_{\perp}}\left[1 + \mu^2\left(\frac{1}{\FAP^2} - 1\right)\right]^{1/2},
\label{eq:scaling1}\\
\mu' &= \frac{\mu}{\FAP}\left[1 + \mu^2\left(\frac{1}{\FAP^2} - 1\right)\right]^{-1/2}
\label{eq:scaling2}
\end{align}
with $F = \alpha_{\parallel}/\alpha_{\perp}$~\citep{Ballinger:1996cd}. The multipole power spectrum, including the BAO radial dilation and warping, can be written as
\begin{align}
P_{\ell}(k) &= \left(\frac{r_s^{\rm fid}}{r_s}\right)^3\frac{(2\ell + 1)}{2\alpha^2_{\perp}\alpha_{\parallel}}\int^1_{-1}d\mu\; P_{\rm g}\left[k'(k, \mu), \mu'(\mu)\right]\mathcal{L}_{\ell}(\mu),
\label{eq:multi}
\end{align}
where $\left(\frac{r_s^{\rm fid}}{r_s}\right)^3\frac{1}{\alpha^2_{\perp}\alpha_{\parallel}}$ accounts for the difference in the cosmic volume in different cosmologies. The ratio of sound horizons is needed to compensate for the sound horizons included in the definitions of the $\alpha$ values. However, since this term is degenerate with our free amplitude parameters, it has no effect on our BAO analysis.

\subsection{The isotropic case}
\label{sec:isotropicmodel}

We also constrain the angle average BAO dilation scale using only the monopole power spectrum, which ignores the Alcock-Paczynski effect by holding the 
Alcock-Paczynski shape $D_AH$ fixed at the fiducial shape, when spherically averaging the clustering information (i.e. we are assuming the radial and transverse distance scales to be the same). In that case we cannot separately constrain $D_A$ and $H$, but only the radial BAO dilation in a spherically averaged clustering, which is traditionally defined as
\begin{equation}
D_V(z) = \left[(1+z)^2D^2_A(z)\frac{cz}{H(z)}\right]^{1/3},
\end{equation}
Our model for the isotropic (monopole only) analysis is a simplified version of the model used in the anisotropic case. Since $\beta$ and $B$ are degenerate when fitting only the monopole power spectrum before reconstruction, we remove the $(1 + \beta\mu^2)^2$ term. The oscillation damping term simplifies to $\Sigma_{\rm nl} \sim \sqrt{(\Sigma^2_{\parallel} + 2\Sigma^2_{\perp})/3}$, and we remove the $\mu$ dependence in eq.~\ref{eq:lorents}. We therefore have
\begin{equation}
P(k) = P_{\rm sm}(k)\left[ 1 + \left(O_{\rm lin}(k) - 1\right)e^{-\left[k^2\Sigma_{\rm nl}^2\right]/2}\right],
\label{eq:isoall}
\end{equation}
and
\begin{equation}
P_{\rm sm}(k) = B^2 P_{\rm sm,lin}(k)F_{\rm fog}(k,\Sigma_s)
\label{eq:smiso}
 \end{equation}
with $P_{\rm sm,lin}$ as given in eq.~\ref{eq:smani}.
 The velocity damping term is given by
\begin{equation}
F_{\rm fog}(k,\Sigma_s) = \frac{1}{(1 + k^2\Sigma_s^2/2)^2}.
\end{equation}
The effect of the radial dilation of the BAO is included as
\begin{align}
P_0(k) &= \left(\frac{r_s^{\rm fid}}{r_s}\right)^3\frac{(2\ell + 1)}{2\alpha^3}\int^1_{-1}d\mu\; P_{\rm g}\left(k'=k/\alpha, \mu\right)\mathcal{L}_{0}(\mu),
\end{align}
where
\begin{align}
\alpha =  \frac{D_V(z)r^{\rm fid}_s(z_d)}{D^{\rm fid}_V(z)r_s(z_d)} .
\end{align}
In total, we have $10$ free parameters in the isotropic case ($B^{\rm NGC}, B^{\rm SGC}, a_{0,1-5}, \alpha, \Sigma_{\rm nl}, \Sigma_s$).

We expect the constraint on $\alpha$ in the isotropic case to be tighter than the constraint on $\alpha$ in the anisotropic case, since in the latter, $\alpha$ (in eq~\ref{eq:ep_alpha}) is marginalised over the warping effect while in the former analysis it is not. We will consider the anisotropic constraints as the our main result, since the anisotropic analysis depends on fewer assumptions. We will show constraints on $D_V$ from the isotropic analysis only for comparison. 

\subsection{Correction for the irregular $\mu$ distribution}
\label{sec:binning}

\begin{figure}
\begin{center}
\epsfig{file=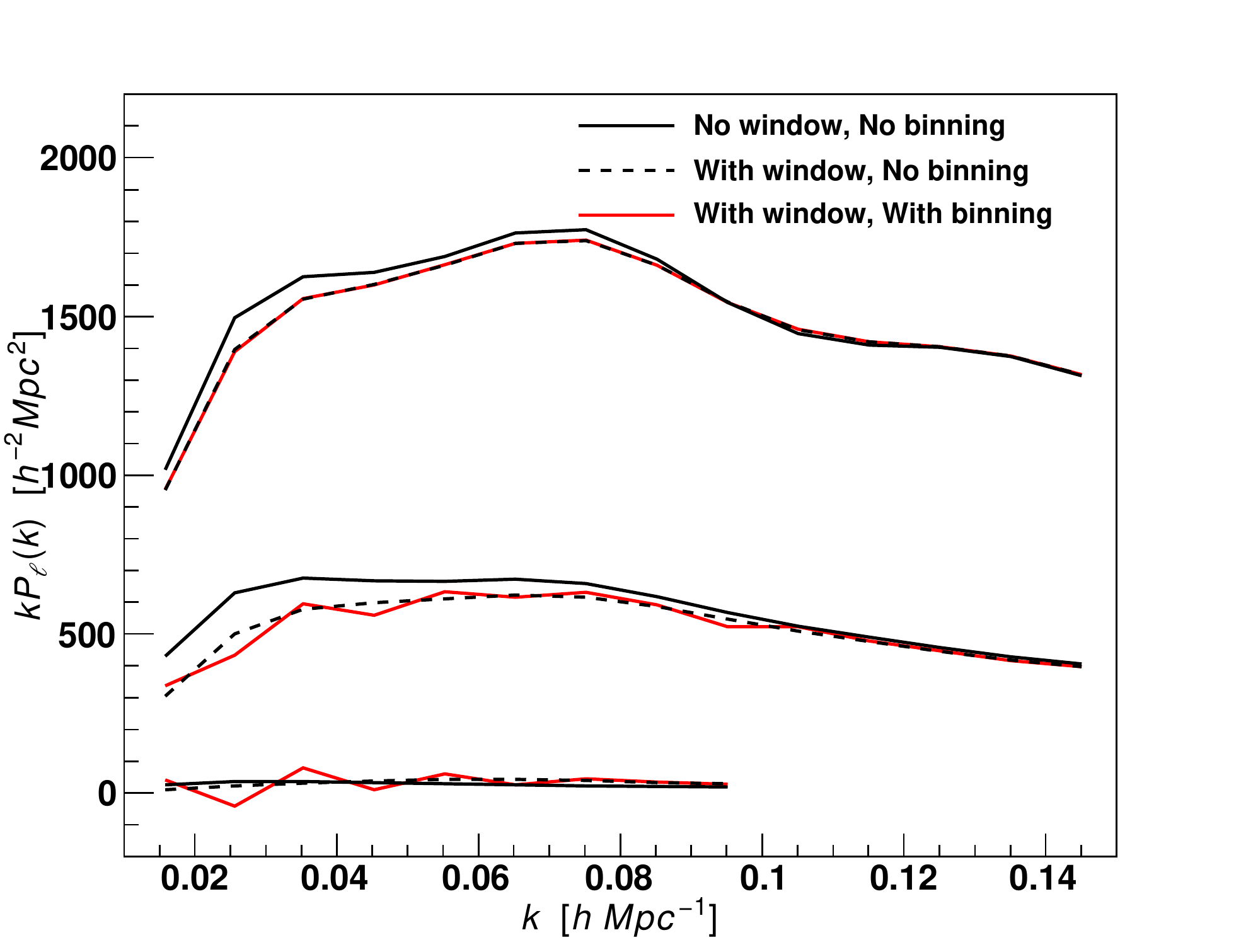,width=8cm}
\caption{The window function and discreteness effects for the lowest redshift bin in the South Galactic Cap (SGC). The three lines show the raw power spectrum model (black solid line), the same model including the convolution with the window function (black dashed line) and including the discreteness effect of \S~\ref{sec:binning} (red solid line). The SGC in the lowest redshift bin has the smallest volume and therefore both the window function and discreteness effects are expected to be largest in this case.}
\label{fig:binning}
\end{center}
\end{figure}

Because the survey volume is not infinite, the power spectra are estimated on a finite and discrete k-space grid. Performing FFTs in a Cartesian lattice makes the angular distribution of the Fourier modes irregular and causes deviation from the isotropic distribution, more so at smaller $k$. As a result, we see small fluctuation-like deviations in the measured power spectrum multipoles that are not caught by the window function, as shown in Figure \ref{fig:binning}. The effect is larger for the quadrupole than the monopole since the quadrupole is more sensitive to an anisotropy of the mode distribution. Given that the SGC in the lowest redshift bin has the smallest volume, we expect this effect to be greatest for this case. In this paper, we include this effect in our power spectrum monopole and quadrupole model. When calculating multipoles, we weight each $\mu$ bin by the normalised number of modes $N(k,\mu)$ counted on a k-space grid that is the same as the grid used to estimate the measured power spectrum. More details of the correction method is discussed in B16. This effect, being apparent only at small $k$, does not influence the result of our analysis.

\subsection{Fitting preparation}

Using the covariance matrix we perform a $\chi^2$ minimisation to find the best fitting parameters. In addition to the scaling of the inverse covariance matrix of eq.~\ref{eq:hartlap}, we must propagate the error in the covariance matrix to the error on the estimated parameters; this is done by scaling the variance for each parameter by~\citep{Percival:2013}
\begin{equation}
M_1 = \sqrt{\frac{1 + B(n_b - n_p)}{1 + A + B(n_p + 1)}}
\label{eq:percival}
\end{equation}
where $n_p$ is the number of parameters and
\begin{align}
A &= \frac{2}{(N_s - n_b - 1)(N_s - n_b - 4)}, \\
B &= \frac{N_s - n_b - 2}{(N_s - n_b -1)(N_s - n_b - 4)}. 
\end{align}
Using our post-reconstruction values of $N_s = 999$ for SGC and $996$ for the NGC, $n_b = 58$ and $n_p = 16$, we obtain a correction of $M_1 \approx 1.013$. When dealing with the variance or standard deviation of a distribution of finite mock results, which has also been fitted with a covariance matrix derived from the same mock results, the standard deviation from these mocks needs to be corrected as 
\begin{equation}
M_2 = M_1\sqrt{\frac{N_s-1}{N_s-n_b-2}}.
\end{equation}  

\section{Testing the model}
\label{sec:sys}

\subsection{Theoretical systematics}
\label{sec:theo}

While perturbation theory can attempt to provide a model for the non-linear power spectrum on quasi-linear scales ($k \leq 0.2\hMpc$), most observed modes are outside the realm of perturbation theory and it has proven very difficult to extract information from these modes.  When focusing on the BAO feature, however, the two main non-linear effects are non-linear damping and an additional small-scale power due to mode-coupling~\citep{Eisenstein:2006nj,Crocce:2007dt,Matsubara:2007wj,Seo:2008yx,Seo:2009fp}: 
\begin{equation}
P_{\rm g}(k,\mu) = G^2(k,\mu,z)P_{\rm lin}(k,\mu) + P_{\rm MC}.
\label{eq:MC}
\end{equation}
Here the propagator $G$ describes the cross-correlation between the initial and final density field, which is responsible for the damping of the BAO. In the high-$k$ limit the dominant behaviour of the propagator can be predicted using perturbation theory~\citep[e.g.,][]{Crocce:2005xz,Matsubara:2007wj} as
\begin{equation}
G \sim \exp\left(-\frac{1}{2}k^2\Sigma^2\right).
\end{equation}
$N$-body simulations have demonstrated that this form is a good approximation over the wave modes that are relevant to the BAO feature before reconstruction, and often even after reconstruction~\citep[e.g.,][]{Seo:2009,Seo:2015eyw}\footnote{\citet{White:2015,Seo:2015eyw} show that it depends on the convention and the details used in the reconstruction procedure.}.
 
The mode-coupling term in eq.~\ref{eq:MC} can be written in standard PT ~\citep{Jain94} as 
\begin{equation}
P_{\rm MC}(k) \simeq 2\int \left[F_2(\vc{k}-\vc{q},\vc{q})\right]^2P_{\rm lin}(\vc{q})P_{\rm lin}(|\vc{k}-\vc{q}|)d\vc{q} + \dots
\end{equation}
with the second-order PT kernel
\begin{equation}
F_{2}(\vc{k}_1,\vc{k}_2) = \frac{5}{7}+\frac{2}{7}\left(\frac{\vc{k}_1\cdot\vc{k}_2}{k_1k_2}\right)^2 + \frac{\vc{k}_1\cdot\vc{k}_2}{2}\left(\frac{1}{k_1^2} + \frac{1}{k_2^2}\right).
\end{equation}
The $F_2$-kernel divides the mode-coupling term into three parts, where the first describes the growth of perturbations, the second represents the transport of matter by the velocity field, and the last term describes the impact of tidal gravitational fields in the growth of structure. The leading contribution to the shift in the BAO scale results from the product of the first and second term~\citep{Crocce:2007dt,Sherwin:2012nh}. The amplitude of this shift depends on redshift (through the growth factor) and galaxy bias~\citep{Seo:2008yx, Seo:2009fp, Padmanabhan:2009yr, Mehta:2011xf} and is approximately described by 
\begin{equation}
\alpha - 1 \approx 0.5\%\left(1 + \frac{3b_2}{2b_1}\right)\left[D(z)/D(0)\right]^2,
\end{equation}
where $D$ is the growth factor. Using the effective redshifts $z_{\rm eff} = 0.38$, $0.51$ and $0.61$ and assuming $b_1 = 2$, $b_2 = 0.2$ and $\Omega_m = 0.3$ within a flat $\Lambda$CDM cosmology, the equation above predicts systematic shifts of $\Delta\alpha = 0.39$, $0.36$ and $0.33\%$, respectively, without accounting for redshift-space distortions. These values are more than a factor of two times smaller than our best measurement uncertainties.

Furthermore, the technique of density field reconstruction, which we describe in section~\ref{sec:recon}, has been shown to substantially undo the non-linear damping as well as remove the mode coupling bias~\citep{Seo:2008yx,Padmanabhan:2009yr, Sherwin:2012nh}. While the efficiency of density field reconstruction depends on the noise level of the galaxy density field as well as various details used during reconstruction, \citet{Seo:2008yx} demonstrated that the shifts are reduced to less than $0.1\%$ even in the presence of non-negligible shot noise, implying the mode-coupling term is quite robustly removed. In our analysis, the BAO constraints clearly improve after reconstruction to the degree that is consistent with the effect seen in the mock catalogues. This result suggests that reconstruction is working and therefore the mode-coupling term should be removed. We therefore proceed without any treatment for a potential systematic bias due to mode coupling.

It has been suggested that the supersonic streaming velocity of baryons relative to dark matter at high redshift may have left an imprint in the low redshift galaxy distribution such that the BAO scale shrinks or stretches relative to the conventional, zero-streaming velocity prediction~\citep[e.g.,][]{Tseliakhovich10,Dalal10,Yoo11,Beutler:2015tla}. \citet{Blazek15} predicts that a level of 1\% effect on the density fluctuation (i.e., streaming velocity bias) will induce a $\sim 0.5\%$ shift in the BAO scale. With little information on the magnitude and sign of the streaming velocity bias and its effect on the reconstruction process, we ignore a possible systematic bias due to this effect.

\subsection{Tests on N-body simulations}

\begin{figure*}
\begin{center}
\epsfig{file=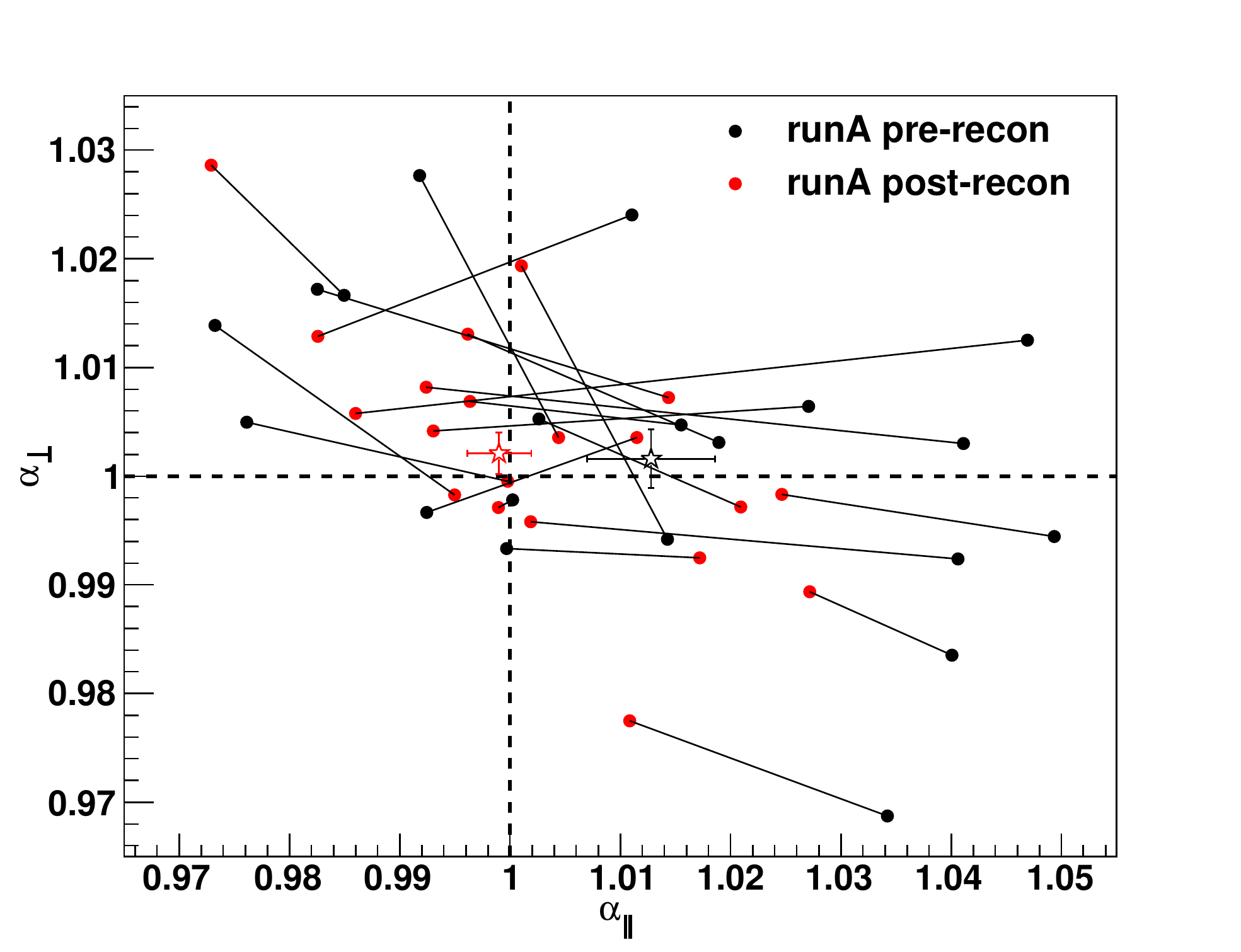,width=8cm}
\epsfig{file=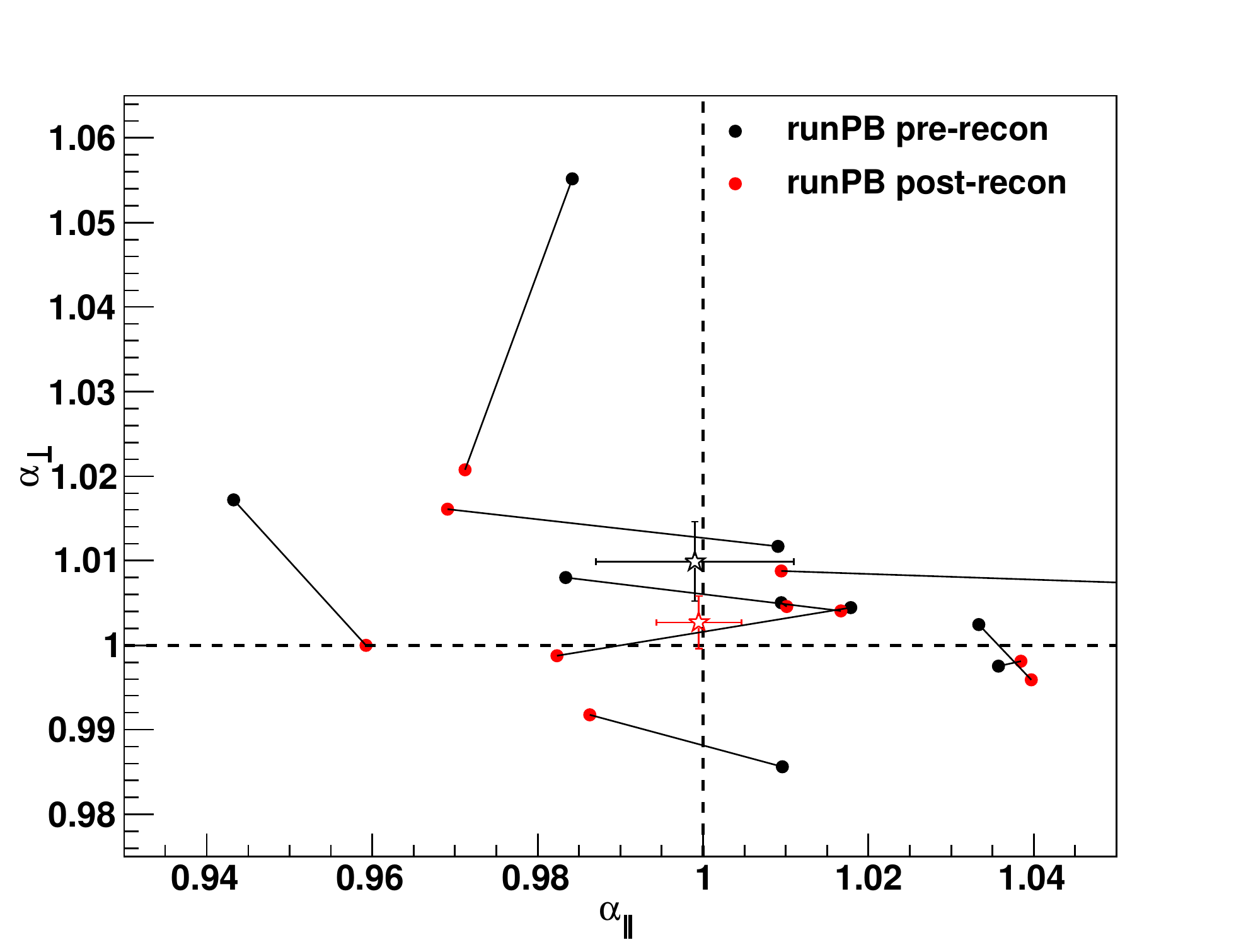,width=8cm}
\caption{The distribution of $\alpha_{\parallel}$ and $\alpha_{\perp}$ for the $20$ realisations of the runA simulation (left) and the $10$ realisations of the runPB simulation (right) before density field reconstruction (black) and after reconstruction (red). The star data points with error bars show the results of the fit to the mean of the simulation boxes again before (black) and after (red) reconstruction (the individual realisations are connected by black lines). The fact that the results before reconstruction (black) are biased to $\alpha > 1$ is consistent with the mode-coupling term. 
Mode-coupling is removed after reconstruction. One of the $10$ runPB 
realisations has $\alpha_{\parallel}=1.28$ before reconstruction, which 
indicates that for this realisation there is no BAO detection along the line 
of sight.}
\label{fig:scatter}
\end{center}
\end{figure*}

\begin{table*}
\begin{center}
\caption{Results for the fit to the mean of the runA and runPB simulations (in periodic boxes). The upper section of the table presents the fitting result using the power spectrum template of the correct cosmology, so that the different scaling parameters $\alpha$ should agree with unity and $\epsilon$ should agree with zero. The lower section of the table uses the runPB cosmology to define the power spectrum template when fitting the runA simulations and the runA cosmology for the fit to runPB simulations. Using a power spectrum template with a different sound horizon results in a shift of the different $\alpha$s, given by the ratio of the true sound horizon to the the used sound horizon. The rows with the label ``(scaled)'' account for the difference in the sound horizon of the two templates, so that these values again should agree with unity. This represents a test of the scaling formalism used to retrieve the BAO scale. The sound horizon for runA is $104.503\hMpc$ ($149.29\,$Mpc), while for runPB it is $102.3477\hMpc$ ($148.33\,$Mpc). The parameters $\alpha$ and $\epsilon$ are derived from 
$\alpha_{\parallel}$ and $\alpha_{\perp}$. The fact that some values of 
$\alpha$ before reconstruction are larger than unity is consisted with the 
mode-coupling term, which predicts a sub-\% level shift to larger $\alpha$ 
(see section~\ref{sec:theo}). Mode-coupling is removed after reconstruction.
}
	\begin{tabular}{lllll}
     		\hline
		  & \multicolumn{2}{c}{runA} & \multicolumn{2}{c}{runPB}\\
		  & pre-recon & post-recon & pre-recon & post-recon\\
		  \hline
		 \multicolumn{5}{c}{anisotropic fit}\\
		\hline
		$\alpha_{\parallel}$ & $1.0128\pm0.0058$ & $0.9973\pm0.0029$ & $0.999\pm0.012$ & $0.9975\pm0.0049$ \\
		$\alpha_{\perp}$ & $1.0016\pm0.0027$ & $1.0013\pm0.0018$ & $1.0099\pm0.0047$ & $1.0017\pm0.0031$ \\
		\hline
		$\alpha$ & $1.0053\pm0.0026$ & $1.0000\pm0.0015$ & $1.0061\pm0.0050$ & $1.0003\pm0.0026$ \\
		$\epsilon$ & $0.0037\pm0.0021$ & $-0.0013\pm0.0011$ & $-0.0037\pm0.0042$ & $-0.0014\pm0.0019$ \\
		\hline
		\multicolumn{5}{c}{isotropic fit}\\
		\hline
		$\alpha$ & $1.0065\pm0.0023$ & $1.0003\pm0.0013$ & $1.0085\pm0.0041$ & $1.0015\pm0.0022$ \\
		\hline
		\hline
		  \multicolumn{5}{c}{anisotropic fit (switched template)}\\
		\hline
		$\alpha_{\parallel}$ & $0.9929\pm0.0058$ & $0.9777\pm0.0030$ & $1.019\pm0.012$ & $1.0176\pm0.0053$ \\
		$\alpha_{\perp}$ & $0.9816\pm0.0026$ & $0.9810\pm0.0018$ & $1.0306\pm0.0048$ & $1.0224\pm0.0031$ \\
		$\alpha_{\parallel}$ (scaled) & $1.0138\pm0.0059$ & $0.9983\pm0.0030$ & $0.998\pm0.012$ & $0.9966\pm0.0052$ \\
		$\alpha_{\perp}$ (scaled) & $1.0023\pm0.0027$ & $1.0017\pm0.0018$ & $1.0093\pm0.0047$ & $1.0013\pm0.0030$ \\
		\hline
		$\alpha$ & $0.9854\pm0.0026$ & $0.9799\pm0.0012$ & $1.0268\pm0.0051$ & $1.0208\pm0.0027$ \\
		$\alpha$ (scaled) & $1.0062\pm0.0027$ & $1.0005\pm0.0012$ & $1.0056\pm0.0050$ & $0.9997\pm0.0026$ \\
		$\epsilon$ & $0.0038\pm0.0021$ & $-0.0011\pm0.0012$ & $-0.0038\pm0.0041$ & $-0.0016\pm0.0020$ \\
		\hline
		\multicolumn{5}{c}{isotropic fit (switched template)}\\
		\hline
		$\alpha$ & $0.9868\pm0.0024$ & $0.9807\pm0.0014$ & $1.0287\pm0.0039$ & $1.0219\pm0.0022$ \\
		$\alpha$ (scaled) & $1.0076\pm0.0025$ & $1.0014\pm0.0014$ & $1.0074\pm0.0038$ & $1.0008\pm0.0022$ \\
		\hline
	  \end{tabular}
	  \label{tab:periodic}
\end{center}
\end{table*}

To test our fitting technique we use two different sets of N-body simulations, designated as runA and runPB. The runA simulations are $20$ halo catalogues of size $[1500\hMpc]^3$ with $1500^3$ particles using the fiducial cosmology of $\Omega_m = 0.274$, $\Omega_{\Lambda} = 0.726$, $n_s = 0.95$, $\Omega_b = 0.0457$, $H_0 = 70\,$km\,s$^{-1}$Mpc$^{-1}$ and $r_s(z_d) = 104.503\hMpc$. The runPB simulations are $10$ galaxy catalogues of size $[1380\hMpc]^3$ with $\Omega_m = 0.292$, $\Omega_{\Lambda} = 0.708$, $n_s = 0.965$, $\Omega_b = 0.0462$, $H_0 = 69\,$km\,s$^{-1}$Mpc$^{-1}$ and $r_s(z_d) = 102.3477\hMpc$. The runPB simulations make use of a CMASS-like halo occupation distribution (HOD) model to populate dark matter halos with galaxies (see~\citealt{Reid:2014iaa} for details).

We calculate the power spectra for the runA and runPB simulations and fit the individual power spectra using the model described in section~\ref{sec:model}. The results are summarised in Table~\ref{tab:periodic} and displayed in Figure~\ref{fig:scatter}. 

In the case of the runA simulations, the pre-reconstruction results indicate a $2\sigma$ bias towards larger values of $\alpha_{\parallel}$. This bias is not statistically significant but might be related to the mode-coupling shift as discussed in section~\ref{sec:theo}. Our power spectrum model does not account for the mode-coupling term and therefore the presence of bias is expected. However, the mode coupling term should be removed after applying density field reconstruction. Our post reconstruction results are indeed consistent with $\alpha = 1$, indicating no systematic bias in our measurements. 

The results for the runPB simulations (Figure~\ref{fig:scatter}, right) are quite similar, even though instead of having a $2\sigma$ bias in $\alpha_{\parallel}$ pre-reconstruction we now find a $2\sigma$ bias in $\alpha_{\perp}$. Again our post-reconstruction results are unbiased.

We also performed tests where we switched the input power spectrum model using the runPB cosmology for the fit to runA and the other way around. These results are included in Table~\ref{tab:periodic} with the label ``switched template''. For these fits an unbiased result does not mean agreement with $\alpha=1$, since the cosmology assumed in the model is different to the true cosmology of the simulation. The results with the label `(scaled)' show the fitting results that are adjusted with the ratio between the cosmology in the template and the input cosmology of the simulations, where now these results can be compared to unity. We found the results are consistent with our previous findings i.e., no bias on the measured BAO scale. 

\subsection{Tests on the MultiDark-Patchy mock catalogues}
\label{sec:patchy}

\begin{figure*}
\begin{center}
\epsfig{file=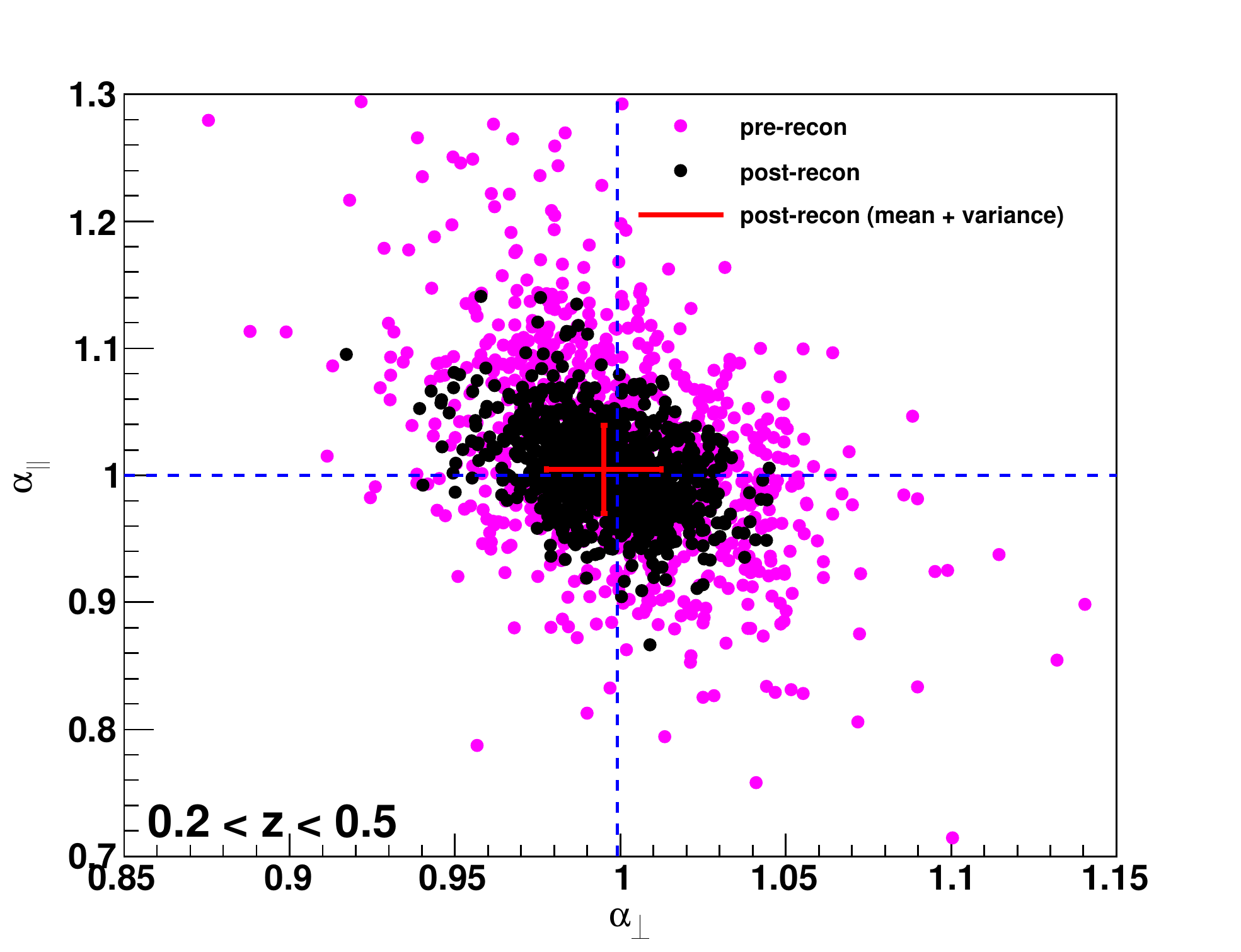,width=8cm}
\epsfig{file=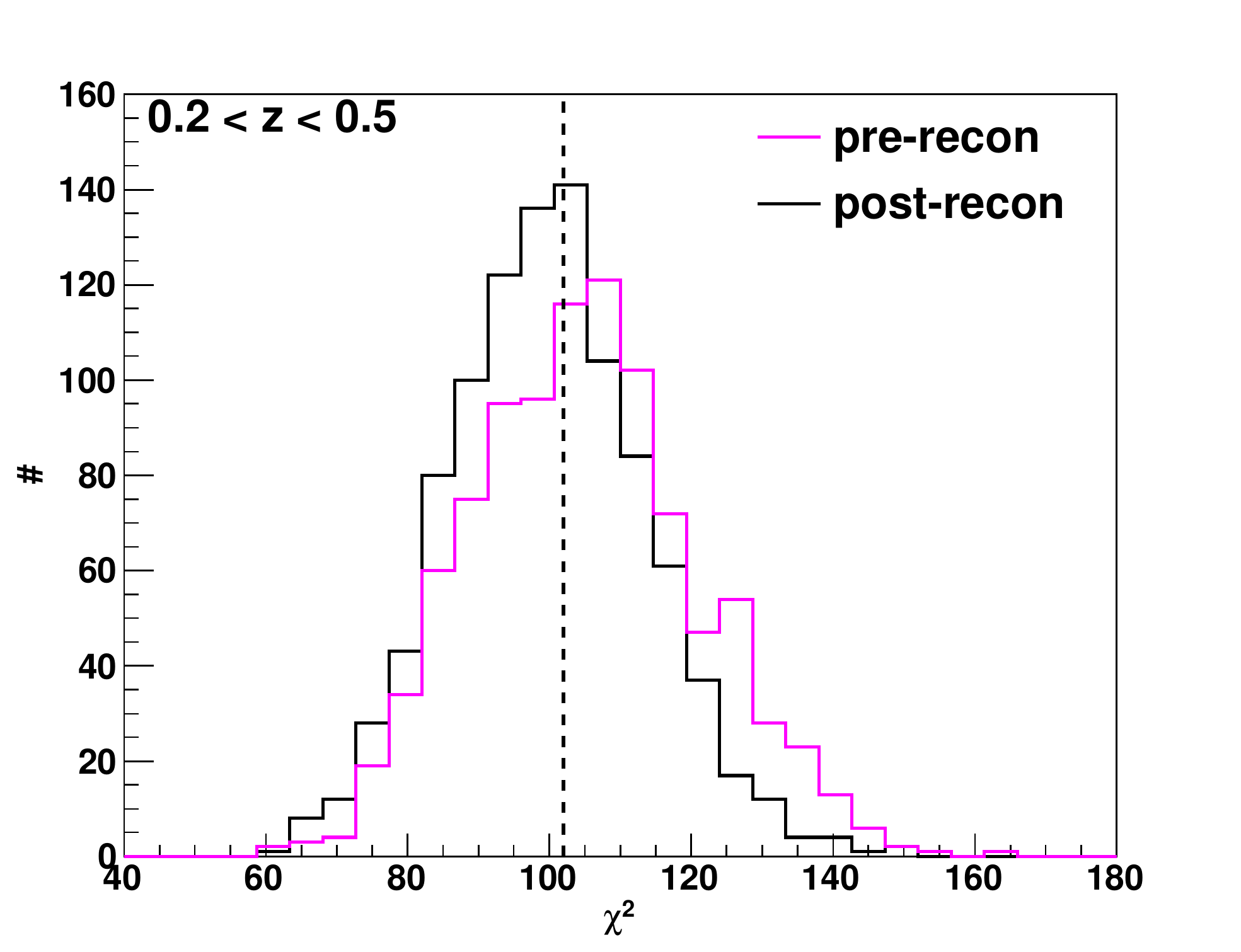,width=8cm}\\
\epsfig{file=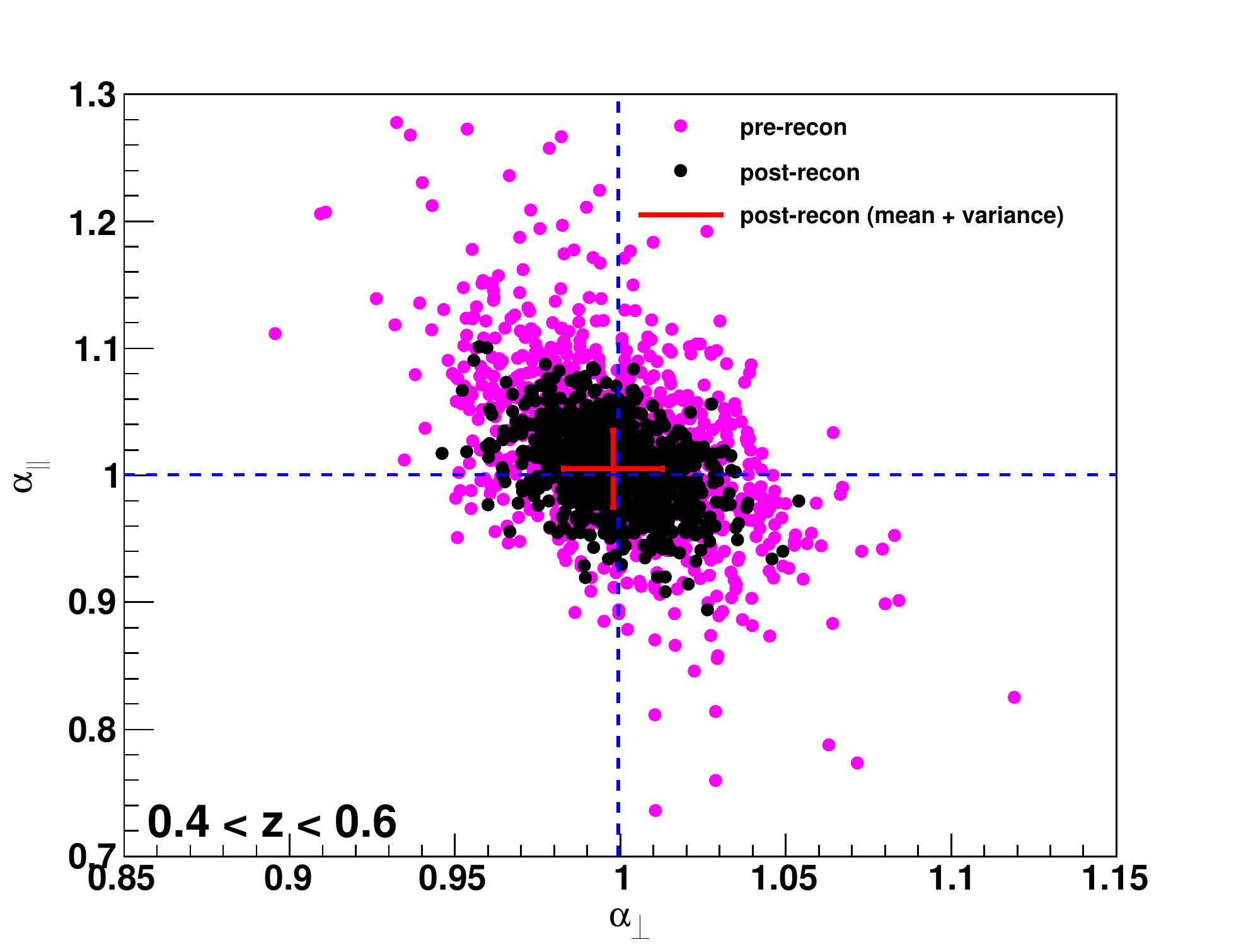,width=8cm}
\epsfig{file=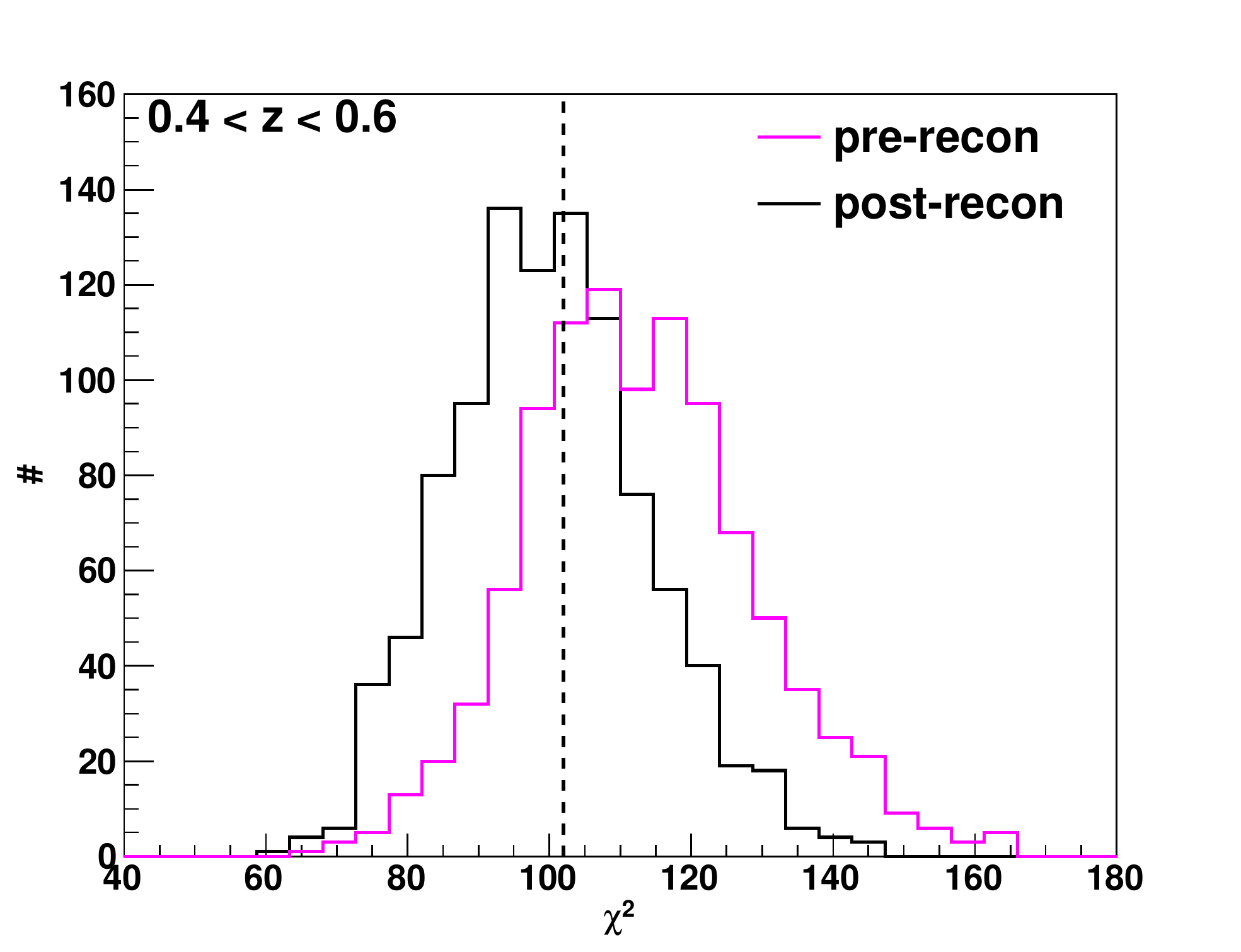,width=8cm}\\
\epsfig{file=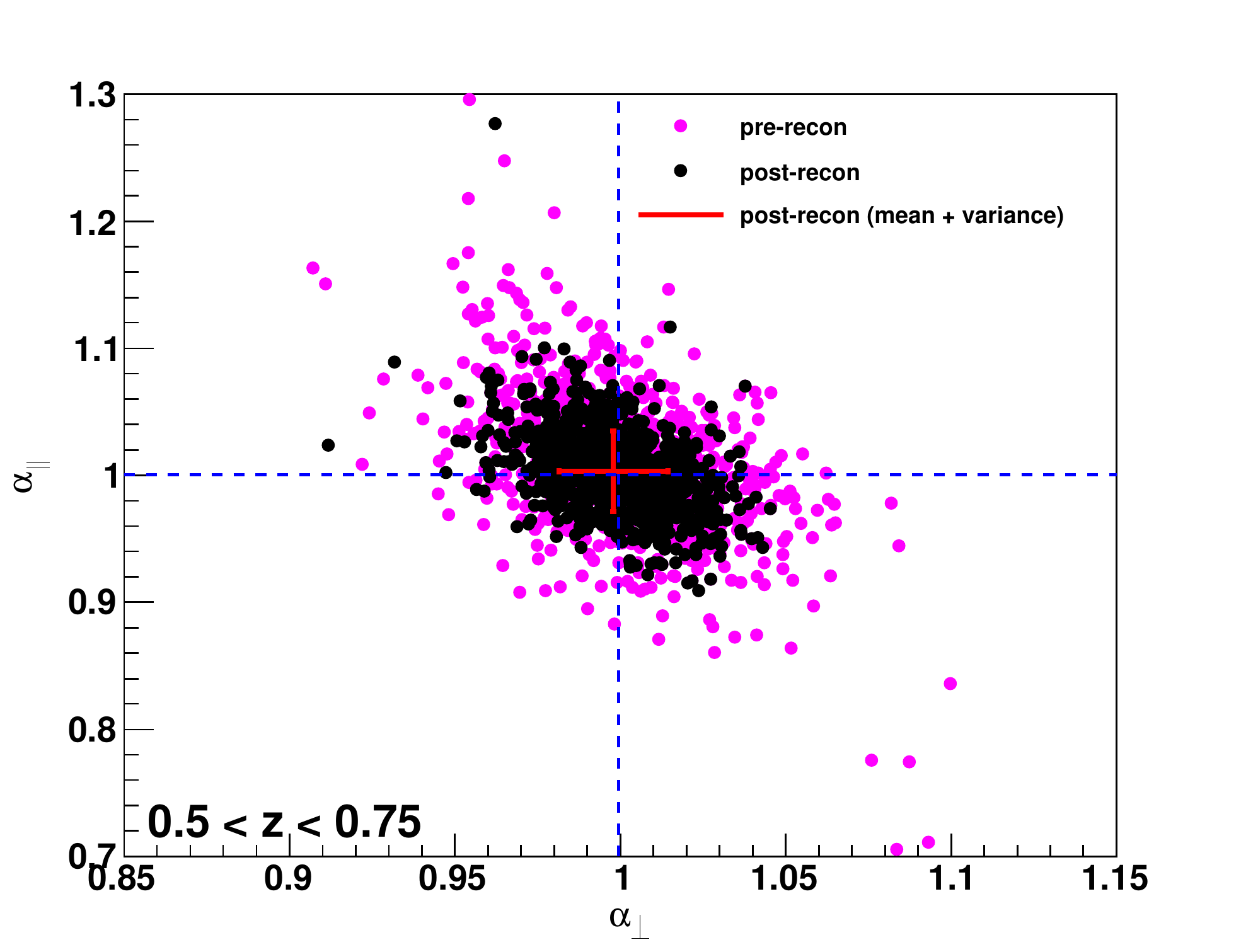,width=8cm}
\epsfig{file=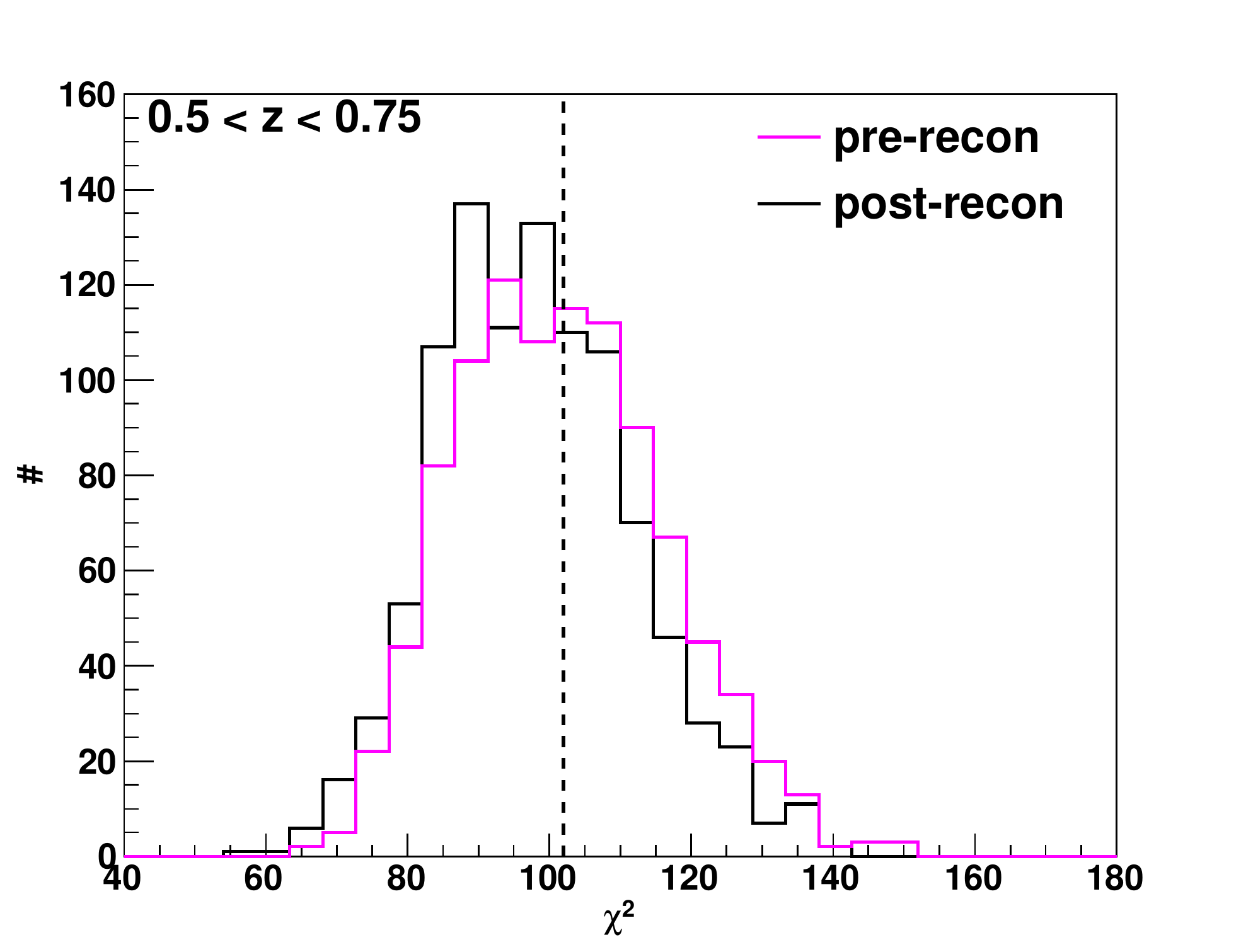,width=8cm}
\caption{Maximum likelihood values for the MultiDark-Patchy mock catalogues (left) and the corresponding minimum, $\chi^2$ distribution (right) for the three redshift bins used in this analysis. The magenta data points on the left and magenta solid line on the right show the pre-reconstruction results, while the black points (left) and the black line (right) present the post-reconstruction results. The red crosses in the left panels are the mean and variance for the mock catalogues post-reconstruction. The black dashed line on the right indicates the degrees of freedom. The results are summarised in Table~\ref{tab:patchy}.}
\label{fig:patchysctatter}
\end{center}
\end{figure*}

\begin{table*}
\begin{center}
\caption{The results for the fits to the MultiDark-Patchy mock catalogues for the three redshift bins used in this analysis before and after density field reconstruction. For each bin we present the anisotropic results ($\alpha_{\parallel}$, $\alpha_{\perp}$) and the isotropic result ($\alpha$). The anisotropic results also show the correlation coefficient $r$ between the two $\alpha$-parameters. The fiducial BOSS cosmology is used when analysing the mock data, which means that the $\alpha$ values do not have to agree with unity. The expectation value for each redshift bin is given in brackets. The uncertainties represent the variance between all mock catalogues (not the error on the mean).}
	\begin{tabular}{lllll}
     		\hline
		  & pre-recon & $r_{\rm pre-recon}$ & post-recon & $r_{\rm post-recon}$\\
		  \hline
		  & \multicolumn{4}{c}{$0.2 < z < 0.5$}\\
		  \hline
		$\alpha_{\parallel}$ & $1.018\pm0.076\;\; (0.9999)$ & \multirow{2}{*}{-0.455} & $1.005\pm0.036\;\; (0.9999)$ & \multirow{2}{*}{-0.398}\\
		$\alpha_{\perp}$ & $0.999\pm0.031\;\; (0.9991)$ & & $0.995\pm0.018\;\; (0.9991)$ &  \\
		$\alpha$ & $1.010\pm0.022\;\;(0.9993)$ & --- & $1.002\pm0.013\;\;(0.9993)$ & ---  \\
		\hline		  
		& \multicolumn{4}{c}{$0.4 < z < 0.6$}\\
		  \hline
		$\alpha_{\parallel}$ & $1.019\pm0.066\;\; (1.0003)$ & \multirow{2}{*}{-0.482} & $1.005\pm0.032\;\; (1.0003)$ & \multirow{2}{*}{-0.397} \\
		$\alpha_{\perp}$ & $0.999\pm0.027\;\; (0.9993)$ & & $0.998\pm0.016\;\; (0.9993)$ &  \\
		$\alpha$ & $1.010\pm0.019\;\;(0.9996)$ & --- & $1.003\pm0.012\;\; (0.9996)$ & --- \\
		\hline
		& \multicolumn{4}{c}{$0.5 < z < 0.75$}\\
		  \hline
		$\alpha_{\parallel}$ & $1.010\pm0.053\;\; (1.0006)$ & \multirow{2}{*}{-0.464} & $1.003\pm0.033\;\; (1.0006)$ & \multirow{2}{*}{-0.413} \\
		$\alpha_{\perp}$ & $1.000\pm0.024\;\; (0.9995)$ & & $0.998\pm0.017\;\; (0.9995)$ &  \\
		$\alpha$ & $1.008\pm0.019\;\;(0.9999)$ & --- & $1.003\pm0.012\;\;(0.9999)$ & --- \\
		\hline
	  \end{tabular}
	  \label{tab:patchy}
\end{center}
\end{table*}

The tests in the last section have been performed on simulations with periodic boundary conditions, which do not take into account the survey geometry of BOSS. Here we use the MultiDark-Patchy mock catalogues, introduced in section~\ref{sec:mocks}, which incorporate the BOSS survey geometry. The mean of the MultiDark-Patchy power spectra (we have $2045$ pre-reconstruction power spectra and $996$ post reconstruction power spectra for the NGC and $2048$ pre-reconstruction and $999$ post-reconstruction power spectra for the SGC) is included in Figure~\ref{fig:psNGC} and~\ref{fig:psSGC} for comparison with the data measurements.

We fitted each individual mock catalogue and included the maximum likelihood value in Figure~\ref{fig:patchysctatter} (left). The black data points correspond to the post-reconstruction results, while the magenta points show the pre-reconstruction results. The results are also included in Table~\ref{tab:patchy}, where we list the mean and variance between the mock results. The largest offset between our mean post-recon results and the true underlying cosmology is $0.5\%$, less than $1/6$ of the standard deviation expected in these measurements.

The distribution of maximum likelihood results for the MultiDark-Patchy mock catalogues indicates a correlation between $\alpha_{\perp}$ ($\propto D_A$) and $\alpha_{\parallel}$ ($\propto H^{-1}$) of $\sim -0.47$ pre-reconstruction, while this value increases to $\sim -0.4$ in our post-reconstruction results. 
Fisher matrix forecasts predict a correlation value of $\approx  -0.41$ (i.e., anti-correlated $D_A$ and $H^{-1}$) pre- and post reconstruction~\citep{Seo:2003pu,Seo:2007ns} for the BAO-only analysis, i.e., when we marginalise over any redshift-space distortion effects. Our post-reconstruction values are in good agreement with the Fisher matrix predictions, while for pre-reconstruction the correlation is smaller (i.e., more negative) than expected. In the limit of a pure 
Alcock-Paczynski test, i.e, when we have a constraint only on $D_AH$, we expect a correlation of unity between $\alpha_{\perp}$ and $\alpha_{\parallel}$; using less information from the BAO scale will therefore increase the contribution from the Alcock-Paczynski test and push the correlation towards $1$ from $-0.4$ ~\citep{Seo:2003pu,Shoji:2009}. The pre-reconstruction correlation coefficient we observe is therefore not easily explained even if we assume a potential inclusion of non-BAO information. Note, however, that the Fisher matrix forecast assumes complete information of $P(k,\mu)$, while our data include only the monopole and quadrupole (excluding the hexadecapole).
The configuration space analysis of~\citet{Ross2016} found $r \sim -0.49$ pre-reconstruction and $r \sim -0.4$ post-reconstruction. These values agree well with our findings.
The consensus result of the BOSS DR11 analysis~\citep{Anderson:2013zyy} had $r = -0.54$ post-reconstruction, significantly more negative than our DR12 correlation as well as the Fisher prediction.~\footnote{The DR11 analysis used 
$R = 1$ in eq~\ref{eq:smani} for the post-reconstruction power spectrum model. 
We find this old fitting model indeed tends to lead to more negative 
correlation.}

\begin{figure}
\begin{center}
\epsfig{file=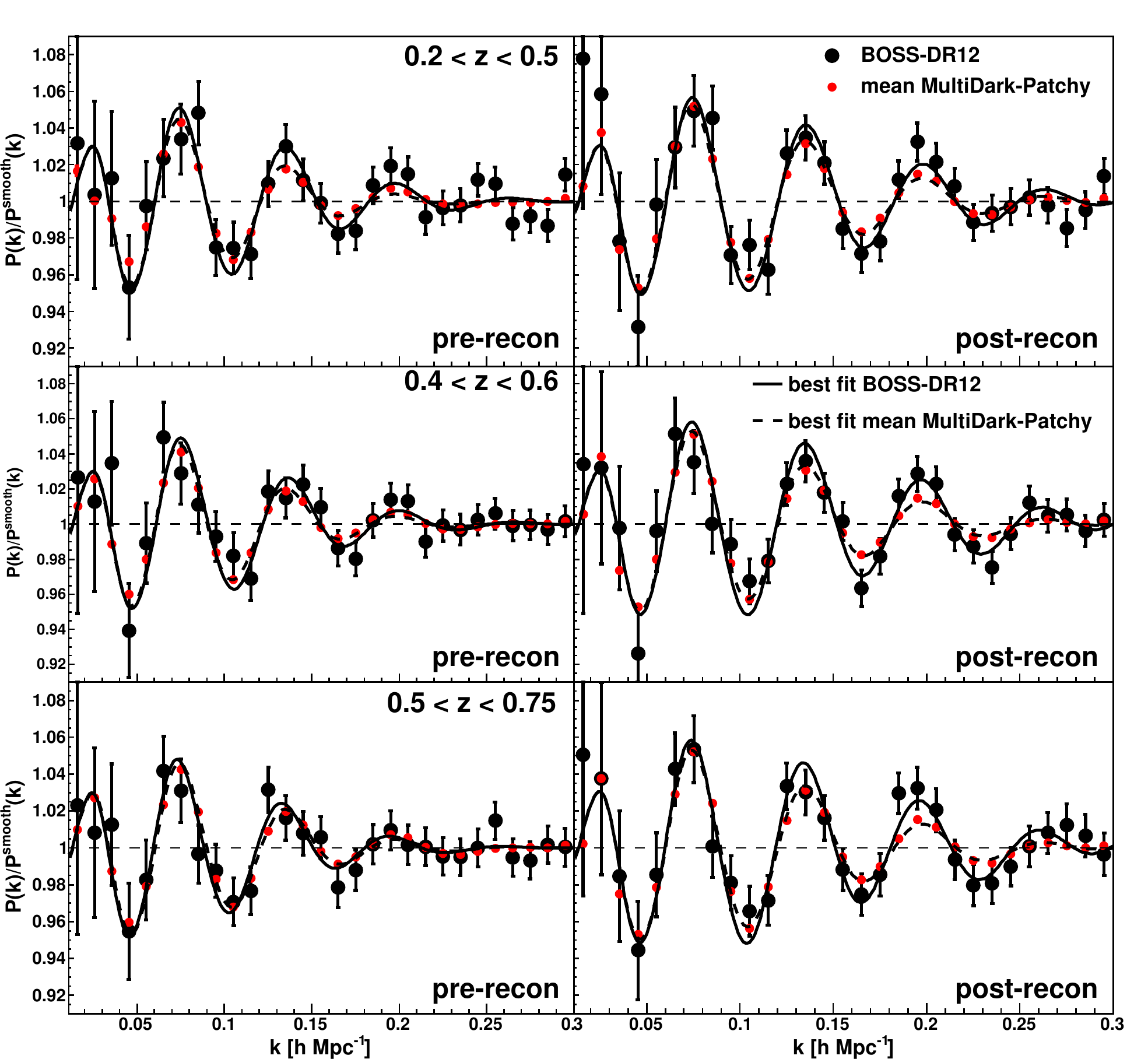,width=8cm}
\caption{The monopole power spectrum measurements (black data points) relative to the best fitting smooth power spectrum of eq.~\ref{eq:smiso}. The left panel shows the pre-reconstruction results and the right panel presents the post-reconstruction results. The black solid line represents the best fitting model to the data. The red data points are the mean power spectrum monopole of the MultiDark-Patchy mock catalogues and the black dashed line shows the best fit to the red data points. We observe a larger damping of the BAO signal in the MultiDark-Patchy mock catalogues compared to the data (see discussion \S~\ref{sec:patchy} for details).}
\label{fig:best_BAO_ratio_combined_patchy}
\end{center}
\end{figure}

\begin{figure*}
\begin{center}
\epsfig{file=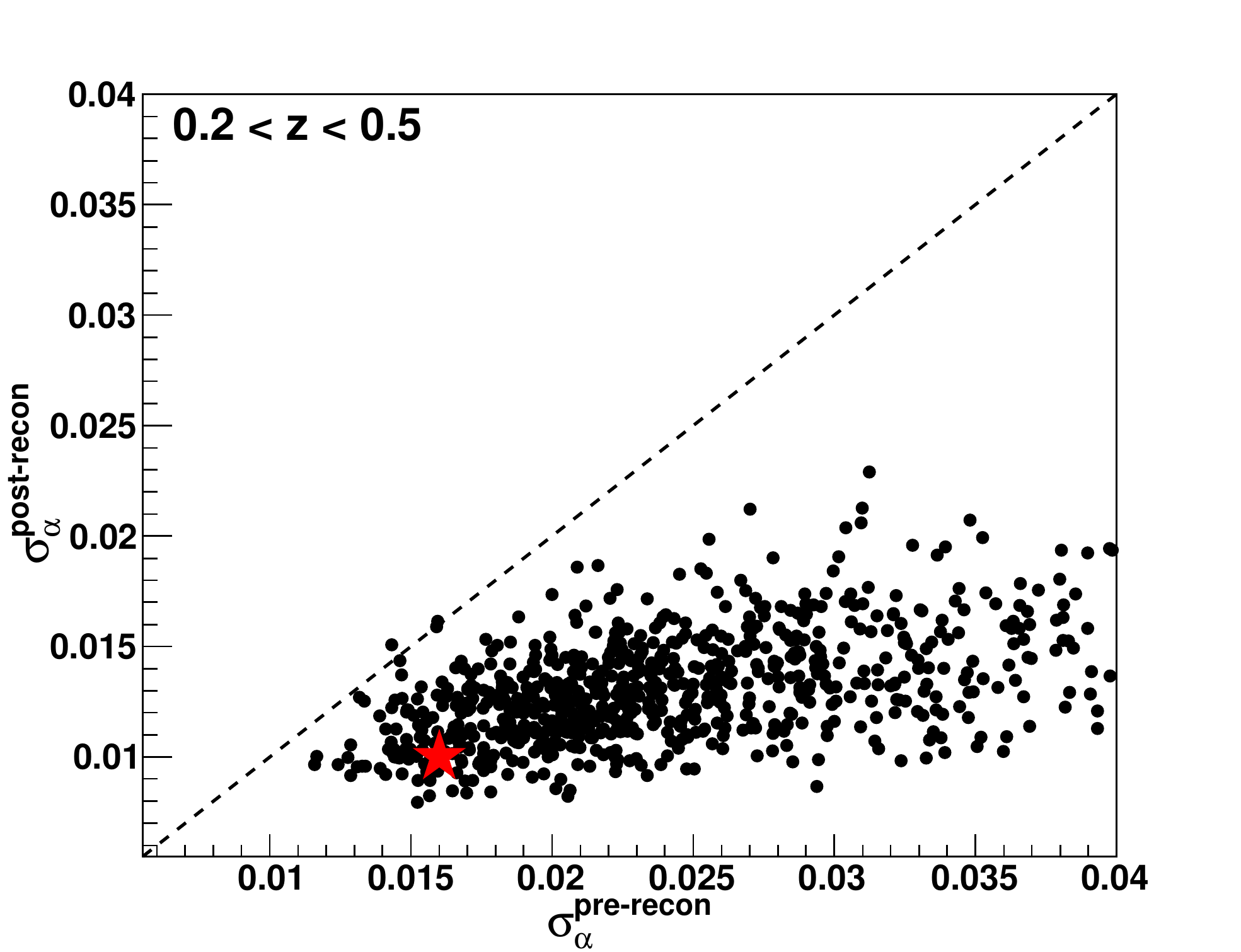,width=5.6cm}
\epsfig{file=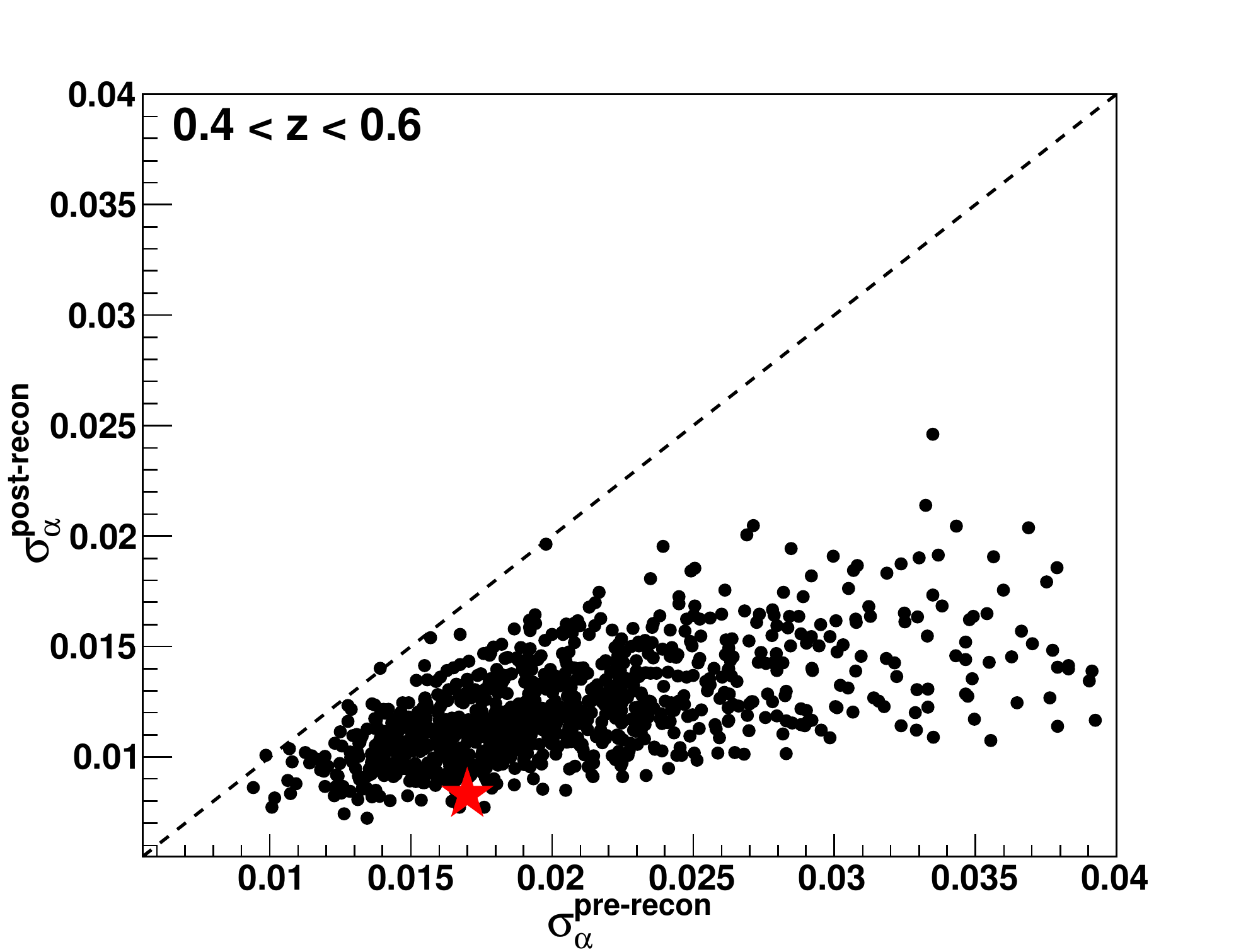,width=5.6cm}
\epsfig{file=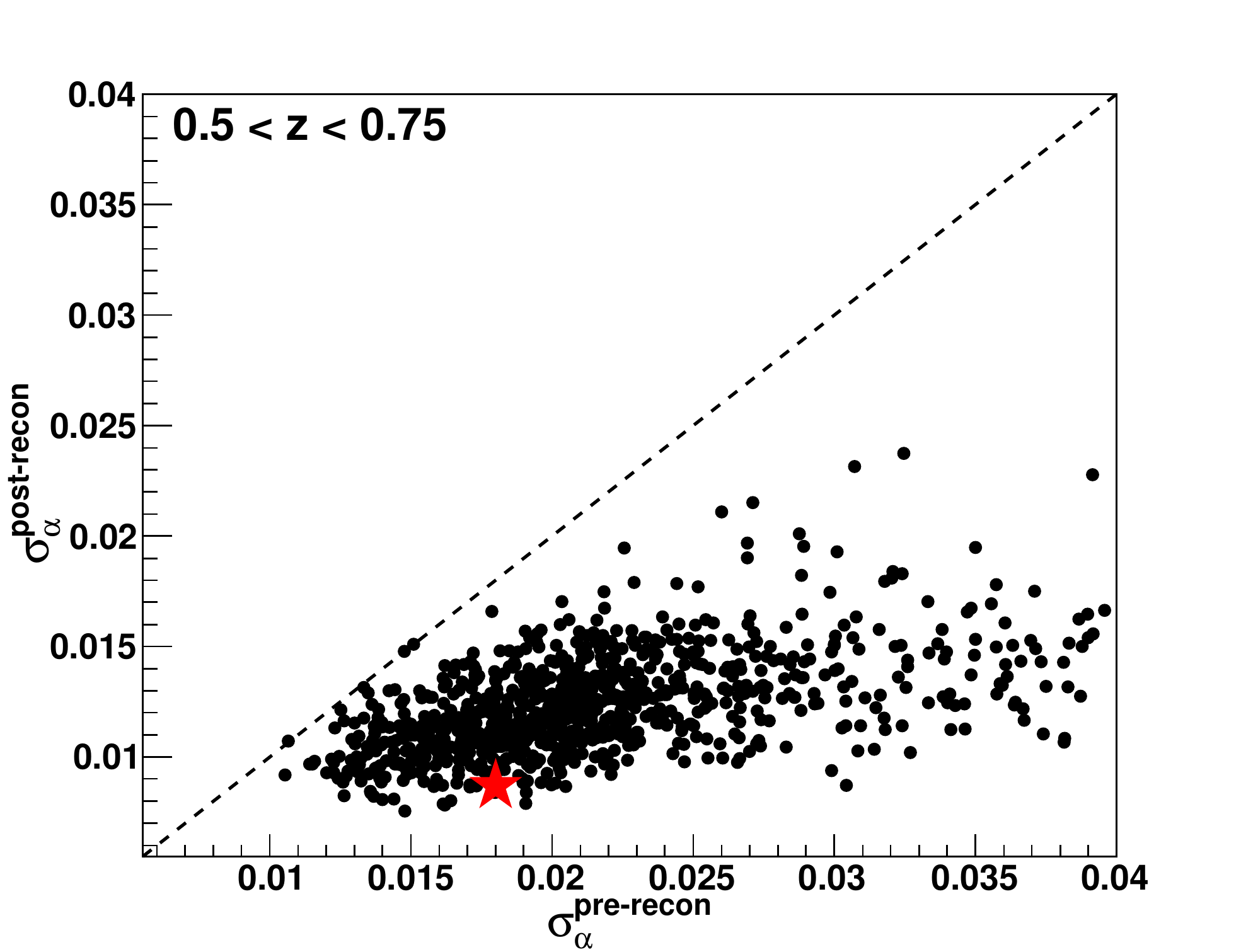,width=5.6cm}
\caption{The uncertainties on the angular average distance scale parameter $\alpha$ before and after density field reconstruction. The red stars show the measurement in BOSS-DR12, while the black points indicate the results for the MultiDark-Patchy mock catalogues.}
\label{fig:sig_patchy}
\end{center}
\end{figure*}

The r.m.s. between the maximum likelihood results of the MultiDark-Patchy mock catalogues is larger than the constraints we observe in the BOSS-DR12 data catalogues by $\sim 30\%$ (compare Table~\ref{tab:patchy} with Table~\ref{tab:results}). To understand this discrepancy, we now investigate how well the MultiDark-Patchy mock catalogues represent the data in terms of the BAO signal. Figure~\ref{fig:best_BAO_ratio_combined_patchy} presents the isotropic BAO signal in the data (black data points) compared to the mean of the MultiDark-Patchy mock catalogues (red data points). The mock catalogues show a larger damping of the BAO signal compared to the data, which is more prominent post-reconstruction. 

Post-reconstruction only $1.9\%$ of the mock catalogues in the high-redshift 
bin and $7.7\%$ of the mocks in the low-redshift bin have smaller uncertainties
than the data. To investigate this tension we first look at the measured 
damping scale in the data, where for simplicity we focus on the isotropic case.
The best fitting damping scales pre-reconstruction are 
$\Sigma_{\rm NL} = 8.3\pm1.4$, $8.8^{+1.6}_{-1.3}$ and 
$9.8^{+2.1}_{-1.6}\hMpc$ for the low, middle and high redshift bins, 
respectively. After density field reconstruction we get 
$\Sigma_{\rm NL} = 5.0\pm1.2$, $3.4^{+1.4}_{-3.0}$ and 
$3.2^{+1.6}_{-4.3}\hMpc$. We can compare these measurements with the 
expectations given by the Zel'dovich 
approximation~\citep[e.g.,]{Matsubara:2007wj}:
\begin{align}
\Sigma_{\rm xy}^2 &= \frac{1}{3\pi^2} \int dp P_{\rm lin}(p, z),\\
\Sigma_{\rm z} &= (1 + f)\Sigma_{\rm xy}.\\
\end{align}
We approximate the damping of the spherically averaged power spectrum to be
\begin{align}
\Sigma_{\rm NL} &= \sqrt{\frac{2}{3}\Sigma^2_{xy} + \frac{1}{3}\Sigma^2_{z}}.
\end{align}

This predicts $\Sigma_{\rm NL} = 8.8\,\hMpc$ at $z=0.38$, 
$\Sigma_{\rm NL} = 8.4\,\hMpc$ at $z=0.51$ and 
$\Sigma_{\rm NL} = 8.1\,\hMpc$ at $z=0.61$ before reconstruction. These values 
are slightly smaller than our measurements but consistent within the 
measurement uncertainties. The expected damping scale post-reconstruction does 
depend on the effectiveness of reconstruction, which depends on e.g. survey 
geometry. \citet{Seo:2015eyw} measures $\Sigma_{\rm NL} \sim 4.3\,\hMpc$ at 
redshift $z \sim 0.57$ post reconstructing, which agrees with our measurements.
The PTHalo mock catalogues~\citep{Manera:2012sc,Manera:2014cpa} used in the 
BOSS DR9~\citep{Anderson:2012sa} and DR10/11~\citep{Anderson:2013zyy} analysis 
showed a damping of $\Sigma_{\rm NL} \sim 4.6\,\hMpc$ and 
$\Sigma_{\rm NL} \sim 4.8\,\hMpc$ at redshift $0.57$ and $0.32$, respectively, 
again agreeing with out measurements. Meanwhile, the MultiDark-Patchy mocks 
show $\Sigma_{\rm NL}\sim 7\hMpc$ post-reconstruction. 
We therefore conclude that the BAO signal from the BOSS DR12 dataset is 
consistent with the expectation, while the MultiDark Patchy mocks tend to 
underestimate the BAO signal. 
We believe this excess damping is a limitation of the 2LPT approximation used
in the mock production (for more details about the MultiDark-Patchy mock 
production see~\citealt{Kitaura:2015}). The effect is presumably more apparent 
post-reconstruction because it is hidden by the larger intrinsic damping 
pre-reconstruction.  

The larger damping scale in the MultiDark-Patchy catalogues is clearly visible 
when fitting the mock power spectra and comparing to the BOSS-DR12 results (see
Figure~\ref{fig:sig_patchy}). Given that the main purpose of the mock 
catalogues is to estimate the band power precision of our power spectrum 
measurements, these effects do not impact our analysis. However, one should 
keep these effects in mind when using these catalogues to study the BAO signal.
We therefore consider our pipeline tests with the N-body simulations as more 
robust.

\section{DR12 Data Analysis}
\label{sec:analysis}

\begin{table*}
\begin{center}
\caption{The constraints from the BOSS DR12 data analysis, representing the 
main results of this paper. The upper section presents the anisotropic fits, 
while the lower part lists the isotropic results. The fitting range is 
$k = 0.01$ - $0.3\ihMpc$ for both the monopole and quadrupole. We include the 
result before (pre-recon) and after (post-recon) density field reconstruction. 
The $\alpha$ and $\epsilon$ values are derived from the $\alpha_{\parallel}$ 
and $\alpha_{\perp}$ values using eq.~\ref{eq:ep_alpha}. We also show the 
constraints on the Alcock-Paczynski parameter 
$F_{\rm AP}(z) = (1 + z) D_A(z) H(z)/c$ and the isotropic distance scale 
$D_V(z) = [(1+z)^2D^2_A(z)cz/H(z)]^{1/3}$. The uncertainties are derived from 
the $68\%$ confidence levels. The results are displayed in 
Figure~\ref{fig:best_fit},~\ref{fig:BAO_contours} and~\ref{fig:best_fit_rel}.
The correlation between $\alpha_{\perp}$ and $\alpha_{\parallel}$ as derived 
from the measurement likelihood is $-0.378$, $-0.389$ and $-0.464$ for the 
low, middle and high redshift bins, respectively.
}
	\begin{tabular}{llllllll}
     		\hline
		 & & \multicolumn{2}{c}{$0.2 < z < 0.5$} & \multicolumn{2}{c}{$0.4 < z < 0.6$} & \multicolumn{2}{c}{$0.5 < z < 0.75$}\\
		  & & \multicolumn{1}{c}{pre-recon} & \multicolumn{1}{c}{post-recon} & \multicolumn{1}{c}{pre-recon} & \multicolumn{1}{c}{post-recon} & \multicolumn{1}{c}{pre-recon} & \multicolumn{1}{c}{post-recon}\\
		  \hline
		 \multicolumn{8}{c}{anisotropic fit}\\
		\hline
		$\alpha_{\parallel}$ & & $1.047\pm0.037$ & $1.028\pm0.030$ & $1.013\pm0.049$ & $0.988\pm0.022$ & $0.944\pm0.041$ & $0.964\pm0.022$ \\
		$\alpha_{\perp}$ & & $0.981\pm0.021$ & $0.984\pm0.016$ & $1.008\pm0.023$ & $0.997\pm0.013$ & $1.009\pm0.025$ & $1.000\pm0.015$ \\
		$\chi^2/$d.o.f. & & $109.9/(116-16)$ & $101.2/(116-16)$ & $105.6/(116-16)$ & $68.0/(116-16)$ & $96.8/(116-16)$ & $97.2/(116-16)$ \\
		\hline
		$\alpha$ & & $1.002\pm0.015$ & $0.999\pm0.011$ & $1.009\pm0.016$ & $0.993\pm0.0091$ & $0.987\pm0.017$ & $0.988\pm0.0090$ \\
		$\epsilon$ & & $0.022\pm0.016$ & $0.015\pm0.013$ & $0.0016\pm0.021$ & $-0.0030\pm0.0099$ & $-0.022\pm0.019$ & $-0.013\pm0.011$ \\
		$F_{\rm AP}(z)$ & & $0.397\pm0.018$ & $0.406\pm0.016$ & $0.591\pm0.037$ & $0.598\pm0.018$ & $0.786\pm0.046$ & $0.761\pm0.025$ \\
		$D_V(z)r_s^{\rm fid}/r_s$ & [Mpc] & $1479\pm23$ & $1474\pm17$ & $1903\pm30$ & $1873\pm17$ & $2141\pm36$ & $2144\pm20$ \\
		$H(z)r_s/r_s^{\rm fid}$ & [km\,s$^{-1}$Mpc$^{-1}$] & $79.3\pm2.8$ & $80.7\pm2.4$ & $88.7\pm4.3$ & $90.8\pm2.0$ & $101.1\pm4.4$ & $98.9\pm2.3$ \\
		$D_A(z)r_s^{\rm fid}/r_s$ & [Mpc] & $1088\pm23$ & $1092\pm18$ & $1323\pm30$ & $1308\pm18$ & $1446\pm36$ & $1433\pm21$ \\
		\hline
		\multicolumn{8}{c}{isotropic fit}\\
		\hline
		$\alpha$ & & $1.006\pm0.016$ & $1.000\pm0.010$ & $1.016\pm0.017$ & $0.9936\pm0.0082$ & $0.991\pm0.019$ & $0.9887\pm0.0087$ \\
		$D_V(z)r_s^{\rm fid}/r_s$ & [Mpc] & $1485\pm24$ & $1476\pm15$ & $1916\pm32$ & $1874\pm16$ & $2150\pm42$ & $2146\pm19$ \\
		$\chi^2/$d.o.f. & & $48.5/(58-10)$ & $43.9/(58-10)$ & $64.8/(58-10)$ & $32.8/(58-10)$ & $49.8/(58-10)$ & $47.0/(58-10)$ \\
		\hline
		\hline
	  \end{tabular}
	  \label{tab:results}
\end{center}
\end{table*}

\begin{figure*}
\begin{center}
\epsfig{file=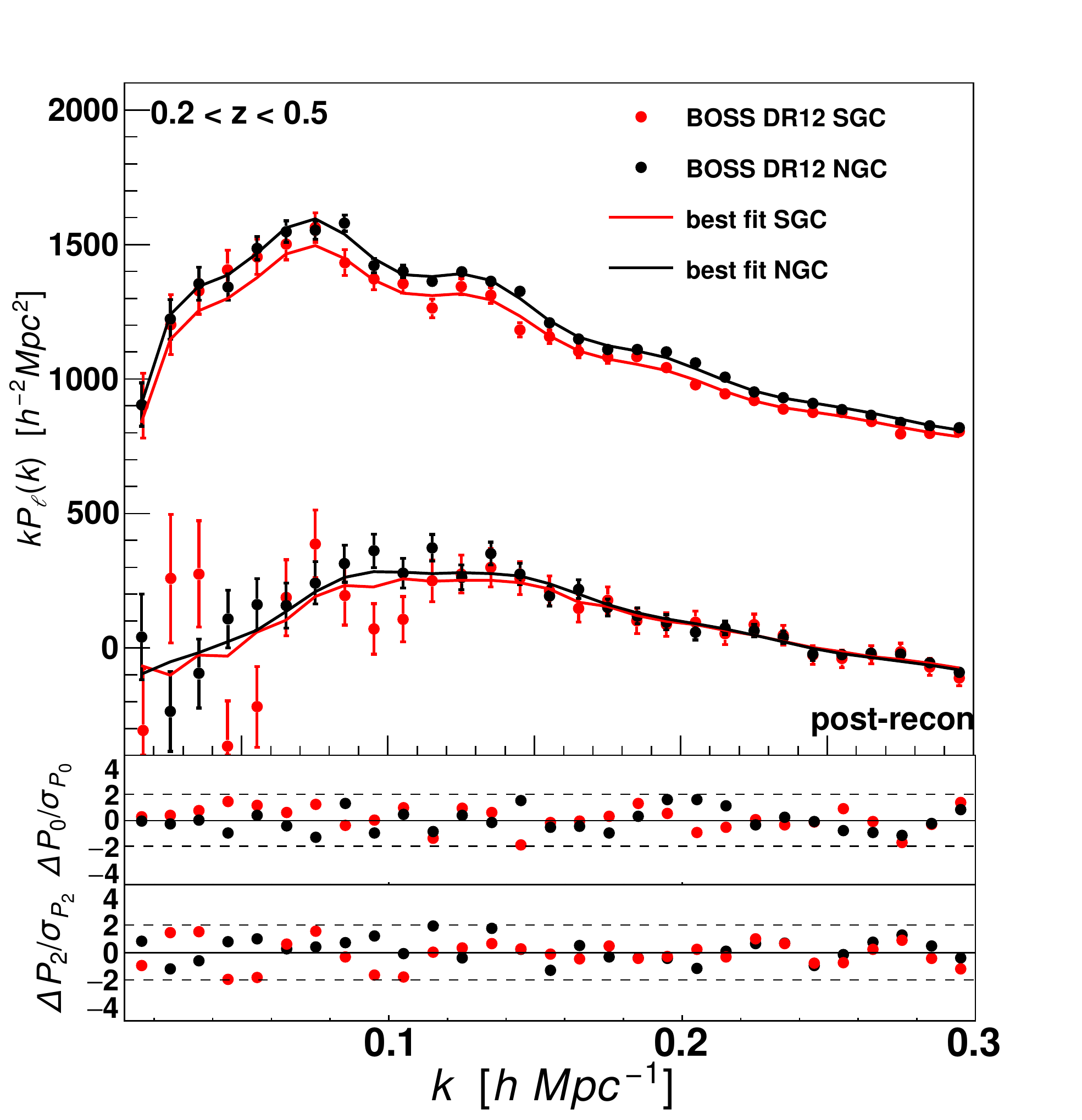,width=5.8cm}
\epsfig{file=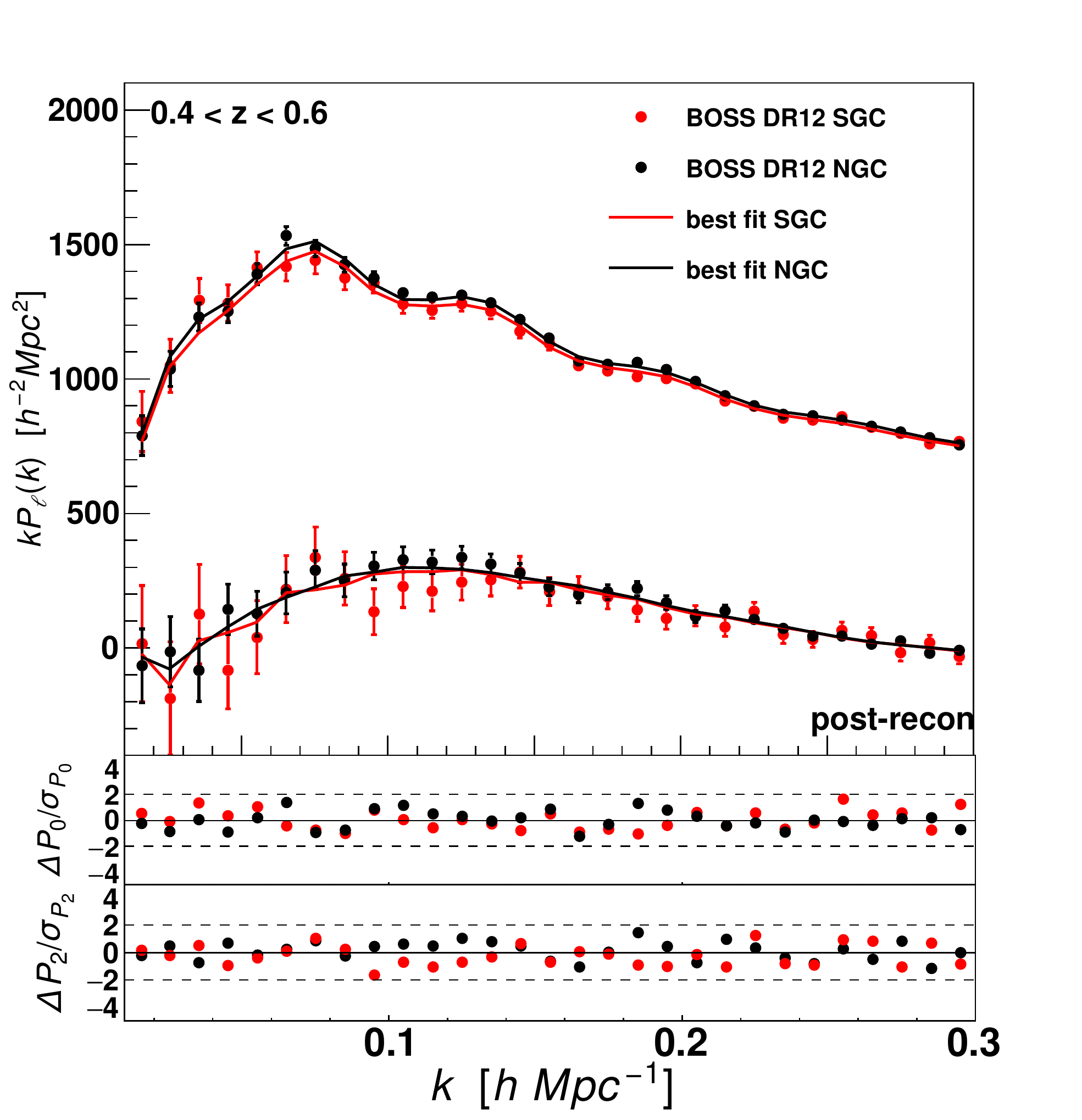,width=5.8cm}
\epsfig{file=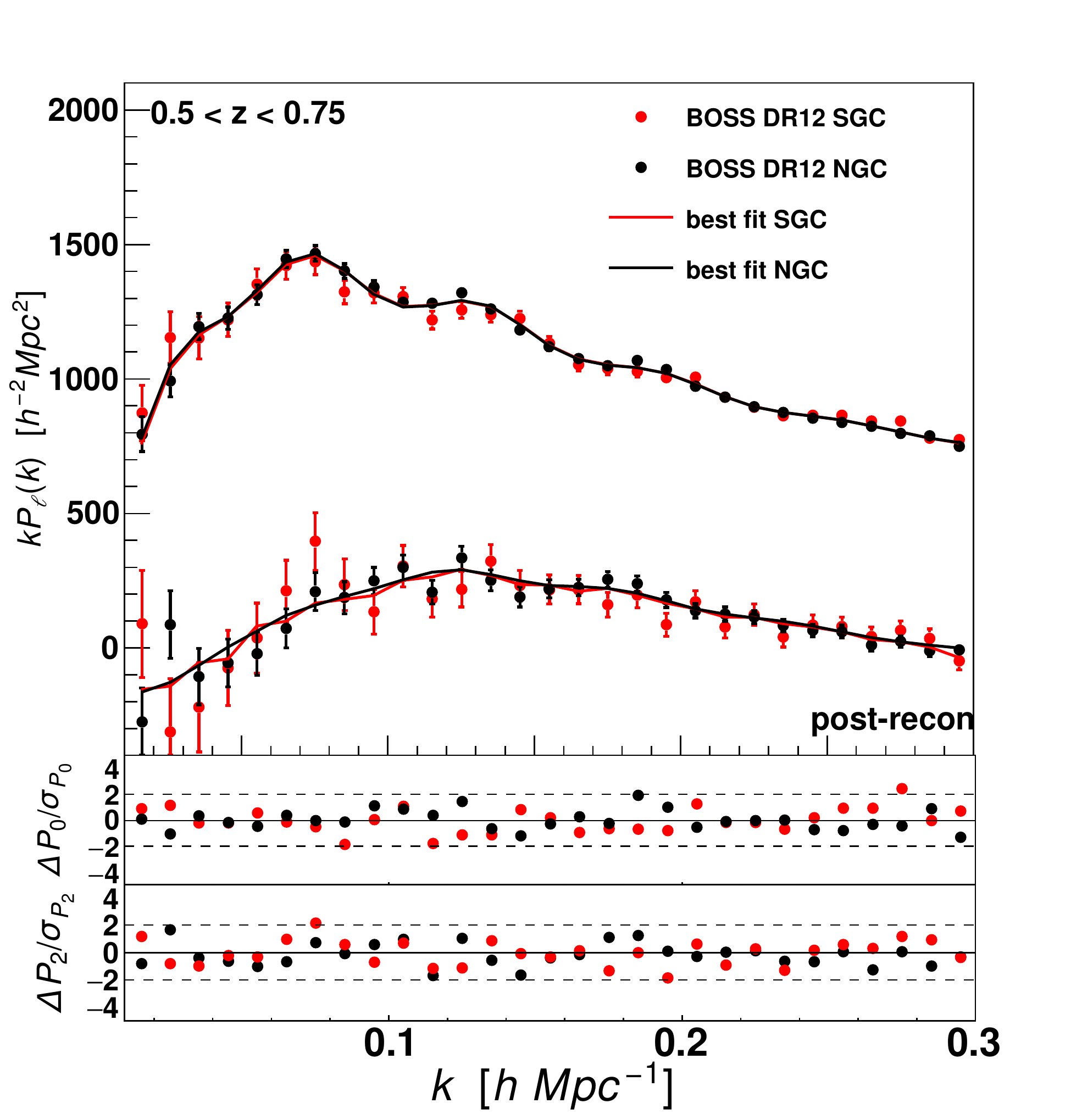,width=5.8cm}
\caption{Comparison between the best fitting model and the BOSS DR12 measurements in the three redshift bins used in this analysis. The errors on the data points are the diagonal of the corresponding covariance matrix. The red line represents the best fitting model to the SGC, while the black line shows the best fitting model for the NGC. The SGC best fitting model includes a small discreteness effect mainly visible at small $k$. The NGC and SGC have been fit simultaneously, using the same cosmological fitting parameters. However, the SGC and NGC have a separate amplitude nuisance parameter and different window functions, which leads to the difference between the red and black line. The reason for having separate nuisance parameters for NGC and SGC are slight differences in the galaxy sample selection (see section~\ref{sec:data} and~\citealt{Alam2016}). See Table~\ref{tab:results} for more details.}
\label{fig:best_fit}
\end{center}
\end{figure*}

\begin{figure}
\begin{center}
\includegraphics[height=5.8cm]{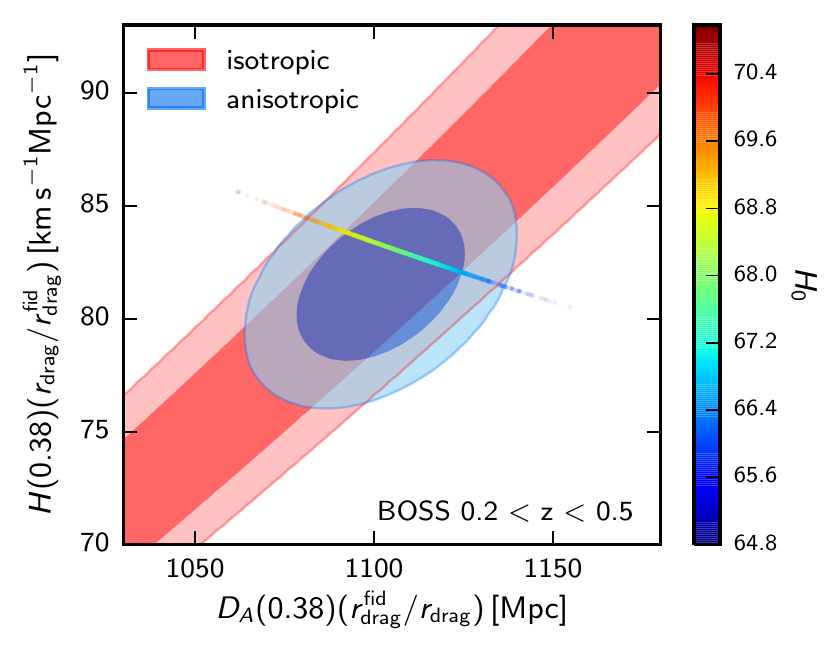}\\
\includegraphics[height=5.8cm]{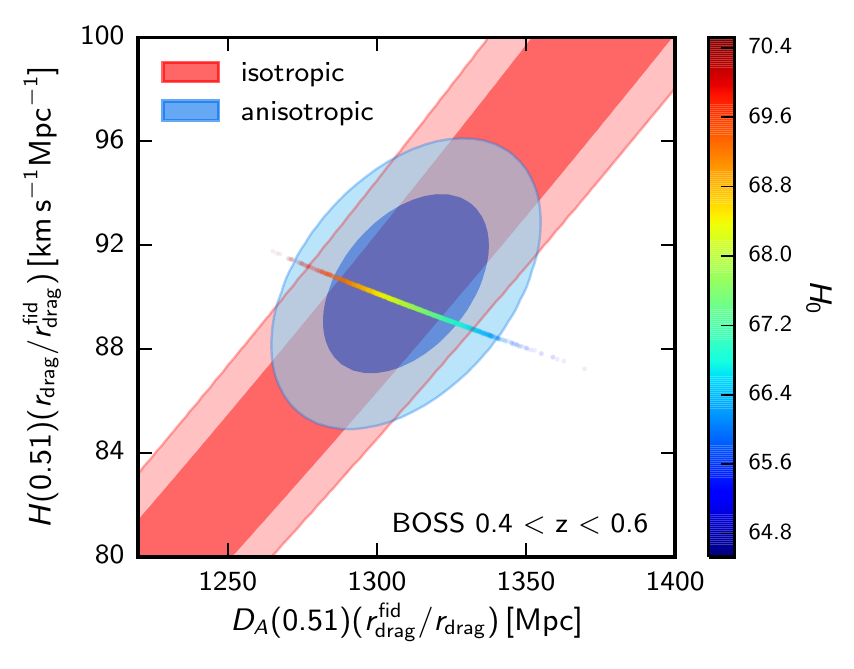}\\
\includegraphics[height=5.8cm]{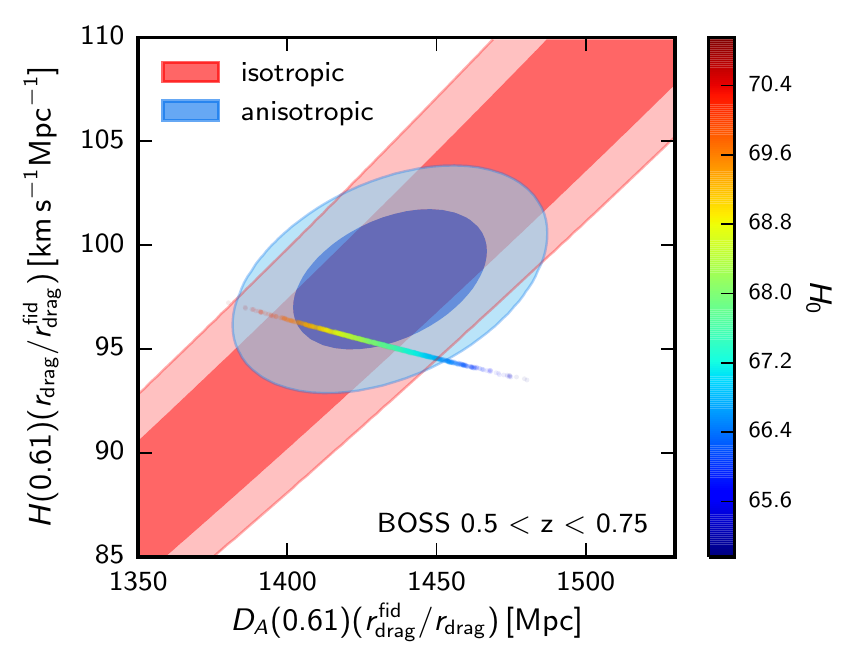}
\caption{The likelihood distribution for the fit to the BOSS power spectrum multipoles for the three redshift bins. The red contours show the isotropic (monopole only) fit, while the blue contours present the anisotropic (monopole + quadrupole) fit. We added the likelihood distribution for $H_0$ measured by Planck (Planck2015+lensing) where we assumed a $\Lambda$CDM model to extrapolate from the redshift of decoupling to the effective redshifts of the three samples. See Table~\ref{tab:results} for more details.}
\label{fig:BAO_contours}
\end{center}
\end{figure}

\begin{figure*}
\begin{center}
\epsfig{file=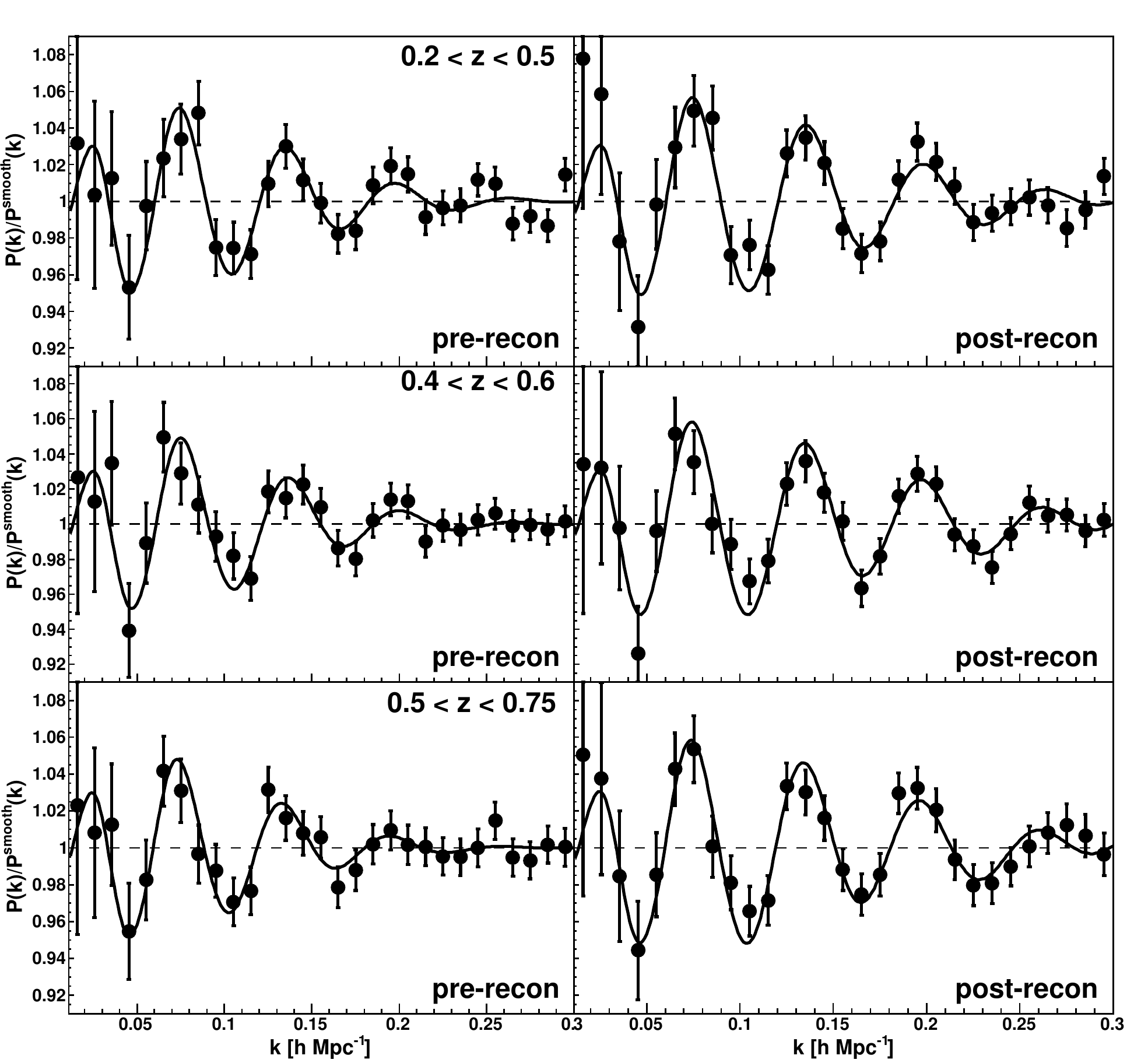,width=12cm}
\caption{The best fitting models (black solid line) of the isotropic BAO analysis compared to the power spectrum monopole measurements (data points). Both the model and the data have been plotted relative to the smooth model, and the data points for NGC and SGC have been combined using the corresponding covariance matrices (see appendix~\ref{app:combining}). The left panel shows the pre-reconstruction result, while the right panel presents the post reconstruction result. Similar plots for the NGC and SGC separately are included in appendix~\ref{app:isotropic}. See Table~\ref{tab:results} for more details.}
\label{fig:best_fit_rel}
\end{center}
\end{figure*}

\subsection{The anisotropic fit}

Figure~\ref{fig:best_fit} compares the best fitting power spectrum multipole models with the measurements. The best fit for the SGC (red solid line) and the NGC (black solid line) use different amplitude parameters $B$ to account for potential differences due to target selection. Except for the amplitude all parameters are identical and we obtain these results by fitting the NGC and SGC simultaneously. The lower two panels in Figure~\ref{fig:best_fit} show the residuals for the monopole and quadrupole separately, indicating a good fit on all scales. The best fitting $\chi^2/$d.o.f. is $101.2/(116-16)$, $68.0/(116-16)$ and $97.2/(116-16)$. The probabilities of having reduced $\chi^2$ values that exceed these values are $44.8\%$, $99.4\%$ and $56.1\%$. The middle redshift bin has a $\chi^2$ below expectation (with a significance $<3\sigma$), which might be related to the higher amplitude of the MultiDark-Patchy mocks for this redshift bin, which overestimates the uncertainties (see Figure~\ref{fig:psNGC}).

The correlation between $\alpha_{\perp}$ and $\alpha_{\parallel}$ as derived 
from the measurement likelihood is $-0.378$, $-0.389$ and $-0.464$ for the low,
middle and high redshift bins, respectively. These values are very similar to 
$-0.398$, $-0.397$ and $-0.413$, which are the corresponding values from the 
Multidark-Patchy mock catalogues.

The best fitting BAO scale parameters for the three redshift bins are shown in Table~\ref{tab:results}. For all redshift bins we found significant (up to a factor of two) improvements after applying density field reconstruction. Post reconstruction we have $2.9\%$, $2.2\%$ and $2.3\%$ constraints on $\alpha_{\parallel}$ for the low ($\zeff=0.38$), middle ($\zeff=0.51$) and high redshift bin ($\zeff=0.61$), respectively. For $\alpha_{\perp}$, the post-reconstruction constraints are $1.6\%$, $1.3\%$ and $1.5\%$ at $\zeff=0.38$, 0.51, and 0.61, respectively. 

\subsection{The isotropic fit}

The best fitting results in case of the isotropic analysis (monopole only) are shown in Figure~\ref{fig:best_fit_rel}. The best post-reconstruction constraints are $\alpha = 1.000\pm0.010$, $0.9936\pm0.0082$ and $0.9887\pm0.0087$ at $\zeff=0.38$, 0.51, and 0.61, respectively. Thus we have two independent (the middle redshift bin is correlated with the other two) $\sim 1\%$ distance constraints at the low and at high redshift bins: $0.88\%$ at $z=0.61$ and $1\%$ at $\zeff=0.38$. The best fitting reduced $\chi^2$ in the isotropic case is $\chi^2/$d.o.f. is $48.5/(58-10)$, $64.8/(58-10)$ and $49.8/(58-10)$ for pre-reconstruction and $43.9/(58-10)$, $32.8/(58-10)$ and $47.0/(58-10)$ post-reconstruction fits. The second redshift bin shows a large $\chi^2$ pre-reconstruction and a fairly small $\chi^2$ post-reconstruction. The probability to have a reduced $\chi^2$ value that exceed this value is $6.5\%$ for the pre-reconstruction value and $96.4\%$ for the post-reconstruction result, meaning that these results are $2\sigma$ fluctuations from expectation.

\section{Discussion}
\label{sec:dis}

The likelihood distribution for our best fitting isotropic and anisotropic results in terms of $D_A(z)$ and $H(z)$ are displayed in Figure~\ref{fig:BAO_contours}, together with the likelihood distribution for Planck within $\Lambda$CDM. Including the quadrupole our anisotropic analysis can break the degeneracy between $D_A$ and $H$ and constrain both parameters separately. Our constraints for all redshift bins are in good agreement with the Planck prediction within $\Lambda$CDM.

\subsection{Comparison to other DR12 BAO measurements}

\citet{Ross2016} analysed the same BOSS DR12 data as used in this analysis in configuration space, finding overall good agreement (any difference is $< 0.5\%$) for the best fitting values and for the measurement uncertainties. We find a correlation of $\sim 0.9$ between our BAO constraints and the constraints of~\citet{Ross2016}. Our companion paper~\citep{Alam2016} combines the BAO constraint derived in this paper with the results in~\citet{Ross2016} to obtain a combined likelihood, representing the final BOSS BAO result.

In~\citet{Gil-Marin:2015nqa} the BOSS DR12 sample has been analysed using the two redshift bins of CMASS and LOWZ.
While the CMASS results agree quite well with our high redshift bin, there are some differences between our low redshift bin and the LOWZ result. The LOWZ sample is defined by the redshift range $0.15$ - $0.43$, while our low redshift sample covers $0.2 < z < 0.5$. The combined sample used in our analysis also includes some new data based on the `early regions' (see \S~\ref{sec:data}), which have been excluded in LOWZ.  

\subsection{Significance of the BAO detection}

We can test the significance of the detection of the BAO signal in BOSS by comparing our results to the limit $\Sigma_{\rm NL} \rightarrow \infty$, which corresponds to a power spectrum without a BAO signal. We focus here on the isotropic analysis. Before applying density field reconstruction, our no-BAO fits result in a $\chi^2$ of $60.2$, $82.1$, $67.4$ for the low, middle and high redshift bin, respectively. Comparing to the best fitting $\chi^2$ in Table~\ref{tab:results} we have $\Delta\chi^2 = 11.7$, $17.3$ and $17.6$, which indicate detection significances of $3.4$, $4.2$ and $4.2\sigma$.
After applying density field reconstruction the $\chi^2$ for the no-BAO fits is $106.2$, $97.3$ and $113.6$ for the low, middle and high redshift bins, respectively. Again comparing to Table~\ref{tab:results} we have $\Delta\chi^2 = 62.3$, $64.5$ and $66.6$, which indicate detection significances of $7.9$, $8.0$ and $8.2\sigma$.

\subsection{Comparison to Fisher matrix forecasts}
\label{sec:Fisher}

As a check that our results are close to optimal, and to assess the 
reliability of
predictions for the future, it is useful to compare our results to pre-survey 
predictions
of BOSS's BAO measuring power \citep{Eisenstein:2011sa} based on
the code of \citet{Seo:2007ns}. The 
details take us too far afield so are given in 
Appendix \ref{app:Fisher}, with the
following summary:

\begin{itemize}

\item Our measurements of $\alpha$, which is approximately
equivalent to the dilation factor $R$ of \citet{Seo:2007ns}, can be 
aggregated to an overall $0.70\%$ distance error.

\item \citet{Eisenstein:2011sa} predicted an error 1.44 times smaller,
$0.48\%$. 

\item Our measured BAO errors are typical given the measured band power 
covariance, i.e., there is no evidence that the measurement was unlucky
in its error bars. The aggregated error averaged over many MultiDark Patchy mock catalogues is $0.88\%$, an even worse
$1.82$ times expected, but we believe this is because the BAO signal is 
smaller than it should be in these mocks (see section~\ref{sec:patchy}), not that this is evidence that the 
measurement on the data was lucky.

\item Fisher predictions for the measured band power errors, using the 
measured number of galaxies and their measured bias,
propagated to BAO errors, predict very close to the observed error ($0.68\%$). Here we use the BOSS fiducial cosmology (see last paragraph of section~\ref{sec:intro}), which is close to the Planck cosmology.

\item Fisher predictions for the measured band power errors, using the 
pre-survey expected number of galaxies and bias (instead of measured),
still using the Planck cosmology, 
predict an aggregated error of $0.59\%$, with most of the difference between this and
$0.68\%$ being due to the measured bias being lower than expected. 

\item Fisher predictions for the measured band power errors, using the 
pre-survey expected number of galaxies and bias,
and the WMAP3 cosmology which is hard-coded into the \citet{Seo:2007ns} code, 
predict aggregated error $0.51\%$, approximately equivalent to the $0.48\%$ 
expectation based on the \citet{Seo:2007ns} code, i.e., approximately half 
the difference between the pre-survey expectation and the achieved measurement
($\sim 0.59/0.51$)
is due to the Planck cosmology being less favourable for BAO than WMAP3 
(e.g., lower baryon/CDM ratio) with another half ($\sim 0.68/0.59$) being 
lower than expected bias. 
\end{itemize} 

Note that contributing effects do not add linearly, and there are various 
smaller effects not mentioned in this summary.
There is no sign of sub-optimality in the data analysis.

\section{Conclusion}
\label{sec:conclusion}

We have measured the power spectrum multipoles from the final BOSS DR12 dataset in three (overlapping) redshift bins, covering the total redshift range $0.2 < z < 0.75$. Our analysis focuses on measuring the isotropic and anisotropic Baryon Acoustic Oscillation signal in Fourier-space. Our main results are:
\begin{enumerate}
\item We measure the power spectrum monopole and quadrupole, accounting for window function, aliasing and discreteness effects and extract the BAO information by marginalising over the broadband shape of the power spectrum. We validate our analysis pipeline using two sets of N-body simulations as well as the MultiDark-Patchy mock catalogues.
\item Fitting the monopole and quadrupole between $k = 0.01$ - $0.30\hMpc$ produces a constraint on the Hubble parameter of $H(z)r_s/r_s^{\rm fid} = 79.3\pm2.8\,$km\,s$^{-1}$Mpc$^{-1}$ and a constraint on the angular diameter distance of $D_A(z)r_s^{\rm fid} = 1088\pm23\,$Mpc for the low-redshift bin and $H(z)r_s/r_s^{\rm fid} = 98.9\pm2.3\,$km\,s$^{-1}$Mpc$^{-1}$ and $D_A(z)r_s^{\rm fid} = 1433\pm21\,$Mpc for the high-redshift bin (see Table~\ref{tab:results} for a complete summary of the results). While the high-redshift bin is in good agreement with previous results from the CMASS sample, our low-redshift constraint is significantly improved compared to previous studies. Our results are included in~\citet{Alam2016}, where a detailed study of the cosmological implications is performed.
\item Ignoring the Alcock-Paczynski effect we can constrain the angular averaged distance $D_V$, for which we obtain a $1\%$ and a $0.88\%$ constraints at the effective redshifts of $z_{\rm eff} = 0.38$ and $0.61$, respectively.  
\item The detection significances of the BAO signal are $3.4$, $4.2$ and $4.2\sigma$ before applying density field reconstruction for the low, middle and high redshift bins, respectively, and increases to $7.9$, $8.0$ and $8.2\sigma$ after density field reconstruction.
\end{enumerate} 
\citet{Alam2016} combines our measurements with the corresponding correlation function measurements of~\citet{Ross2016} and the growth of structure measurements of~\citet{Beutleretal2},~\citet{Grieb:2016},~\citet{Sanchez2016} and~\citet{Satpathy2016} into a final BOSS likelihood and investigates the cosmological implications. 


\section*{Acknowledgments}

FB acknowledges support from the UK Space Agency through grant ST/N00180X/1.

Funding for SDSS-III has been provided by the Alfred P. Sloan Foundation, the Participating Institutions, the National Science Foundation, and the U.S. Department of Energy Office of Science. The SDSS-III web site is http://www.sdss3.org/.

SDSS-III is managed by the Astrophysical Research Consortium for the Participating Institutions of the SDSS-III Collaboration including the University of Arizona, the Brazilian Participation Group, Brookhaven National Laboratory, Carnegie Mellon University, University of Florida, the French Participation Group, the German Participation Group, Harvard University, the Instituto de Astrofisica de Canarias, the Michigan State/Notre Dame/JINA Participation Group, Johns Hopkins University, Lawrence Berkeley National Laboratory, Max Planck Institute for Astrophysics, Max Planck Institute for Extraterrestrial Physics, New Mexico State University, New York University, Ohio State University, Pennsylvania State University, University of Portsmouth, Princeton University, the Spanish Participation Group, University of Tokyo, University of Utah, Vanderbilt University, University of Virginia, University of Washington, and Yale University.

This research used resources of the National Energy Research Scientific Computing Center, which is supported by the Office of Science of the U.S. Department of Energy under Contract No. DE-AC02-05CH11231.

H.-J. Seo is supported by the U.S. Department of Energy, Office of Science, Office of High Energy Physics under Award Number DE-SC0014329.
C. C. acknowledges support as a MultiDark Fellow. 
C. C. acknowledges support from the Spanish MICINNs Consolider-Ingenio 2010 Programme under grant MultiDark CSD2009-00064, MINECO Centro de Excelencia Severo Ochoa Programme under grant SEV-2012-0249, and grant AYA2014-60641-C2-1-P. 

\setlength{\bibhang}{2em}
\setlength{\labelwidth}{0pt}

\newpage

\appendix
\numberwithin{equation}{section}

\section{NGC vs. SGC}
\label{app:isotropic}

\begin{figure*}
\begin{center}
\epsfig{file=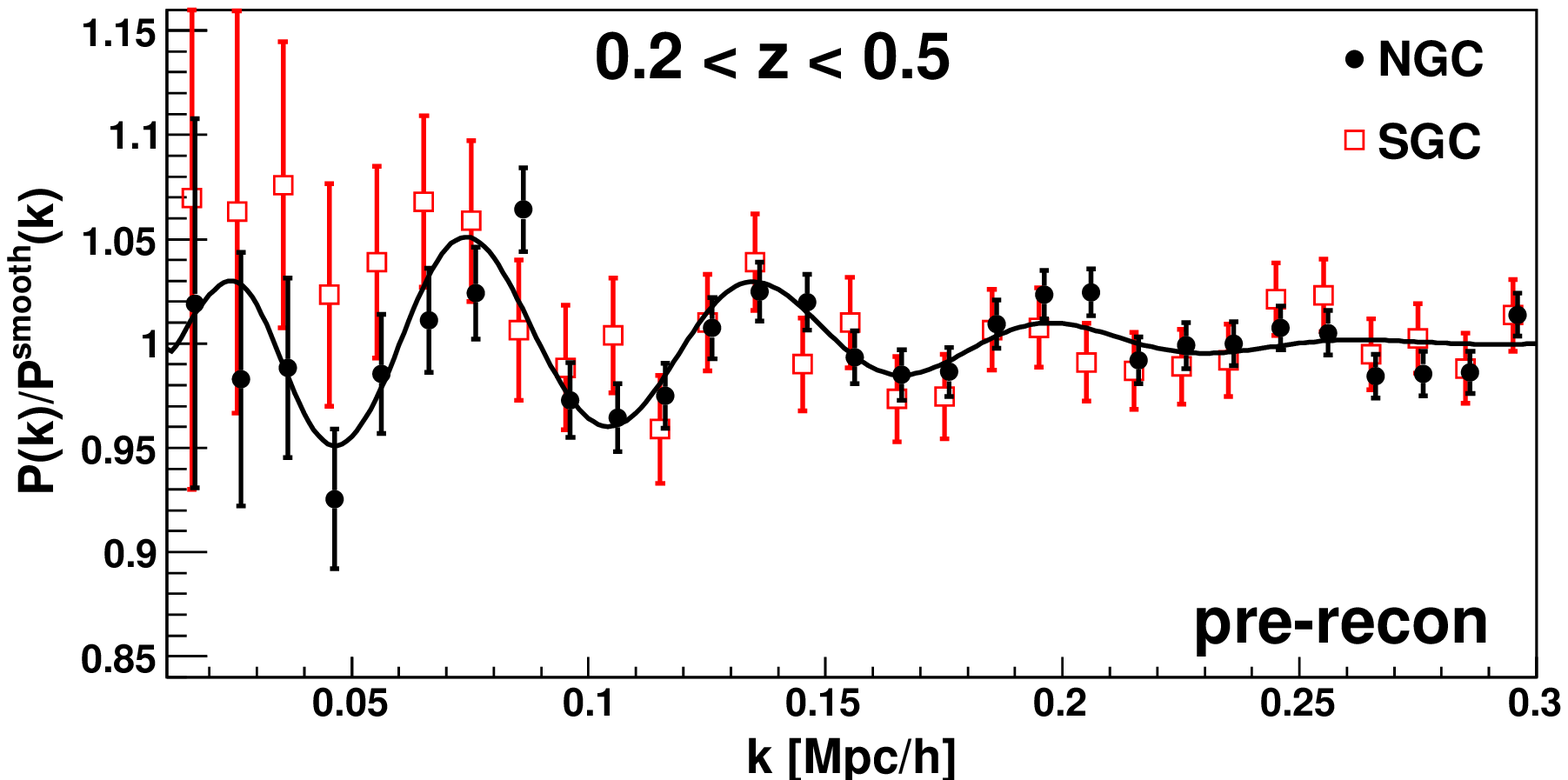,width=8.8cm}
\epsfig{file=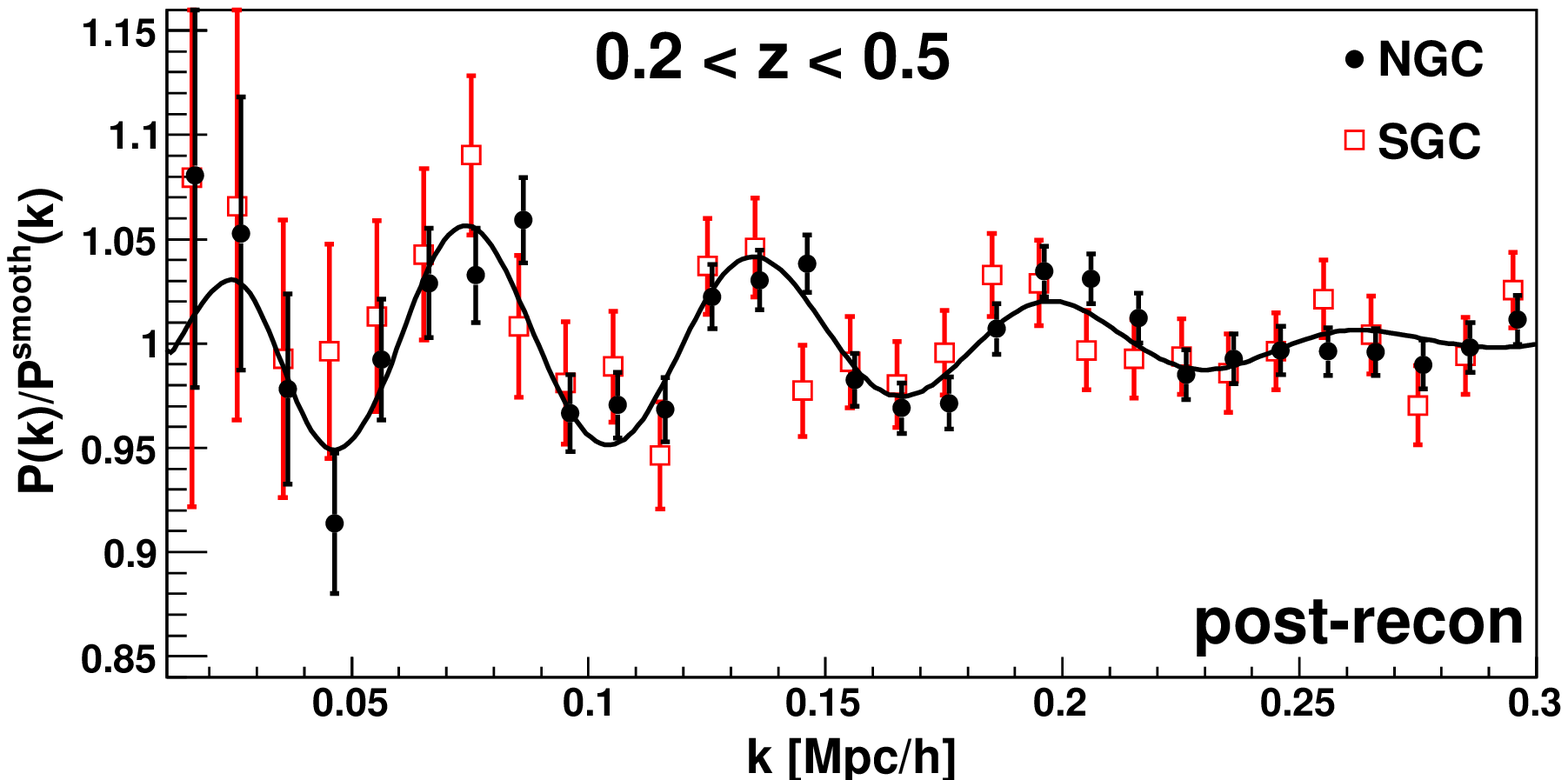,width=8.8cm}\\
\epsfig{file=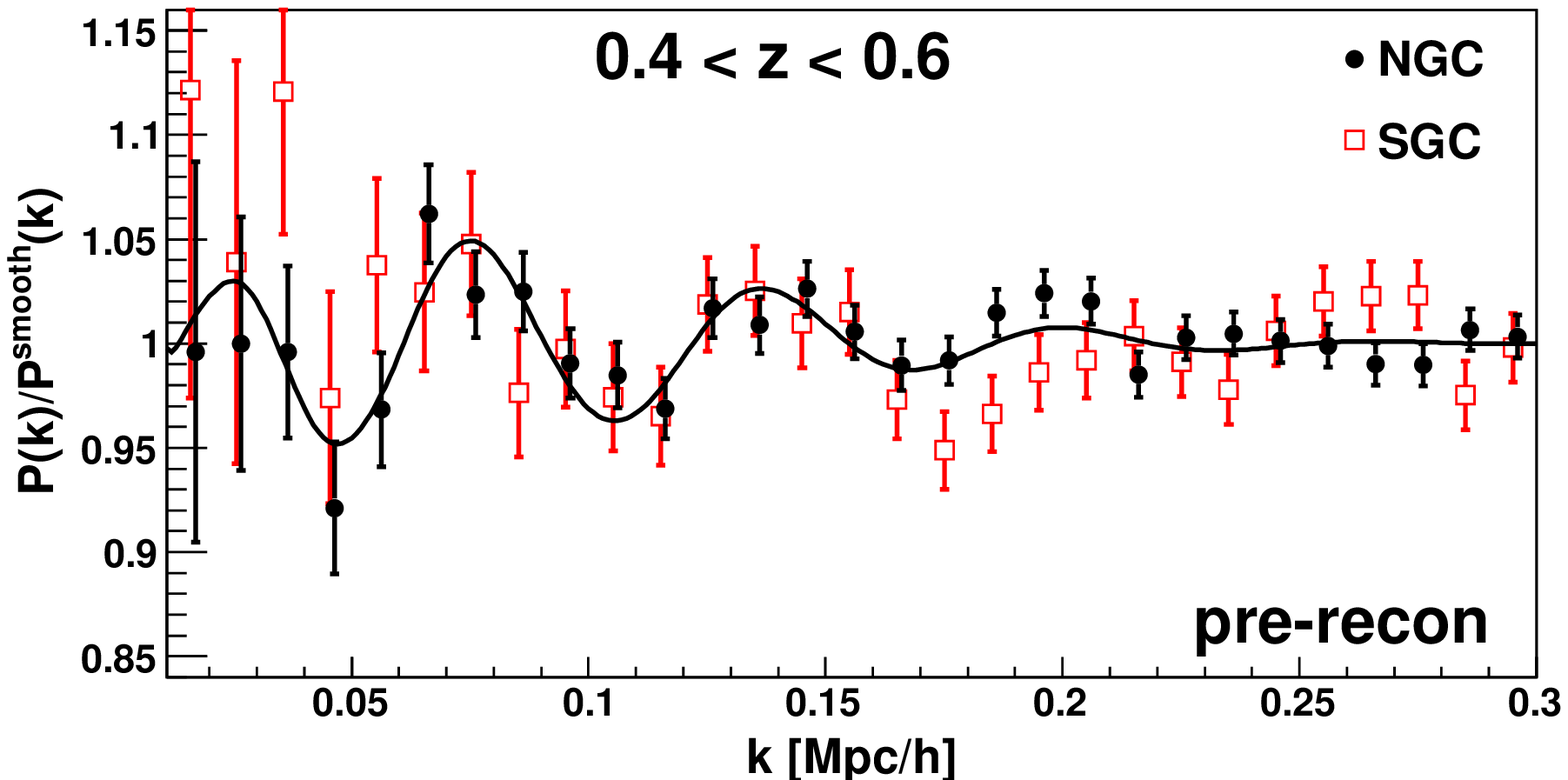,width=8.8cm}
\epsfig{file=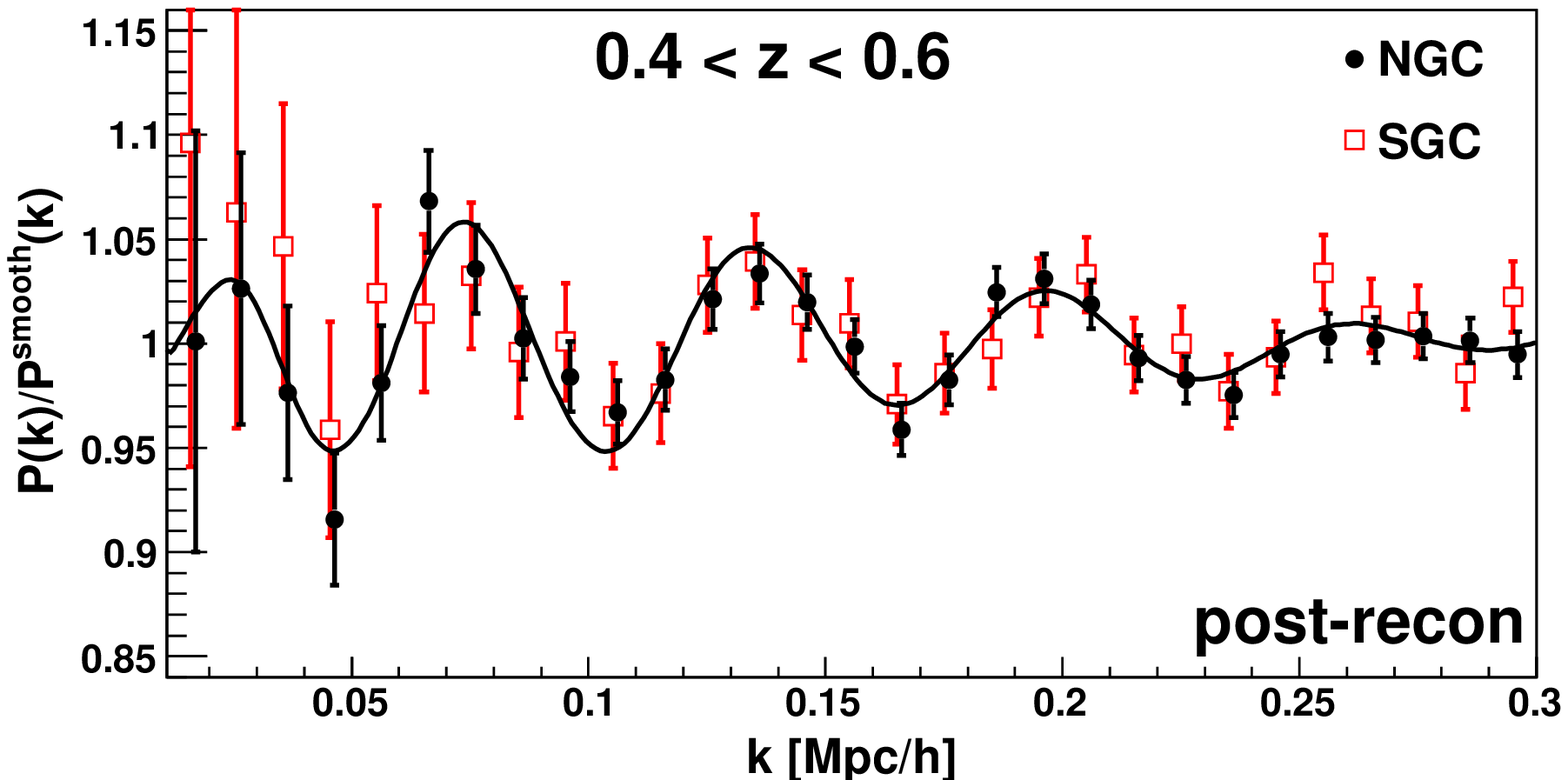,width=8.8cm}\\
\epsfig{file=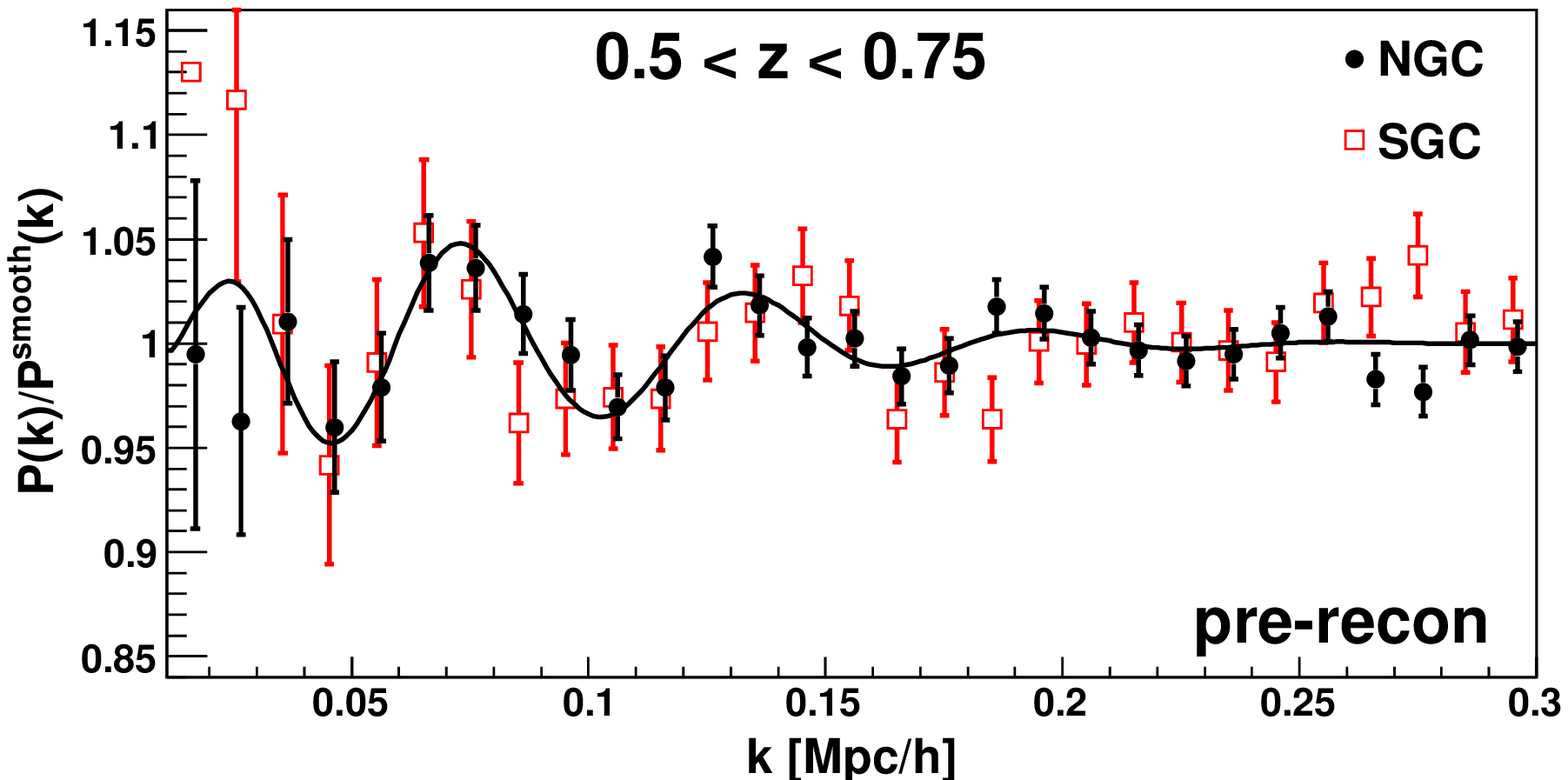,width=8.8cm}
\epsfig{file=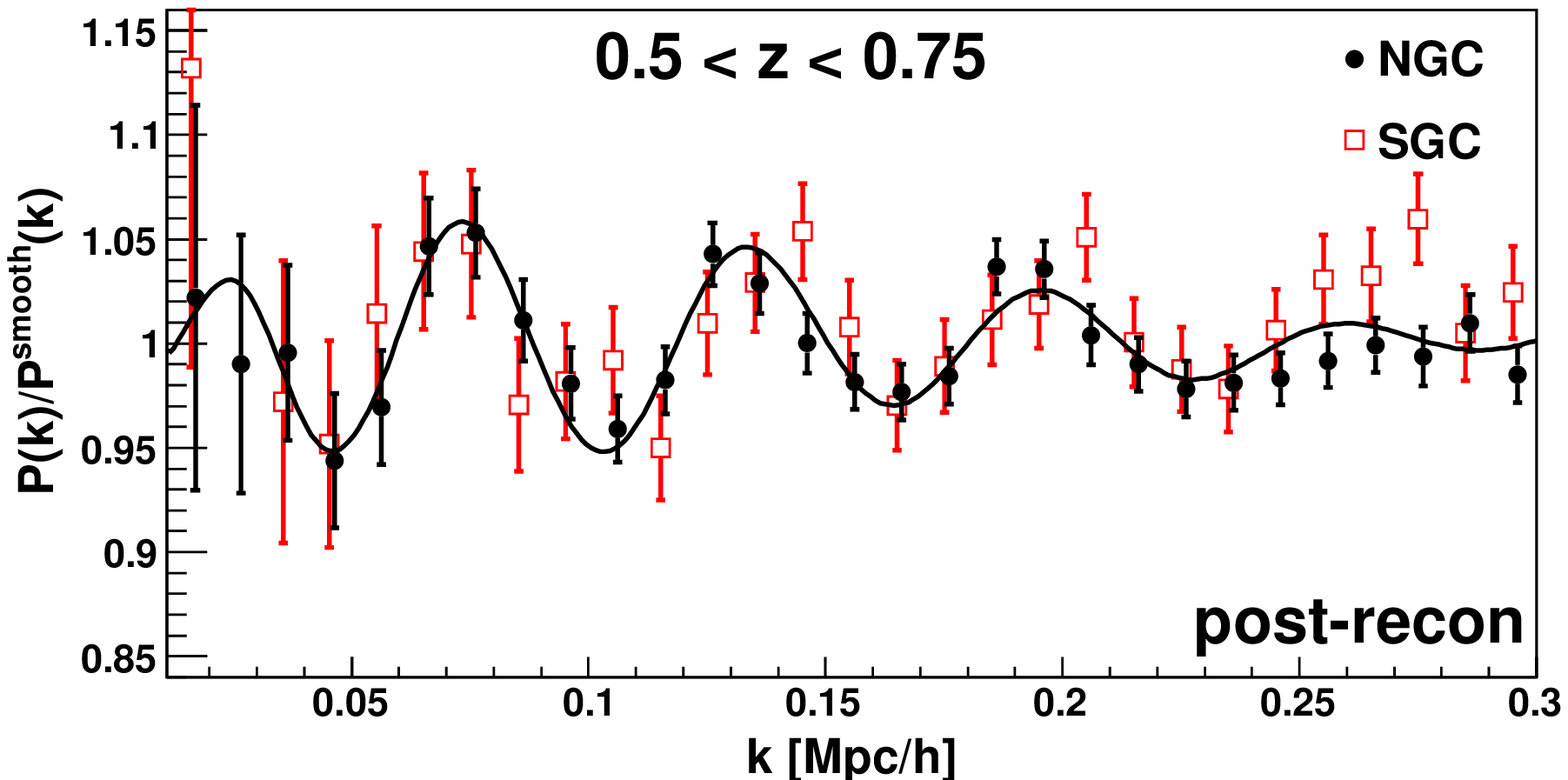,width=8.8cm}
\caption{The best fitting results for the isotropic BAO analysis pre-reconstruction (left) and post reconstruction (right). The NGC (black circles) and SGC (open red squares) measurements are displayed relative to the smooth power spectrum (see eq.~\ref{eq:smiso}). After dividing by the smooth power spectrum, the best fitting model for the NGC and SGC is the same (solid black line). The uncertainties are the diagonal of the covariance matrix. The covariance matrix has significant non-diagonal terms at large wave numbers, which are larger in the pre-reconstruction case. The best fitting parameters and $\chi^2$ are included in the lower part of Table~\ref{tab:results}.}
\label{fig:best_fit_rel2}
\end{center}
\end{figure*}

Figure~\ref{fig:best_fit_rel2} shows the best fitting isotropic power spectrum models compared to the monopole measurements for the three redshift bins pre- and post-reconstruction. These plots are similar to Figure~\ref{fig:best_fit_rel}, but here we separate the SGC and NGC components.

\section{Combining NGC and SGC}
\label{app:combining}

We combine the NGC and SGC measurements in Figure~\ref{fig:best_BAO_ratio_combined_patchy} and~\ref{fig:best_fit_rel} using 
\begin{equation}
C^{-1}_{\rm NGC+SGC} = C_{\rm NGC}^{-1} + C_{\rm SGC}^{-1}
\end{equation}
and 
\begin{equation}
C^{-1}_{\rm NGC+SGC}P(k) = C_{\rm NGC}^{-1}P_{\rm NGC}(k) + C_{\rm SGC}^{-1}P_{\rm SGC}(k),
\label{eq:iso_combine}
\end{equation}
where $C$ is the covariance matrix measured from the MultiDark-Patchy mock catalogues.

\section{Comparison to Fisher matrix forecasts}
\label{app:Fisher}

As summarized in \S \ref{sec:Fisher}, here we compare our BOSS 
measurements with the original Fisher 
forecasts for the survey \citep{Eisenstein:2011sa,Seo:2007ns} 
to see whether BOSS has performed as expected.
We start with the isotropic constraints. Using the 
MultiDark-Patchy mock catalogues we can construct the correlation matrix for 
the angle averaged distance constraint $D_V$  
\begin{equation}
R = \begin{pmatrix} 1 &  & \\
			        0.392 & 1 & \\
			        0.033 & 0.437 & 1\end{pmatrix}.
\end{equation} 
This matrix suggests that there is almost no correlation between the low and high redshift bin. We can now derive a combined isotropic constraint as
\begin{equation}
\sigma_{\alpha, \rm comb~iso} = \frac{1}{\sqrt{\sum C_{ij}^{-1}}},
\end{equation}
where $C^{-1}$ is the inverse covariance matrix, which can be obtained by 
combining the correlation matrix above with the measurement uncertainties in 
Table~\ref{tab:results}.
We obtain a combined constraint of $\sigma_{\alpha, \rm comb~iso} = 0.00643$ 
(post-reconstruction). Ignoring the middle redshift bin and assuming that the 
high and low redshift bins are not correlated, we find 
$\sigma'_{\alpha, \rm comb~ iso} = 1/\sqrt{1/\sigma^2_{z1} + 1/\sigma^2_{z3}} = 
0.00656$, which is close to the former value. This result suggests that the 
middle redshift bin does not contain much additional information. This is the
combined error from our isotropic fits to the monopole -- the 
error on $\alpha$ from the anisotropic fit, similarly computed, is a bit worse
at
$\sigma_{\alpha, \rm comb~ aniso}=0.0070$. Generally we would
expect $\sigma_{\alpha, \rm comb~ aniso}$ to correspond more directly to the 
optimally averaged dilation factor
error coming out of Fisher matrix projections \citep{Seo:2007ns}.
The difference must enter through the isotropic vs. anisotropic 
fitting details, as the  
error on the $D_V$ factor $\alpha_\parallel^{1/3}\alpha_\perp^{2/3}$ is 
only $\sim 1$\% larger than an optimally weighted dilation factor error. 
 
BOSS was originally projected to achieve
measurements of $D_A(z)$
and $H(z)$ to $1.0\%$ and $1.8\%$, respectively, at $z=0.35$, and
$1.0\%$ and $1.7\%$ at $z=0.6$ \citep{Eisenstein:2011sa}.
These original projections combine to $0.48\%$ overall
isotropic error \citep{Font:2014}, which is a factor of $1.44$ smaller than 
our measurement of $0.70\%$ (see above, where we use the aggregated 
anisotropic result).
This is a larger discrepancy than we are happy to accept without explanation
(equivalent to a factor $0.48$ reduction in survey area, for example).

Statistical fluctuations are one potential source of discrepancy between
predicted and achieved errors. For this reason
we might at first think it is better to look at results averaged over many
mocks for a more accurate gauge of the survey performance. By this criteria the
error discrepancy is actually quite a bit worse, at 1.82 times expected
(i.e., combined $\alpha$ error for mocks of 0.88\% in Table \ref{tab:patchy}).
However, as discussed
in \S \ref{sec:patchy}, 
we believe the large errors in the mocks are due to excessively
damped BAO in the mocks, and therefor do not give a realistic survey
expectation.
An alternative way to take randomness out of the achieved BAO errors
is to estimate the errors by taking approximate 2nd derivatives of $\chi^2$
with respect to model parameters
around the pre-survey expected model, in contrast to our standard estimate
which fully marginalises over all of the parameters. 
To be concrete: with
$\chi^2 = \left[\vd - \vt\left(\vtheta\right)\right]^t \vC^{-1}
\left[\vd - \vt\left(\vtheta\right)\right]$,
where $\vd$ is the data vector (band power measurement)
and $\vt(\vtheta)$ the theory predictions for
it as a function of parameter vector $\vtheta$,
our standard way of estimating errors on parameter $\theta_i$ is to use the
likelihood formula $L(\vtheta)\propto
\exp\left[-\chi^2\left(\vtheta\right)/2\right]$ to compute
$\sigma_{\theta_i}=
\left<\left(\theta_i-\bar{\theta}_i\right)^2\right>^{1/2}$,
with $\bar{\theta}_i=\left<\theta_i\right>$, by integrating over all
$\vtheta$ (e.g., by MCMC).
An alternative method is to assume the likelihood is Gaussian
in $\vtheta$ around some Taylor expansion point and invert the 2nd derivative
matrix
$\frac{d^2\chi^2}{d\theta_i d\theta_j}$
to find the implied covariance matrix for the
parameters. The standard choice of expansion point would be the maximum
likelihood point, which is equivalent to the exact likelihood integration in
the small-error limit, however, choosing this point (or doing the exact
integration) makes the results sensitive
to the actual data points (i.e., the maximum likelihood parameter vector is
$\vtheta_{{\rm max} L}=\vtheta_{{\rm max} L}(\vd)$),
which can push you into an area of parameter space with
bigger or smaller
errors (e.g., because of statistical fluctuations in the apparent shape of the
BAO feature). On the other hand, expanding around a prior expected $\vtheta$,
and approximating the 2nd derivative matrix as 
$\frac{d\vt^t}{d\theta_i}\vC^{-1} \frac{d\vt}{d\theta_j}$
gives an error manifestly insensitive to the data points 
(i.e., $\vd$), only reflecting the
covariance matrix and the structure of the theory. I.e., this is like a Fisher
matrix estimate, except with a predicted $\vC$ replaced with the
achieved one.  For this
comparison we use a slightly different broadband model which allows for more
arbitrary
fluctuations in BAO amplitude and damping (in this model, we use 
interpolation points in $k$ as the free parameters, including both an 
additive function and function that multiplies the BAO wiggles -- an advantage
of this model is that the predicted power is linear in all parameters except 
the BAO distance scale, so that we can perform exact marginalisation
over all the nuisance parameters essentially instantly, making the fits
very fast). 
The alternative model gives an error of $0.68\%$ (i.e., very similar
to our quoted $0.66\%$, 
where we use only the upper and lower redshift bins here, and fit the monopole
only -- this is also close to the 0.70\% we found from the anisotropic fit) 
with full 
marginalisation, and also $0.68\%$ when
expanding around the a priori expected model (e.g., with BAO damping given
by $50\%$ reconstruction of the non-linear damping factors of
\cite{Seo:2007ns}). The similarity of these results 
suggests that we
are not seeing significant random fluctuations in the BAO measurement error
relative to expected (note that we do use the lower bias mentioned 
below here, which has a small error driven by the broadband power spectrum and 
therefor should not contribute randomness to the BAO error).

The next level of comparison is to estimate BAO errors
given predictions for Fisher band power errors based on the survey area,
galaxy number
density, and bias as a function of redshift (i.e., not using the 
\citet{Seo:2007ns} code). After computing Fisher band power errors, we derive
BAO errors as discussed above, by taking derivatives of $\chi^2$, using the 
same theory ($\vt(\vtheta)$) that we use to fit the data but now using the 
Fisher band power errors for $\vC$.
Given the pre-survey
nominal bias $b(z)=1.7 D(0)/D(z)$ and number densities
\citep{Font:2014} over $10\,000$ deg$^{2}$., we find that band power
predictions propagate to $0.59\%$ distance error, still $23\%$ over the original
expectation but closer than the achieved $\sim 0.68$\%. 
(To be clear: the 
difference between the original projection of $0.48\%$ and $0.59\%$ is the 
difference between using the \citet{Seo:2007ns} code and a new projection
of the expected Fisher band power errors given the original survey 
parameters, propagated to BAO errors by direct application of our BAO fitting
theory instead of the approximations of \citet{Seo:2007ns}.)

It turns out that the measured bias is $\sim 0.8$ times
expected, producing larger fractional band power errors. Accounting for this in
the Fisher band power prediction results in a predicted BAO measurement 
uncertainty
of $0.68\%$. Further accounting for small differences between expected and
achieved area and number density brings this to $0.71\%$, essentially
equivalent to the measured value. On the other hand, we know that the 
measured errors could be 3\% smaller if we had infinite mocks 
(Eq. \ref{eq:hartlap}), and it appears in Figs. \ref{fig:psNGC} and 
\ref{fig:psSGC} that the power in the mocks is sometimes too high relative to
the measurement, which we estimate could be wrongly adding another $\sim 3$\%
to the measured errors, i.e., running infinite mocks with correct power would
result in measured errors $\sim 6$\% smaller than we quote. Putting these 
pieces together suggests that the band power errors estimated from the mocks
are if anything a little bit better than one would expect based on 
observed survey volume, numbers and bias. 

So with $\sim 20$\% under-achievement (0.71/0.59)
relative to pre-survey expectations accounted for by bias, number of galaxies,
and sky area (mostly the lower than expected bias) we are left to explain
the remaining $\sim 23$\% (0.59/0.48) discrepancy between
predictions based on applying our BAO fitting apparatus to Fisher band power
error predictions and based on the approximations in the
code of~\cite{Seo:2007ns} (i.e., the~\cite{Seo:2007ns} projection is optimistic
relative to both data and Fisher predicted band powers). 
The two projections (\cite{Seo:2007ns} vs. fitting Fisher band powers) 
nominally make essentially the
same assumptions, i.e., are supposed to be just different ways of numerically
evaluating the same basic
Fisher matrix equation, so there is no reason to expect this kind of difference,
but the \cite{Seo:2007ns} code employs approximations and calibration
of the signal strength that make a very direct comparison difficult. 
Surprisingly, however, it seems that the difference is mostly accounted for
by the difference between our fiducial cosmology and the WMAP3 cosmology 
hard-coded in the \cite{Seo:2007ns} code. 
The Fisher band-power BAO error based on the WMAP3 model, at fixed 
total number of observed galaxies and observed power amplitude, is 0.51\% --
within the margin of fine details of the 0.48\% of \cite{Seo:2007ns}.  
A 16\% (0.59/0.51) change between cosmologies may seem surprising, but
recall that the WMAP3 cosmology had $\sim 12$\% higher baryon to CDM 
density ratio than the current standard, with correspondingly larger BAO 
signal amplitude, which very simply accounts for most of the difference. 

Now we want to look at the anisotropic constraints in the form of $\alpha_{\perp}$ and $\alpha_{\parallel}$. The combined covariance matrix for $\alpha_{\perp}$ and $\alpha_{\parallel}$ is 
\begin{align}
C_{z1+z3} &= \left[C^{-1}_{z1} + C^{-1}_{z3}\right]^{-1}\\
&= \begin{pmatrix} \sigma^2_{\alpha_{\parallel}} & r\sigma_{\alpha_{\perp}}\sigma_{\alpha_{\parallel}}\\
		        r\sigma_{\alpha_{\perp}}\sigma_{\alpha_{\parallel}}  & \sigma^2_{\alpha_{\perp}} \end{pmatrix} \\
			    &=    10^{-4}\begin{pmatrix} 3.1 & -0.83\\
			        -0.83 & 1.2\end{pmatrix}.
\label{eq:cov_combined}
\end{align} 
This produces a $1.76\%$ constraint on $\alpha_{\parallel}$ ($\propto 1/H$) 
and a $1.09\%$ constraint on $\alpha_{\perp}$ ($\propto D_A$) with a 
correlation of $r=-0.42$ (post-reconstruction). 
Compared to the two $1\%$ $D_A$ measurements originally predicted by 
\cite{Eisenstein:2011sa}, and $1.7\%$ \& $1.8\%$ predicted $H$ measurements, 
the 
measurement errors are $1.54$ and $1.42$ times expected for $D_A$ and $H$, 
respectively. This is similar to the $1.44$ times expected that we 
found for $\alpha$. We speculate that the reduction in signal due to bias and
cosmology leads to more degradation of the measurement in the transverse than
radial direction because the radial direction is boosted by redshift space
distortions.  

\label{lastpage}

\end{document}